\documentclass[graybox,natbib,norunningheads]{svmult}

\usepackage{mathptmx}       
\usepackage{helvet}         
\usepackage{courier}        
\usepackage{type1cm}        
\usepackage{makeidx}         
\usepackage{graphicx}        
\usepackage{multicol}        
\usepackage[bottom]{footmisc}

\makeindex             

\newcommand{\bea}{\begin{eqnarray}}
\newcommand{\eea}{\end{eqnarray}}
\newcommand{\bc}{\begin{center}}
\newcommand{\ec}{\end{center}}

\newcommand{\dd}{{\rm d}}

\usepackage{mathtools}
\usepackage{amssymb,amsmath}
\usepackage{listings}

\usepackage{fancyhdr}
\renewcommand{\sectionmark}[1]{\markright{\thesection.\ #1}}
\begin{document}

\lstset{language=C, basicstyle=\small\ttfamily,
numbers=left, numberstyle=\tiny, stepnumber=1, numbersep=-5pt}

\title*{High performance computing and numerical modelling}
\titlerunning{Numerical modelling}
\author{Volker Springel}
\authorrunning{Volker Springel}
\institute{Volker Springel \at 
Heidelberg Institute for Theoretical Studies, Schloss-Wolfsbrunnenweg
35, 69118 Heidelberg,\\
and  Heidelberg University, Zentrum f\"{u}r Astronomie, Astronomisches Recheninstitut, 
M\"{o}nchhofstr. 12-14, 69120 Heidelberg, Germany, \email{volker.springel@h-its.org}}

\maketitle

Lecture Notes\\
\ \\
{\bf 43rd Saas-Fee Course}\\ 
{\rm Star formation in galaxy evolution: connecting
models to reality}

\tableofcontents

\section{Preamble}

Numerical methods play an ever more important role in
astrophysics. This can be easily demonstrated through a cursory
comparison of a random sample of paper abstracts from today and 20
years ago, which shows that a growing fraction of studies in astronomy
is based, at least in part, on numerical work. This is especially true
in theoretical works, but of course, even in purely observational
projects, data analysis without massive use of computational methods
has become unthinkable. For example, cosmological inferences of large
CMB experiments routinely use very large Monte-Carlo simulations as
part of their Baysian parameter estimation.

The key utility of computer simulations comes from their ability to
solve complex systems of equations that are either intractable with
analytic techniques or only amenable to highly approximative
treatments. Thanks to the rapid increase of the performance of
computers, the technical limitations faced when attacking the
equations numerically (in terms of calculational time, memory use,
numerical resolution, etc.) become progressively smaller. But it is
important to realize that they will always stay with us at some
level. Computer simulations are therefore best viewed as a powerful
complement to analytic reasoning, and as the method of choice to model
systems that feature enormous physical complexity -- such as star
formation in evolving galaxies, the topic of this \emph{43rd Saas Fee
  Advanced Course}.

The organizers asked me to lecture about \emph{High performance
  computing and numerical modelling} in this winter school, which took
place March 11-16, 2013, in Villars-sur-Ollon, Switzerland.  As my
co-lecturers Ralf Klessen und Nick Gnedin should focus on the physical
processes in the interstellar medium and on galactic scales, my task
was defined as covering the basics of numerically treating gravity and
hydrodynamics, and on making some remarks on the use of high
performance computing techniques in general. In a nutshell, my
lectures hence intend to cover the basic numerical methods necessary
to simulate evolving galaxies. This is still a vast field, and I
necessarily had to make a selection of a subset of the relevant
material. I have tried to strike a compromise between what I
considered most useful for the majority of students and what I could
cover in the available time.

In particular, my lectures concentrate on techniques to compute
gravitational dynamics of collisionless fluids composed of dark matter
and stars in galaxies. I also spend a fair amount of time explaining
basic concepts of various solvers for Eulerian gas dynamics. Due to
lack of time, I am not discussing collisional N-body dynamics as
applicable to star cluster, and I omit a detailed discussion of
different schemes to implement adaptive mesh refinement.

The written notes presented here quite closely follow the lectures as
held in Villars-sur-Ollon, apart from being expanded somewhat in
detail where this seemed adequate. I note that the shear breadth of
the material made it impossible to include detailed mathematical
discussions and proofs of all the methods. The discussion is therefore
often at an introductory level, but hopefully still useful as a
general overview for students working on numerical models of
galaxy evolution and star formation. Interested readers are referred
to some of the references for a more detailed and mathematically sound
exposition of the numerical techniques.

\section{Collisionless N-body dynamics}

According to the $\Lambda$CDM paradigm, the matter density of our
Universe is dominated by {\em dark matter}, which is thought to be
composed of a yet unidentified, non-baryonic elementary particle
\citep[e.g.][]{Bertone2005}. A full description of the dark mass in a
galaxy would hence be based on following the trajectories of each dark
matter particle -- resulting in a gigantic N-body model. This is
clearly impossible due to the large number of particles
involved. Similarly, describing all the stars in a galaxy as point
masses would require of order $10^{11}$ bodies. This may come within
reach in a few years, but at present it is still essentially
infeasible. In this section we discuss why we can nevertheless
describe both of these galactic components as discrete N-body systems,
but composed of far fewer particles than there are in reality.

\subsection{The hierarchy of particle distribution functions}

The state of an $N$-particle ensemble at time $t$ can be specified by
the {\em exact} particle distribution function \citep{Hockney1988}, in the form
\begin{equation}
F(\vec{r}, \vec{v}, t) = \sum_{i=1}^N \delta(\vec{r} - \vec{r}_i(t))\cdot 
\delta(\vec{v} - \vec{v}_i(t)),
\end{equation}
where $\vec{r}_i$ and $\vec{v}_i$ denote the position and velocity of
particle $i$, respectively.  This effectively gives the number density
of particles at phase-space point $(\vec{r}, \vec{v})$ at time $t$. Let
now
\begin{equation}
p(\vec{r}_1, \vec{r}_2, \ldots, \vec{r}_N, \vec{v}_1, \vec{v}_2,
\ldots \vec{v}_N)\, {\rm d}\vec{r}_1\,{\rm d}\vec{r}_2 \cdots {\rm
  d}\vec{r}_N
\, {\rm d}\vec{v}_1\,{\rm d}\vec{v}_2 \cdots {\rm
  d}\vec{v}_N ,
\end{equation}
be the probability that the system is in the given state at time
$t$. Then a reduced statistical description is obtained by {\em
  ensemble averaging}:
\begin{equation}
f_1(\vec{r}, \vec{v}, t)  = \left< F(\vec{r}, \vec{v}, t)\right> =
\int F\cdot p \cdot
{\rm d}\vec{r}_1\,{\rm d}\vec{r}_2 \cdots {\rm
  d}\vec{r}_N
\, {\rm d}\vec{v}_1\,{\rm d}\vec{v}_2 \cdots {\rm
  d}\vec{v}_N .
\end{equation}
We can integrate out one of the Dirac delta-functions in $F$ to obtain
\begin{equation}
f_1(\vec{r}, \vec{v}, t)  =
N \int  
p(\vec{r}, \vec{r}_2, \ldots, \vec{r}_N, \vec{v}, \vec{v}_2,
\ldots \vec{v}_N)\, {\rm d}\vec{r}_2 \cdots {\rm
  d}\vec{r}_N
\, {\rm d}\vec{v}_2 \cdots {\rm
  d}\vec{v}_N .
\label{integrateout}
\end{equation}
Note that as all particles are equivalent we can permute the arguments
in $p$ where $\vec{r}$ and $\vec{v}$ appear.  $f_1(\vec{r}, \vec{v},
t) \,{\rm d}\vec{r}\,{\rm d}\vec{v}$ now gives the {\em mean number}
of particles in a phase-space volume ${\rm d}\vec{r}\,{\rm d}\vec{v}$
around $(\vec{r}, \vec{v})$.

Similarly, the ensemble-averaged two-particle distribution (``the mean
product of the numbers of particles at $(\vec{r}, \vec{v})$ and
$(\vec{r}', \vec{v}')$'') is given by
\begin{eqnarray}
 & & f_2(\vec{r}, \vec{v}, \vec{r}', \vec{v}', t)  =  \left< F(\vec{r},
    \vec{v}, t) F(\vec{r}', \vec{v}', t)\right> \\
& & = 
N (N-1) \int  
p(\vec{r}, \vec{r}', \vec{r}_3, \ldots, \vec{r}_N, \vec{v}, \vec{v}', \vec{v}_3,
\ldots \vec{v}_N)\, {\rm d}\vec{r}_3 \cdots {\rm
  d}\vec{r}_N
\, {\rm d}\vec{v}_3 \cdots {\rm
  d}\vec{v}_N .  \nonumber
\end{eqnarray}
Likewise one may define $f_3, f_4, \ldots$ and so on. This yields the
so-called BBGKY (Bogoliubov-Born-Green-Kirkwood-Yvon) chain
\citep[e.g.][]{Kirkwood1946}, see also \citet{Hockney1988} for a
detailed discussion.

\runinhead{Uncorrelated (collisionless) systems} The simplest closure
for the BBGKY hierarchy is to assume that particles are {\em
  uncorrelated}, i.e.~that we have
\begin{equation}
f_2(\vec{r}, \vec{v}, \vec{r}', \vec{v}', t)
=
f_1(\vec{r}, \vec{v}, t)\, f_1(\vec{r}', \vec{v}', t).
\end{equation}
Physically, this means that a particle at $(\vec{r}, \vec{v})$ is
completely unaffected by one at $(\vec{r}', \vec{v}')$. Systems in
which this is approximately the case include stars in a galaxy, dark
matter particles in the universe, or electrons in a plasma.  We will
later consider in more detail under which conditions a system is
collisionless.

Let's now go back to the probability density $p(\vec{w})$ which
depends on the $N$-particle phase-space state $\vec{w} = (\vec{r}_1,
\vec{r}_2, \ldots, \vec{r}_N, \vec{v}_1, \vec{v}_2, \ldots
\vec{v}_N)$. The conservation of probability in phase-space means that
it fulfills a continuity equation
\begin{equation}
  \frac{\partial p}{\partial t} + \nabla_{\vec{w}} \cdot (p \, \vec{\dot{w}}) = 0.
\end{equation}
We can cast this into
\begin{equation}
\frac{\partial p}{\partial t} + \sum_i \left(
p\frac{\partial \vec{\dot{r}}_i}{\partial \vec{r}_i} 
+ \frac{\partial p}{\partial \vec{r}_i} \vec{\dot{r}}_i
+p\frac{\partial \vec{\dot{v}}_i}{\partial \vec{v}_i} 
+ \frac{\partial p}{\partial \vec{v}_i} \vec{\dot{v}}_i
\right) = 0.
\end{equation}
Because only conservative gravitational fields are involved, the
system is described by classical mechanics as a so-called Hamiltonian
system.  Recalling the equations of motion $\vec{\dot{r}} =
\frac{\partial H}{\partial \vec{p}}$ and $\vec{\dot{p}} =
-\frac{\partial H}{\partial \vec{r}}$ of Hamiltonian dynamics
\citep{Goldstein1950}, we can differentiate them to get
$\frac{\partial \vec{\dot{r}}}{\partial \vec{r}} = \frac{\partial^2
  H}{\partial \vec{r}\partial \vec{p}}$, and $\frac{\partial
  \vec{\dot{p}}}{\partial \vec{p}} = -\frac{\partial^2 H}{\partial
  \vec{r}\partial \vec{p}}$. Hence it follows $ \frac{\partial
  \vec{\dot{r}}}{\partial \vec{r}} = - \frac{\partial
  \vec{\dot{v}}}{\partial \vec{v}}$. Using this we get
\begin{equation}
\frac{\partial p}{\partial t} + \sum_i \left(
\vec{v}_i \frac{\partial p}{\partial \vec{r}_i} 
+ \vec{a}_i \frac{\partial p}{\partial \vec{v}_i} 
\right) = 0,
\end{equation}
where $\vec{a}_i = \vec{\dot{v}}_i = {\vec{F}_i}/{m_i}$ is the
particle acceleration and $m_i$ is the particle mass. This is {\em Liouville's theorem}.

Now, in the collisionless/uncorrelated limit, this directly carries
over to the one-point distribution function $f=f_1$ if we integrate
out all particle coordinates except for one as in
equation~(\ref{integrateout}), yielding the {\em Vlasov equation},
also known as collisionless Boltzmann equation:
\begin{equation}
\frac{\partial f}{\partial t} + 
\vec{v} \frac{\partial f}{\partial \vec{r}} 
+ \vec{a} \frac{\partial f}{\partial \vec{v}}  = 0.
\end{equation}
The close relation to Liouville's equation means that also here the
phase space-density stays constant along characteristics of the system
(i.e.~along orbits of individual particles).

\runinhead{What about the acceleration?}  In the limit of a
collisionless system, the acceleration $\vec{a}$ in the above equation
cannot be due to another single particle, as this would imply local
correlations. However, \emph{collective effects}, for example from the
gravitational field produced by the whole system are still allowed.

For example, the source field of self-gravity (i.e.~the mass density)
can be described as
\begin{equation}
\rho(\vec{r}, t) = m \int f(\vec{r}, \vec{v}, t) \,{\rm d}\vec{v}.
\end{equation}
This then produces a gravitational field through Poisson's equation,
\begin{equation}
\nabla^2 \Phi(\vec{r}, t) = 4\pi G \rho(\vec{r}, t), 
\end{equation}
which gives the accelerations as
\begin{equation}
\vec{a} = - \frac{\partial \Phi}{\partial \vec{r}}. 
\end{equation}
One can also combine these equations to yield the 
Poisson-Vlasov system, given by
\begin{equation}
\frac{\partial f}{\partial t} + 
\vec{v} \frac{\partial f}{\partial \vec{r}} 
- \frac{\partial \Phi}{\partial \vec{r}} \frac{\partial f}{\partial \vec{v}}  = 0,
\end{equation}
\begin{equation}
\nabla^2 \Phi = 4\pi G 
m \int f(\vec{r}, \vec{v}, t) \,{\rm d}\vec{v}.
\end{equation}
This holds in an analogous way also for a plasma where the mass
density is replaced by a charge density.

It is interesting to note that in this description the particles have
basically completely vanished and have been replaced with a continuum
fluid description. Later, for the purpose of solving the equations, we
will have to reintroduce particles as a means of discretizing the
equations -- but these are then not the real physical particles any
more, rather they are fiducial macro particles that sample the
phase-space in a Monte-Carlo fashion.

\subsection{The relaxation time -- When is a system collisionless?}

Consider a system of size $R$ containing $N$ particles. The time for
one crossing of a particle through the system is of order
\begin{equation}
t_{\rm cross} = \frac{R}{v},
\end{equation}
where $v$ is the typical particle velocity \citep{Binney1987,
  Binney2008}. For a self-gravitating system of that size we expect
\begin{equation}
v^2 \simeq \frac{G N m}{R} = \frac{GM}{R},
\label{eqnv2sqrd}
\end{equation}
where $M = N\, m$ is the total mass.

\begin{figure}[t]
  \sidecaption
  \resizebox{7cm}{!}{\includegraphics{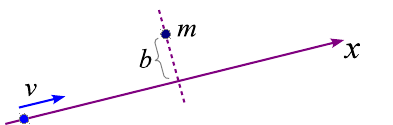}}
  \caption{Sketch of a two-body encounter, in which a particle passes
    another particle (assumed to be at rest) with impact parameter $b$
    and velocity $v$.}
\label{fig_sketch_crossingtime}
\end{figure}

We now want to estimate the rate at which a particle experiences weak
deflections by other particles, which is the process that violates
perfect collisionless behavior and which induces relaxation. We
calculate the deflection in the impulse approximation where the
particle's orbit is taken as a straight path, as sketched in
Fig.~\ref{fig_sketch_crossingtime}.

To get the deflection, we compute the transverse momentum acquired by
the particle as it flies by the perturber (assumed to be stationary
for simplicity):
\begin{equation}
\Delta p = m \Delta v = \int F_{\perp} {\rm d} t = \int \frac{G
  m^2}{x^2 + b^2}\frac{b}{\sqrt{x^2 + b^2}} \frac{{\rm d}x}{v}
= \frac{2Gm^2}{bv}.
\label{eqndefl}
\end{equation}

How many encounters do we expect in one crossing? For impact
parameters between $[b, b+{\rm d}b]$ we have
\begin{equation}
{\rm d}n = N \frac{2\pi b\, {\rm d}b}{\pi R^2}
\end{equation}
targets. The velocity perturbations from each encounter have random
orientations, so they add up in quadrature. Per crossing we hence have
for the quadratic velocity perturbation:
\begin{equation}
(\Delta v)^2 = \int \left(
\frac{2Gm}{b v}\right)^2 {\rm d}n =  8 N \left(\frac{Gm }{R
  v}\right)^2 \ln \Lambda,
\end{equation}
where
\begin{equation}
\ln \Lambda = \ln \frac{b_{\rm max}}{b_{\rm min}}
\end{equation}
is the so-called Coulomb logarithm, and $b_{\rm max}$ and $b_{\rm
  min}$ are the adopted integration limits. We can now define the relaxation
time as 
\begin{equation}
t_{\rm relax} \equiv  \frac{v^2}{(\Delta v)^2 / t_{\rm cross}},
\end{equation}
i.e.~after this time the individual perturbations have reached $\sim
100\%$ of the typical squared velocity, and one can certainly not
neglect the interactions any more. With our result for $(\Delta v)^2$,
and using equation~(\ref{eqnv2sqrd}) this now becomes
\begin{equation}
t_{\rm relax} = \frac{N}{8 \ln \Lambda}\, t_{\rm cross}.
\end{equation}
But we still have to clarify what we can sensibly use for $b_{\rm
  min}$ and $b_{\rm max}$ in the Coulomb logarithm. For $b_{\rm max}$,
we can set the size of the system, i.e.~$b_{\rm max} \simeq R$. For
$b_{\rm min}$, we can use as a lower limit the $b$ where very strong
deflections ensue, which is given by
\begin{equation}
\frac{2 G m}{b_{\rm min} v}
\simeq v ,
\end{equation}
i.e~where the transverse velocity perturbation becomes as large as the
velocity itself (see equation~\ref{eqndefl}). This then yields $b_{\rm
  min} = 2 R / N$. We hence get for the Coulomb logarithm $\ln \Lambda
\simeq \ln (N/2)$. But a factor of 2 in the logarithm might as well be
neglected in this coarse estimate, so that we obtain $\ln \Lambda \sim
N$. We hence arrive at the final result \citep{Chandrasekhar1943}:
\begin{equation}
t_{\rm relax} = \frac{N}{8 \ln N}\, t_{\rm cross}.
\end{equation}

A system can be viewed as collisionless if $t_{\rm relax} \gg t_{\rm
  age}$, where $t_{\rm age}$ is the time of interest. We note that
$t_{\rm cross}$ depends only on the size and mass of the system, but
{\em not} on the particle number $N$ or the individual masses of the
N-body particles. We therefore clearly see that the primary
requirement to obtain a collisionless system is to use a sufficiently
large $N$.

\runinhead{Examples:}
\begin{itemize}
\item globular star clusters have $N\sim 10^5$, $t_{\rm cross} \sim
  \frac{3\,{\rm pc}}{6\,{\rm km/sec}} \simeq 0.5 \,{\rm Myr}$.  This
  implies that such systems are strongly affected by collisions over
  the age of the Universe, $t_{\rm age} = \frac{1}{H_0} \sim 10\,{\rm
    Gyr}$, where $H_0$ is the Hubble constant.
\item stars in a typical galaxy: Here we have $N\sim 10^{11}$ and
  $t_{\rm cross} \sim \frac{1}{100\,H_0}$. This means that these large
  stellar systems are collisionless over the age of the Universe to
  extremely good approximation.
\item dark matter in a galaxy: Here we have $N\sim 10^{77}$ if the
  dark matter is composed of a $\sim 100\,{\rm GeV}$ weakly
  interacting massive particle (WIMP). In addition, the crossing time
  is longer than for the stars, $t_{\rm cross} \sim
  \frac{1}{10\,H_0}$, due to the larger size of the `halo' relative to
  the embedded stellar system. Clearly, dark matter represents the
  cr\`{e}me de la cr\`{e}me of collisionless systems.
\end{itemize}

\subsection{N-body models and gravitational softening}

We now reintroduce particles in order to discretize the collisionless
fluid described by the Poisson-Vlasov system. We use however {\em far
  fewer} particles than in real physical systems, and we
correspondingly give them a higher mass. These are hence fiducial
macro-particles. Their equations of motion in the case of gravity take
on the form:
\begin{equation}
\vec{\ddot{x}}_i = -\nabla_i \Phi(\vec{r}_i),
\end{equation}
\begin{equation}
\Phi(\vec{r}) = -G \sum_{j=1}^N \frac{m_j}{[ (\vec{r} - \vec{r}_j)^2 +
  \epsilon^2]^{1/2}} . \label{eqnPoisson}
\end{equation}
A few comments are in order here:
\begin{itemize}
\item Provided we can ensure $t_{\rm relax} \gg t_{\rm sim}$, where
  $t_{\rm sim}$ is the simulated time-space, the numerical model keeps
  behaving as a collisionless system over $t_{\rm sim}$ despite a
  smaller $N$ than in the real physical system. In this limit, the
  collective gravitational potential is sufficiently smooth.
\item Note that the mass of a macro-particle used to discretize the
  collision system drops out from its equation of motion (because
  there is no self-force). Provided there are enough particles to
  describe the gravitational potential accurately, the orbits of the
  macro-particles will be just as valid as the orbits of the real
  physical particles.
\item The N-body model gives only one (quite noisy) realization of the
  one-point function. It does not give the ensemble average directly
  (this would require multiple simulations).
\item The equations of motion contain a {\bf softening length}
  $\epsilon$.  The purpose of the force softening is to avoid large
  angle scatterings and the numerical expense that would be needed to
  integrate the orbits with sufficient accuracy in singular
  potentials. Also, we would like to prevent the possibility of the
  formation of bound particle pairs -- they would obviously be highly
  correlated and hence strongly violate collisionless behavior. We
  don't get bound pairs if
\begin{equation}
 \left<v^2\right> \gg \frac{G m }{\epsilon},
\end{equation}
which can be viewed as a necessary (but not in general sufficient)
condition on reasonable softening settings \citep{Power2003}. The
adoption of a softening length also implies the introduction of a
smallest resolved length-scale. The specific softening choice one
makes ultimately represents a compromise between spatial resolution,
discreteness noise in the orbits and the gravitational potential,
computational cost, and the relaxation effects that adversely
influence results.

\end{itemize}

\subsection{N-body equations in cosmology}

In cosmological simulations, it is customary to use comoving
coordinates $\vec{x}$ instead of physical coordinates $\vec{r}$. The
two are related by
\begin{equation}
\vec{r} = a(t)\, \vec{x},
\end{equation}
where $a = 1 / (1+z)$ is the cosmological scale factor. Its evolution
is governed by the Hubble rate
\begin{equation}
\frac{\dot{a}}{a} = H(a),
\end{equation}
which in turn is given by $H(a) = [\Omega_0 a^{-3} + (1-\Omega_0 -
\Omega_\Lambda) a^{-2} + \Omega_\Lambda]^{1/2}$ in standard
Friedmann-Lemaitre models \citep[e.g.][]{Peacock1999,Mo2010}.

In an (infinite) expanding space, modelled through period replication
of a box of size $L$, one can then show
\citep[e.g.][]{Springel2001gadget} that the Newtonian equations of
motion in comoving coordinates can be written as
\begin{equation}
\frac{{\rm d}}{{\rm d}t} ( a^2 \vec{\dot{x}}) = - \frac{1}{a}\nabla_i \phi(\vec{x}_i),
\end{equation}
\begin{equation}
\nabla^2\phi(\vec{x}) = 4 \pi G \sum_i m_i \left[-\frac{1}{L^3} +
  \sum_{\vec{n}} \delta(\vec{x} -\vec{x}_i - \vec{n}L)\right], 
\end{equation}
where the sum over $i$ extends over $N$ particles in the box, and
$\phi$ is the \emph{peculiar gravitational potential}. It corresponds
to the Newtonian potential of density deviations around a constant
mean background density. Note that the sum over all particles for
calculating the potential extends also over all of their period
images, with $\vec{n} =(n_1,n_2, n_3)$ being a vector of integer
triples. The term $-1/L^3$ is simply needed to ensure that the mean
density sourcing the Poisson equation vanishes, otherwise there would
be no solution for an infinite space.

\subsection{Calculating the dynamics of an N-body system}

Once we have discretized a collisionless fluid in terms of an N-body
system, two questions come up:
\begin{enumerate}
\item How do we integrate the equations of motion in time?
\item How do we compute the right hand side of the equations of
  motion, i.e.~the gravitational forces?
\end{enumerate}

For the first point, we can use an integration scheme for ordinary
differential equations, preferably a symplectic one since we are
dealing with a Hamiltonian system. We shall briefly discuss elementary
aspects of these time integration methods in the following
section. 

The second point seems also straightforward at first, as the
accelerations (forces) can be readily calculated through {\em direct
  summation}. In the isolated case this reads as
\begin{equation}
\vec{\ddot{r}}_i = -G \sum_{j=1}^N \frac{m_j}{[ (\vec{r}_i - \vec{r}_j)^{2} +
  \epsilon^2]^{3/2}} (\vec{r}_i - \vec{r}_j).
\end{equation}
For a periodic space, the force kernel is slightly different but in
principle the same summation applies \citep{Hernquist1991}.  This
calculation is {\em exact}, but for each of the $N$ equations we have
to calculate a sum with $N$ partial forces, yielding a computational
cost of order ${\cal O}(N^2)$. This quickly becomes prohibitive for
large $N$, and causes a conflict with our urgent need to have a large
$N$!

Perhaps a simple example is in order to show how bad the $N^2$ scaling
really is in practice. Suppose you can do $N=10^6$ in a month of
computer time, which is close to the maximum that one may want to do
in practice. A particle number of $N=10^{10}$ would then already take
of order 10 million years.

We hence need faster, \emph{approximative} force calculation
schemes. We shall discuss a number of different possibilities for this
in Section~\ref{SecGrav}, namely:
\begin{itemize}
\item Particle-mesh (PM) algorithms
\item Fourier-transform based solvers of Poisson's equations 
\item Iterative solvers for Poisson's equation (multigrid-methods)
\item Hierarchical multipole methods (``tree-algorithms'')
\item So-called TreePM methods
\end{itemize}
Various combinations of these approaches may also be used, and
sometimes they are also applied together with direct summation on
small scales. The latter may also be accelerated with special-purpose
hardware \citep[e.g.~the GRAPE board;][]{Makino2003}, or with graphics
processing units (GPUs) that are used as fast number-crushers.

\section{Time integration techniques}

We discuss in the following some basic methods for the integration of
{\em ordinary differential equations} (ODEs). These are relations
between an unknown scalar or vector-values function $\vec{y}(t)$ and
its derivatives with respect to an independent variable, $t$ in this
case (the following discussion associates the independent variable
with `time', but this could of course be also any other
quantity). Such equations hence formally take the form
\begin{equation}
\frac{{\rm d}\vec{y}}{{\rm d}t}  =
\vec{f} (\vec{y}, t),
\end{equation}
and we seek the solution $\vec{y}(t)$, subject to boundary conditions.

Many simple dynamical problems can be written in this form, including
ones that involve second or higher derivatives. This is done through a
procedure called \emph{reduction to 1st order}. One does this by
adding the higher derivatives, or combinations of them, as further
rows to the vector $\vec{y}$.

For example, consider a simple pendulum of length $l$ with the
equation of motion
\begin{equation}
\ddot{q} = -\frac{g}{l} \sin(q), 
\end{equation}
where $q$ is the angle with respect to the vertical.  Now define
$p\equiv \dot q$, yielding a state vector
\begin{equation}
\vec{y} \equiv\left(
\begin{array}{c}
q\\
p
\end{array}
\right),
\end{equation}
and a first order ODE of the form:
\begin{equation}
  \frac{{\rm d}\vec{y}}{{\rm d}t}  = \vec{f}(\vec{y})
  = \left(
\begin{array}{c}
p\\
-\frac{g}{l} \sin(q)
\end{array}
\right).
\end{equation}

A numerical approximation to the solution of an ODE is a set of values
$\{y_0$, $y_1$, $y_2$, $\ldots\}$ at discrete times $\{t_0$, $t_1$,
$t_2$, $\ldots\}$, obtained for certain boundary conditions. The most
common boundary condition for ODEs is the \emph{initial value problem}
(IVP), where the state of $\vec{y}$ is known at the beginning of the
integration interval. It is however also possible to have mixed
boundary conditions where $\vec{y}$ is partially known at both ends of
the integration interval.

There are many different methods for obtaining a discrete solution of
an ODE system \citep[e.g.][]{Press1992}. We shall here discuss some of
the most basic ones, restricting ourselves to the IVP problem, for
simplicity, as this is the one naturally appearing in cosmological simulations.

\subsection{Explicit and implicit Euler methods}

\runinhead{Explicit Euler} This solution method, sometimes also called
``forward Euler'', uses the iteration
\begin{equation}
y_{n+1} = y_n + f(y_n) \Delta t,
\end{equation}
where $y$ can also be a vector. $\Delta t$ is the integration step. 
\begin{itemize}
\item 
This approach is the simplest of all.
\item The method is called {\em explicit} because $y_{n+1}$ is
  computed with a right-hand-side that only depends on quantities that are
  already known.
\item The stability of the method can be a sensitive function of the
  step size, and will in general only be obtained for a sufficiently
  small step size.
\item It is recommended to refrain from using this scheme in practice,
  since there are other methods that offer higher accuracy at the same
  or lower computational cost. The reason is that the Euler method is
  only {\em first order accurate}. To see this, note that the
  truncation error in a single step is of order ${\cal O}_s(\Delta
  t^2)$, which follows simply from a Taylor expansion. To integrate
  over a time interval $T$, we need however $N_s = T/\Delta t$ steps,
  producing a total error that scales as $N_s {\cal O}_s(\Delta t^2) =
  {\cal O}_T(\Delta t)$.
\item The method is also not time-symmetric, which makes it prone to
  accumulation of secular integration errors.
\end{itemize}

We remark in passing that for a method to reach a global error that
scales as ${\cal O}_T(\Delta t^n)$ (which is then called an ``$n^{\rm
  th}$ order accurate'' scheme), a local truncation error of one order
higher is required, i.e.~${\cal O}_s(\Delta t^{n+1})$.

\runinhead{Implicit Euler} In a so-called ``backwards Euler'' scheme,
one uses
\begin{equation}
  y_{n+1} = y_n + f(y_{n+1}) \Delta t,
\end{equation}
which seemingly represents only a tiny change compared to the explicit
scheme.
\begin{itemize}
\item This approach has excellent stability properties, and for some
  problems, it is in fact essentially always stable even for extremely
  large timestep. Note however that the accuracy will usually
  nevertheless become very bad when using such large steps.

\item This stability property makes implicit Euler sometimes useful
  for \emph{stiff equations}, where the derivatives (suddenly) can
  become very large.

\item The implicit equation for $y_{n+1}$ that needs to be solved here
  corresponds in many practical applications to a non-linear equation
  that can be complicated to solve for $y_{n+1}$. Often, the root of
  the equation has to be found numerically, for example through an
  iterative technique.
\item
The method is still first order accurate, and also lacks
time-symmetry, just like the explizit Euler scheme.
\end{itemize}

\runinhead{Implicit midpoint rule}
If we use 
\begin{equation}
y_{n+1} = y_n + f\left(\frac{y_n + y_{n+1}}{2}\right) \Delta t,
\end{equation}
we obtain the implicit midpoint rule, which can be viewed as a
symmetrized variant of explicit and implicit Euler.  This is {\em
  second order accurate}, but still implicit, so difficult to use in
practice. Interestingly, it is also time-symmetric, i.e.~one can
formally integrate backwards and recover exactly the same steps
(modulo floating point round-off errors) as in a forward integration.

\subsection{Runge-Kutta methods}

The Runge-Kutta schemes form a whole class of versatile integration
methods \citep[e.g.][]{Aatkinson1978, stoer2002}. Let's derive one of the
simplest Runge-Kutta schemes.
\begin{enumerate}
\item We start from the exact solution,
\begin{equation}
y_{n+1} = y_n + \int_{t_n}^{t_{n+1}} f\left(y(t)\right) {\rm d}t .
\end{equation}

\item Next, we approximate the integral with the (implicit)
  trapezoidal rule:
\begin{equation}
y_{n+1} = y_n + \frac{f(y_n) + f(y_{n+1})}{2} \Delta t .
\end{equation}
\item \citet{Runge1895} proposed to predict the unknown
  $y_{n+1}$ on the right hand side by an Euler step, yielding a {\em
    2nd order accurate Runge-Kutta scheme}, sometimes also called
  predictor-corrector scheme:
\begin{eqnarray}
k_{1} & = & f(y_{n}, t_n) ,  \label{eqnpred} \\ 
k_{2} & = & f(y_{n} + k_1 \Delta t, t_{n+1}) ,
\label{eqncorr} 
  \\
y_{n+1} & = & \frac{k_1 + k_{2}}{2} \Delta t .
\end{eqnarray}
Here the step done with the derivate of equation~(\ref{eqnpred}) is
called the `predictor' and the one done with equation~(\ref{eqncorr})
is the corrector step.
\end{enumerate}

\runinhead{Higher order Runge-Kutta schemes} A variety of further
Runge-Kutta schemes of different order can be defined. Perhaps the
most commonly used is the classical $4^{\rm th}$-order Runge-Kutta
scheme:
\begin{eqnarray}
k_{1} & = & f(y_{n}, t_n) , \\
k_{2} & = & f\left(y_{n} + k_1 \frac{\Delta t}{2}, t_{n}+\frac{\Delta
    t}{2}\right) , \\
k_{3} & = & f\left(y_{n} + k_2 \frac{\Delta t}{2}, t_{n}+\frac{\Delta
    t}{2}\right) , \\
k_{4} & = & f\left(y_{n} + k_3 \Delta t, t_{n}+ \Delta
    t \right) .
\end{eqnarray}
These four function evaluations per step are then combined in a
weighted fashion to carry out the actual update step:
\begin{equation}
y_{n+1} = y_n + \left(\frac{k_1}{6} + \frac{k_2}{3} + \frac{k_3}{3} +
  \frac{k_4}{6}\right)
\Delta t + {\cal O}(\Delta t^5) .
\end{equation}
We note that the use of higher order schemes also entails more
function evaluations per step, i.e.~the individual steps become more
complicated and expensive. Because of this, higher order schemes are
not always better; they usually are up to some point, but sometimes
even a simple second-order accurate scheme can be the best choice for
certain problems.

\subsection{The leapfrog}

Suppose we have a second order differential equation of the type
\begin{equation}
\ddot{x} = f(x).
\end{equation}
This could of course be brought into standard form, $\dot{\vec{y}} =
\vec{\tilde f}(\vec{y})$, by defining something like $\vec{y} = (x,
\dot{x})$ and $\vec{\tilde f} = (\dot x, f(x))$, followed by applying
a Runge-Kutta scheme as introduced above.

However, there is also another approach in this case, which turns out
to be particularly simple and interesting. Let's define $v\equiv \dot
x$. Then the so-called Leapfrog integration scheme is the mapping
$(x_n, v_n) \to (x_{n+1}, v_{n+1})$ defined as:
\begin{eqnarray}
v_{n+\frac{1}{2}} & = & v_n + f(x_n) \frac{\Delta t}{2} ,\\
x_{n+1} & = & x_n + v_{n+\frac{1}{2}}\, \Delta t , \\
v_{n+1} & = & v_{n+\frac{1}{2}} + f(x_{n+1}) \frac{\Delta t}{2} . 
\end{eqnarray}
\begin{itemize}
\item This scheme is 2nd-order accurate (proof through Taylor
  expansion).
\item It requires only 1 evaluation of the right hand side per step
  (note that $f(x_{n+1})$ can be reused in the next step.
\item The method is time-symmetric, i.e.~one can integrate backwards
  in time and arrives at the initial state again, modulo numerical
  round-off errors.
\item The scheme can be written in a number of alternative ways, for
  example by combining the two half-steps of two subsequent steps. One
  then gets:
\begin{eqnarray}
x_{n+1} & = & x_n + v_{n+\frac{1}{2}}\, \Delta t , \\
v_{n+\frac{3}{2}} & = & v_{n+\frac{1}{2}} + f(x_{n+1}) \, \Delta t . 
\end{eqnarray}
One here sees the time-centered nature of the formulation very
clearly, and the interleaved advances of position and velocity give it
the name leapfrog.
\end{itemize}

\begin{figure}
\begin{center}
\begin{minipage}[c]{6cm}%
\resizebox{5.8cm}{!}{\includegraphics{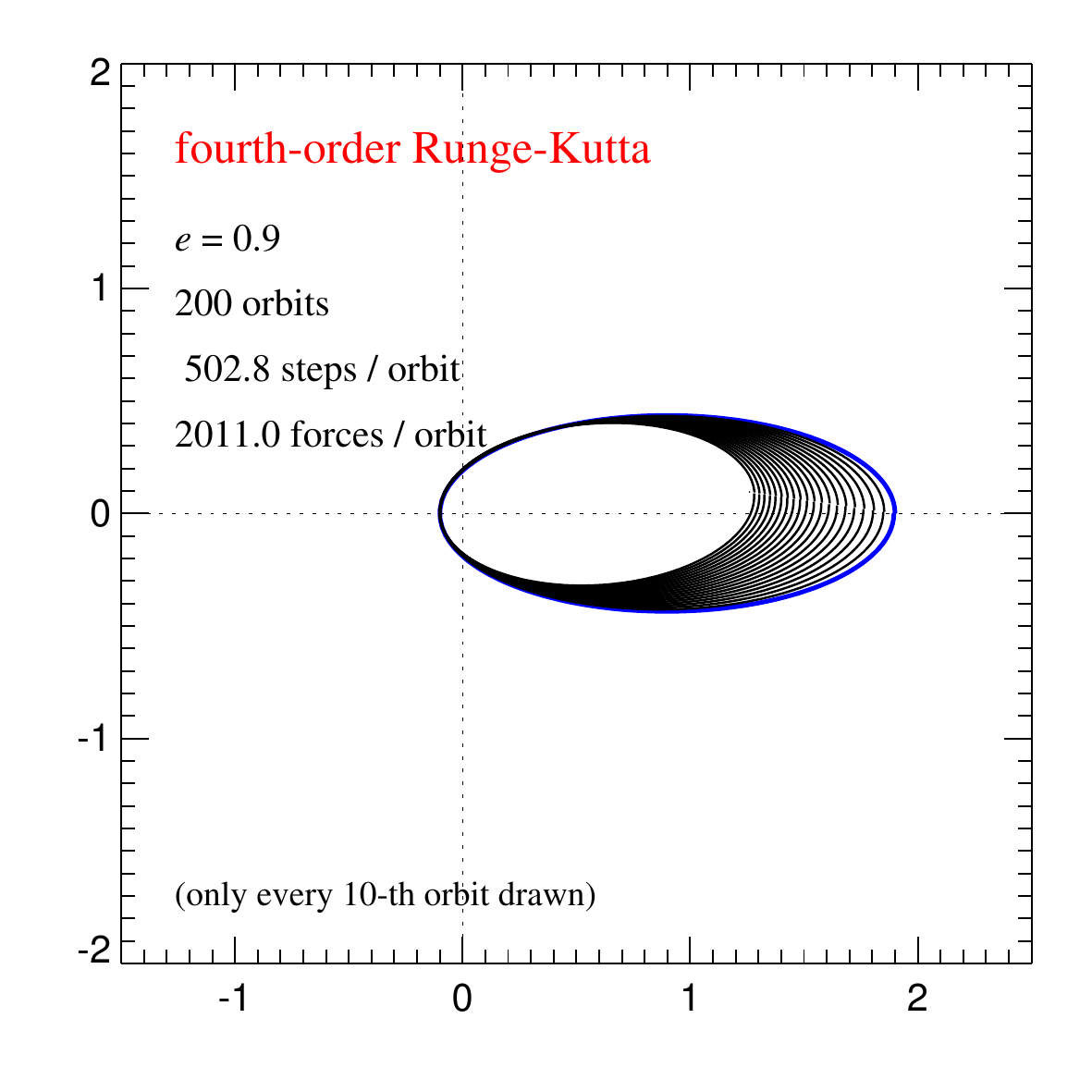}}%
\end{minipage}\hspace*{0.1cm}\begin{minipage}[c]{6cm}
\resizebox{5.8cm}{4.5cm}{\includegraphics{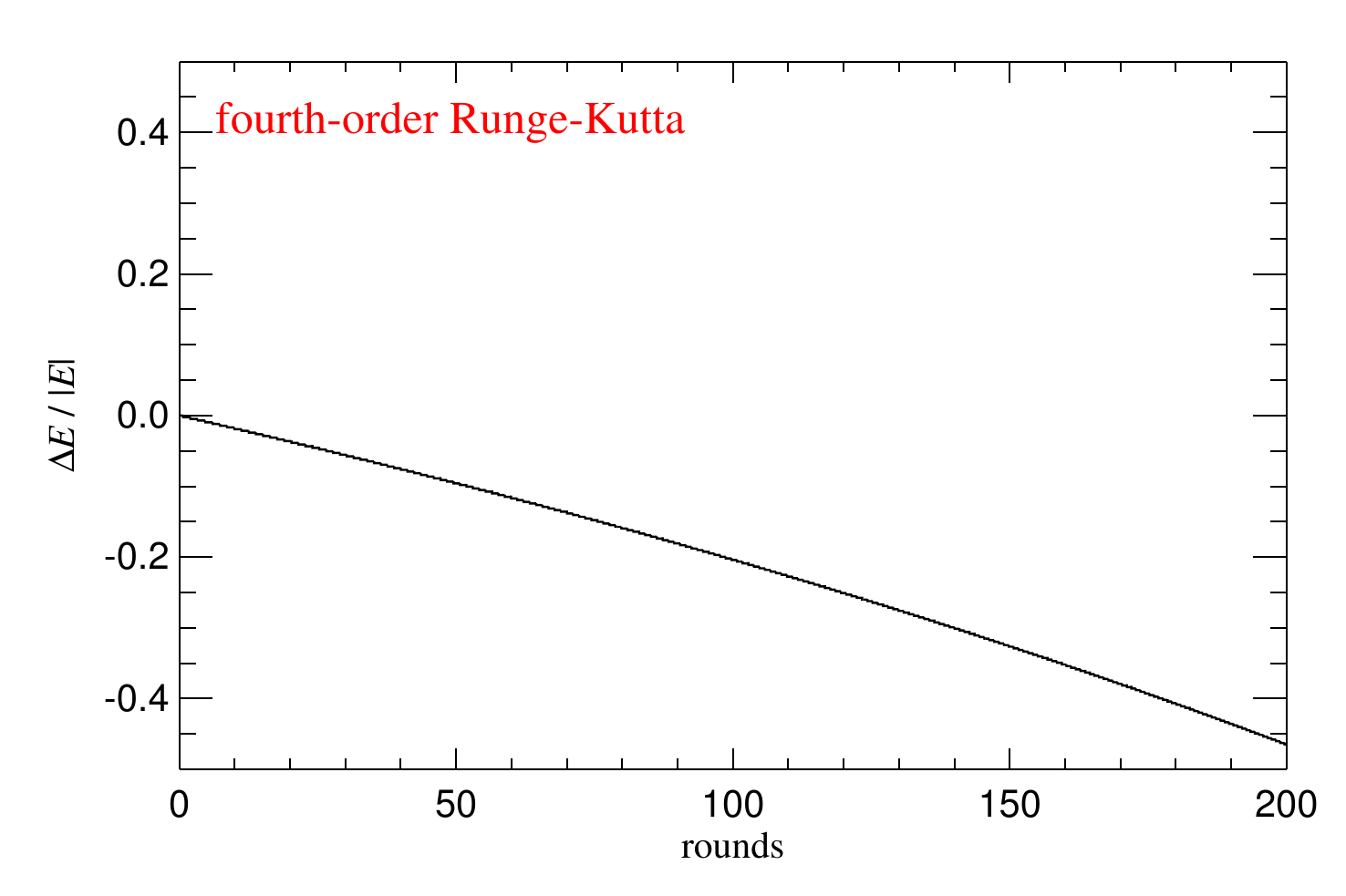}}%
\end{minipage}\\
\begin{minipage}[c]{6cm}%
\resizebox{5.8cm}{!}{\includegraphics{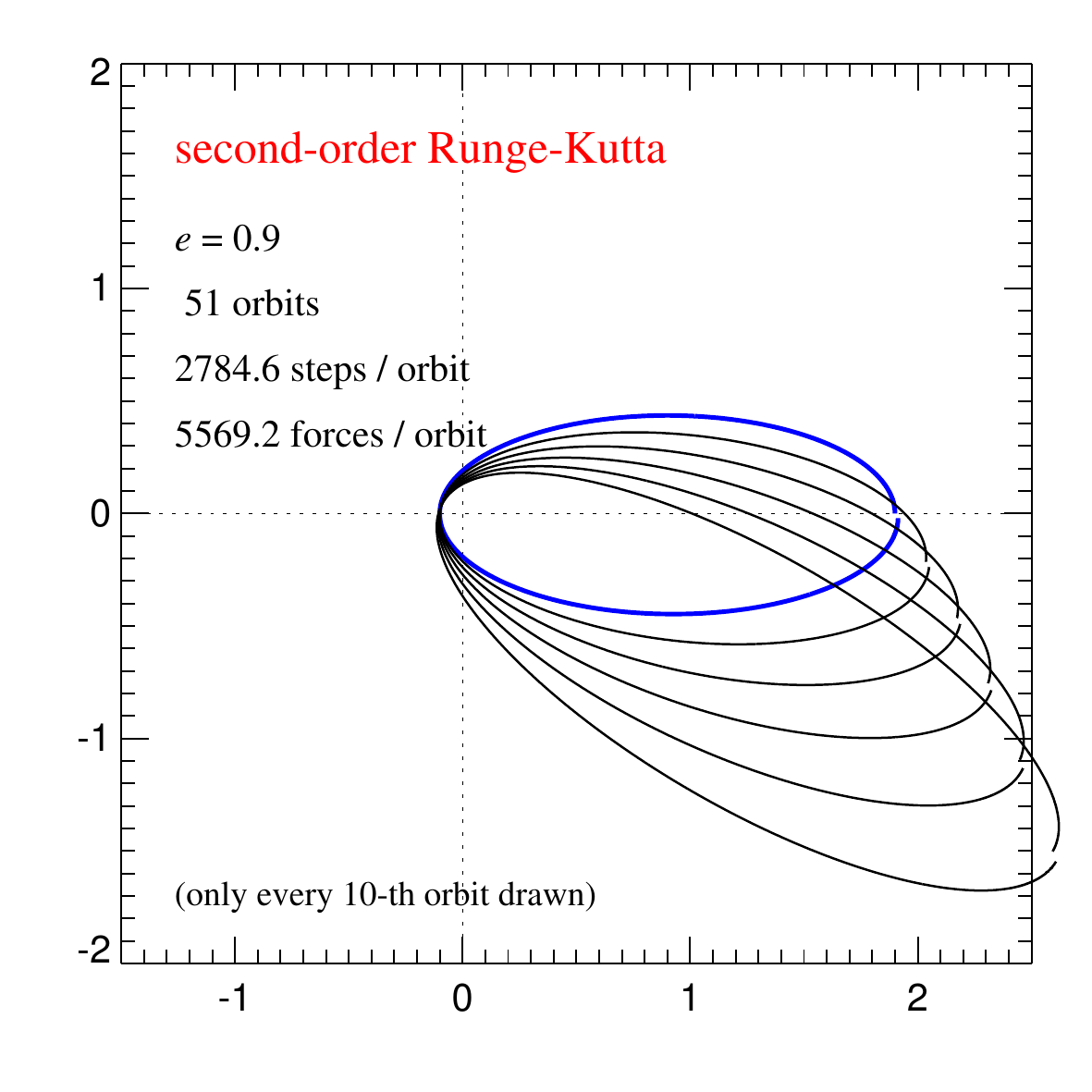}}%
\end{minipage}\hspace*{0.1cm}\begin{minipage}[c]{6cm}
\resizebox{5.8cm}{4.5cm}{\includegraphics{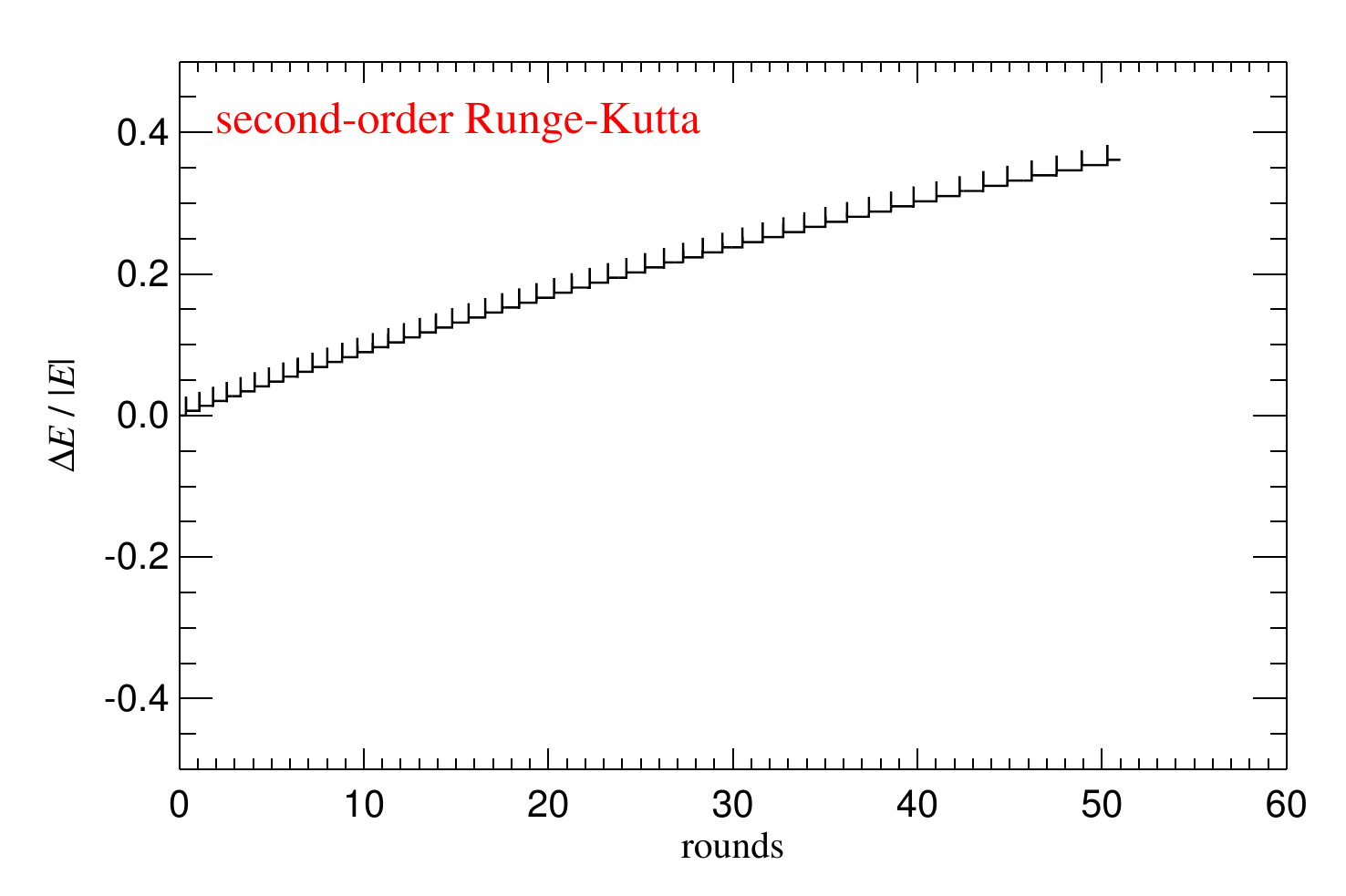}}%
\end{minipage}\\
\begin{minipage}[c]{6cm}%
\resizebox{5.8cm}{!}{\includegraphics{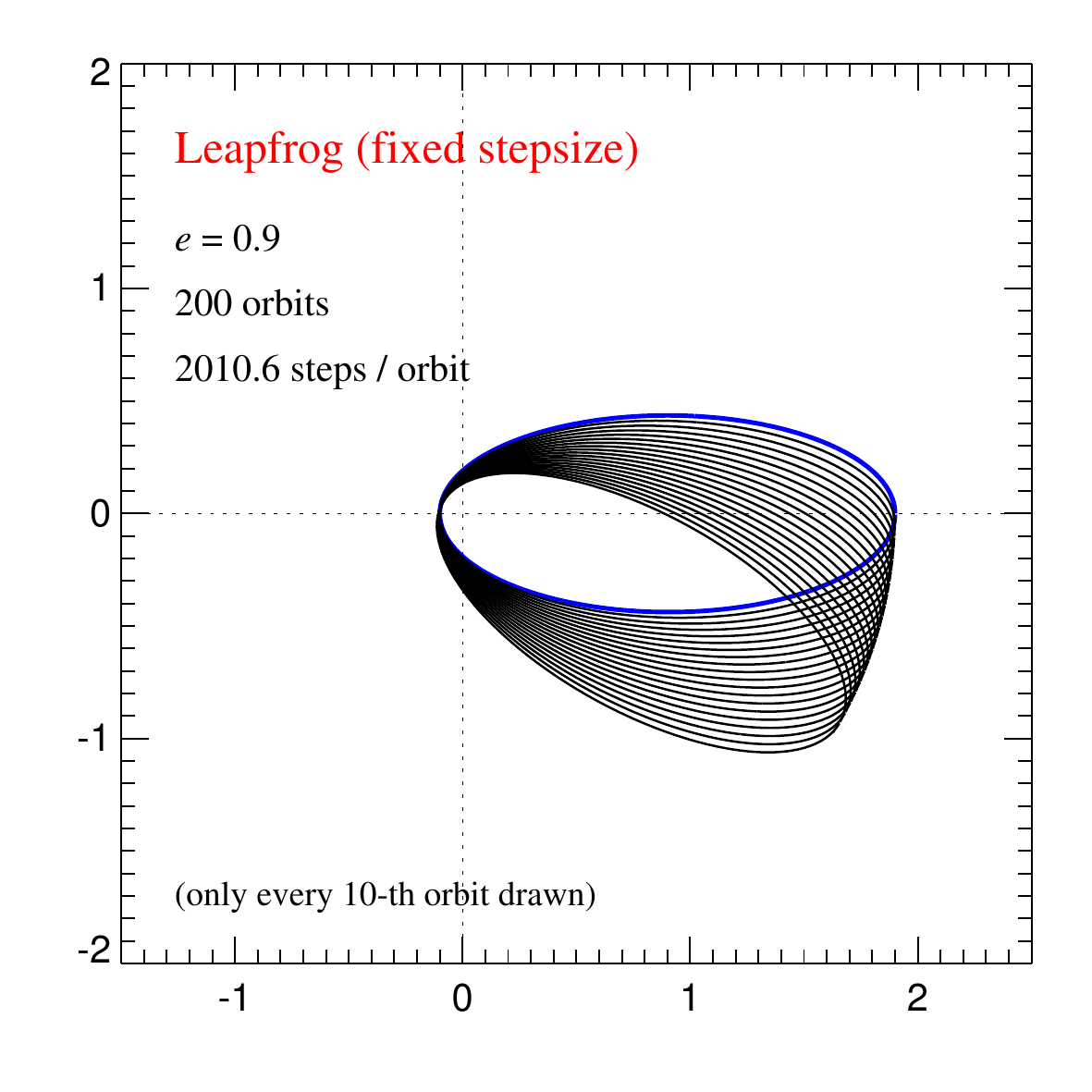}}%
\end{minipage}\hspace*{0.1cm}\begin{minipage}[c]{6cm}
\resizebox{5.8cm}{4.5cm}{\includegraphics{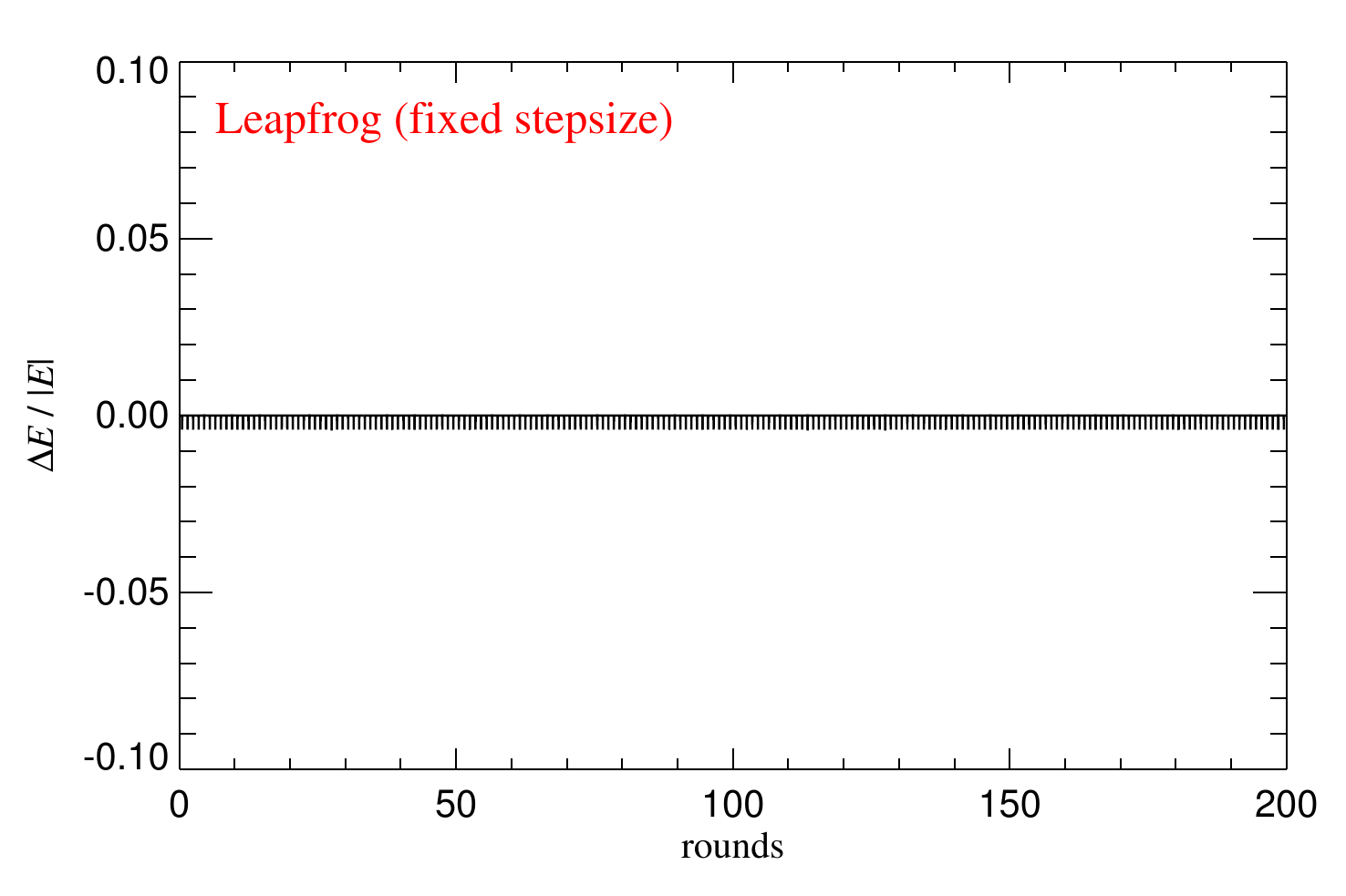}}%
\end{minipage}\\
\end{center}
\caption{Kepler problem integrated with different integration schemes
  \citep{Springel2005}. The panels on top are for a 4th-order Runge
  Kutta scheme, the middle for a 2nd order Runge-Kutta, and the bottom
  for a 2nd-order leapfrog. The leapfrog does not show a secular drift
  of the total energy, and is hence much more suitable for long-term
  integration of this Hamiltonian system.}
\label{FigKepler}
\end{figure}

The performance of the leapfrog in certain problems is found to be
surprisingly good, better than that of other schemes such as
Runge-Kutta which have formally the same or even a better error
order. This is illustrated in Figure~\ref{FigKepler} for the Kepler
problem, i.e.~the integration of the motion of a small point mass in
the gravitational field of a large mass.  We see that the long-term
evolution is entirely different. Unlike the RK schemes, the leapfrog
does not build up a large energy error. So why is the leapfrog
behaving here so much better than other 2nd order or even 4th order
schemes?

\subsection{Symplectic integrators}

The reason for these beneficial properties lies in the fact that the
leapfrog is a so-called symplectic method. These are
structure-preserving integration methods \citep[e.g.][]{Saha1992,
  Hairer2002} that observe important special properties of Hamiltonian
systems: Such systems have first conserved integrals (such as the
energy), they also exhibit phase-space conservation as described by
the Liouville theorem, and more generally, they preserve Poincare's
integral invariants.

\runinhead{Symplectic transformations}
\begin{itemize}
\item A linear map $F: \mathbb{R}^{2d} \to \mathbb{R}^{2d} $ is called
  symplectic if $\omega(F\xi, F\eta) = \omega (\xi, \eta)$ for all
  vectors $\xi, \eta \in \mathbb{R}^{2d}$, where $\omega$ gives the area of
  the parallelogram spanned by the the two vectors.

\item A differentiable map $g: U \to \mathbb{R}^{2d}$ with $U\in \mathbb{R}$ is called
  symplectic if its Jacobian matrix is everywhere symplectic,
  i.e. $\omega(g'\xi, g'\eta) = \omega(\xi, \eta)$.

\item \emph{Poincare's theorem} states that the time evolution
  generated by a Hamiltonian in phase-space is a symplectic
  transformation.

\end{itemize}

The above suggests that there is a close connection between exact
solutions of Hamiltonians and symplectic transformations. Also, two
consecutive symplectic transformations are again symplectic.

\runinhead{Separable Hamiltonians}

Dynamical problems that are described by Hamiltonians of the form 
\begin{equation}
H(p,q) = \frac{p^2}{2m} +U(q) 
\end{equation}
are quite common. These systems have separable Hamiltonians that can
be written as
\begin{equation}
H(p,q) = H_{\rm kin}(p) + H_{\rm pot}(q) .
\end{equation}
Now we will allude to the general idea of {\em operator splitting}
\citep{Strang1968}. Let's try to solve the two parts of the
Hamiltonian individually:

\begin{enumerate}
\item
For the part $H = H_{\rm kin} = \frac{p^2}{2m}$, the equations of
motion are
\begin{equation}
\dot{q} = \frac{\partial H}{\partial p} = \frac{p}{m},
\end{equation} 
\begin{equation}
\dot{p} = -\frac{\partial H}{\partial q} = 0.
\end{equation}
These equations are straightforwardly solved and give
\begin{eqnarray}
q_{n+1} & = & q_n + \frac{p_n}{m} \Delta t, \\
p_{n+1} & = & p_n.
\end{eqnarray}
Note that this solution is exact for the given Hamiltonian, for
arbitrarily long time intervals $\Delta t$. Given that it is a
solution of a Hamiltonian, the solution constitutes a symplectic
mapping.

\item The potential part, $H = H_{\rm pot} = U(q)$, leads to the
equations
\begin{equation}
\dot{q} = \frac{\partial H}{\partial p} = 0 ,
\end{equation} 
\begin{equation}
\dot{p} = -\frac{\partial H}{\partial q} = -\frac{\partial U}{\partial q}.
\end{equation}
This is solved by 
\begin{eqnarray}
q_{n+1} & = & q_n, \\
p_{n+1} & = & p_n - \frac{\partial U}{\partial q} \Delta t .
\end{eqnarray}
Again, this is an exact solution independent of the size of $\Delta
t$, and therefore a symplectic transformation.
\end{enumerate}

Let's now introduce an operator $\varphi_{\Delta t}(H)$ that describes
the mapping of phase-space under a Hamiltonian $H$ that is evolved
over a time interval $\Delta t$. Then it is easy to see that the
leapfrog is given by
\begin{equation}
\varphi_{\Delta t}(H) = \varphi_{\frac{\Delta t}{2}}(H_{\rm pot})
\circ \varphi_{\Delta t}(H_{\rm kin})
\circ\varphi_{\frac{\Delta t}{2}}(H_{\rm pot})
\label{eqnconcat} 
\end{equation}
for a separable Hamiltonian $H = H_{\rm kin} + H_{\rm pot}$. 

\begin{itemize}
\item Since each individual step of the leapfrog is symplectic, the
  concatenation of equation~(\ref{eqnconcat}) is also symplectic.
\item In fact, the leapfrog generates the exact solution of a modified
  Hamiltonian $H_{\rm leap}$, where $H_{\rm leap} = H+ H_{\rm
    err}$. The difference lies in the `error Hamiltonian' $H_{\rm
    err}$, which is given by
\begin{equation}
H_{\rm err} \propto \frac{\Delta t^2}{12} \left\{ \left\{ H_{\rm kin}, H_{\rm
  pot} \right\}, H_{\rm kin} + \frac{1}{2}H_{\rm pot} \right\} + {\cal
O}({\Delta t^3}),
\end{equation}
where the curly brackets are Poisson brackets
\citep[][]{Goldstein1950}. This can be demonstrated by expanding
\begin{equation}
{\rm e}^{(H+H_{\rm err}) \Delta t} = 
{\rm e}^{  H_{\rm pot}\frac{\Delta t}{2} }\;
{\rm e}^{ H_{\rm kin} \Delta t }\;
{\rm e}^{  H_{\rm pot}\frac{\Delta t}{2}}
\end{equation}
with the help of the Baker-Campbell-Hausdorff formula
\citep{Campbell1897, Saha1992}.
\item The above property explains the superior long-term stability of
  the integration of conservative systems with the leapfrog. Because
  it respects phase-space conservation, secular trends are largely
  absent, and the long-term energy error stays bounded and reasonably
  small.
\end{itemize}

\section{Gravitational force calculation} \label{SecGrav}

As mentioned earlier, calculating the gravitational forces exactly for
a large number of bodies becomes computational prohibitive very
quickly. Fortunately, in the case of collisionless systems, this is
also not necessary, because comparatively large force errors can be
tolerated. All they do is to shorten the relaxation time slightly by
an insignificant amount \citep{Hernquist1993}. In this section, we
discuss a number of the most commonly employed approximate force
calculation schemes, beginning with the so-called particle mesh
techniques \citep{White1983, Klypin1983} that were originally pioneered
in plasma physics \citep{Hockney1988}.

\subsection{Particle mesh technique}

An important approach to accelerate the force calculation for an
N-body system lies in the use of an auxiliary mesh. Conceptually, this
so-called particle-mesh (PM) technique involves four steps:

\begin{enumerate}

\item Construction of a density field $\rho$ on a suitable mesh.

\item Computation of the potential on the mesh by solving the Poisson equation.

\item Calculation of the force field from the potential.

\item Calculation of the forces at the original particle positions.

\end{enumerate}

\noindent We shall now discuss these four steps in turn. An excellent
coverage of the material in this section is given by
\citet{Hockney1988}.

\begin{figure}[t]
\sidecaption
\resizebox{6cm}{!}{\includegraphics{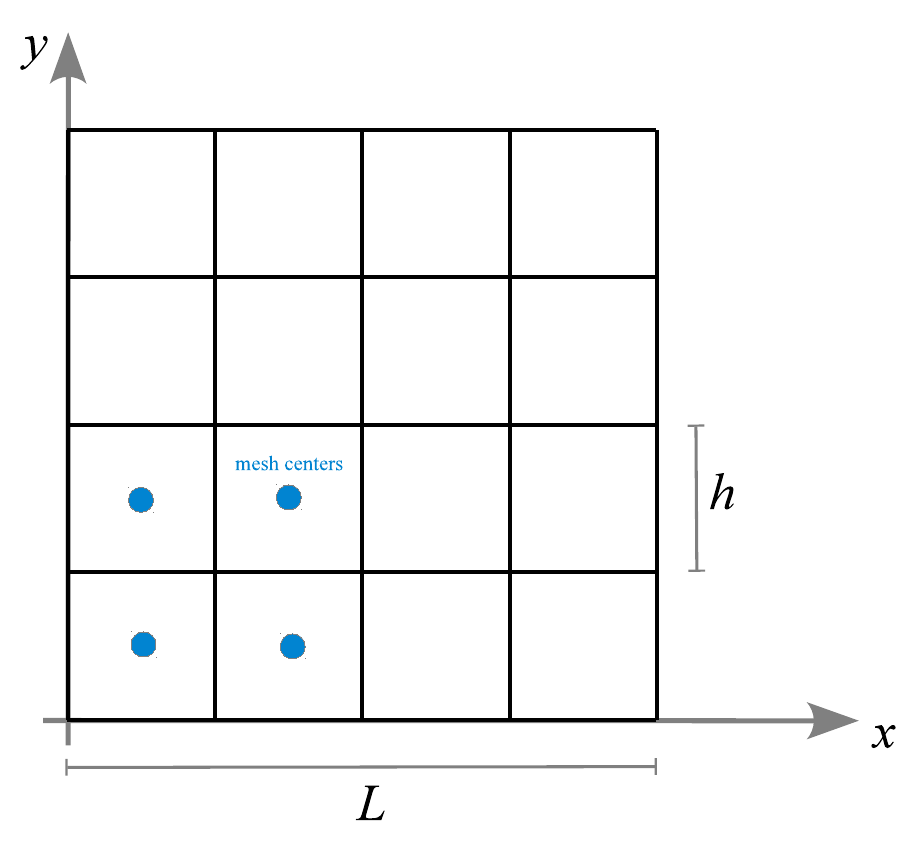}}
\caption{Sketch of the mesh geometry used in typical particle-mesh
  techniques with Cartesian grids.}
\end{figure}

\subsubsection{Mass assignment}

We want to put $N$ particles with mass $m_i$ and coordinates
$\vec{r}_i$ ($i= 1, 2, \ldots, N$) onto a mesh with uniform spacing
$h= L/N_{\rm g}$. For simplicity, we will assume a cubical
calculational domain with extension $L$ and a number of $N_{\rm g}$
grid cells per dimension.  Let $\{\vec{r}_{\vec{p}}\}$ denote the set
of discrete cell-centers, with $\vec{p}=(p_x,p_y,p_z)$ being a
suitable integer index ($0\le p_{x,y,z} < N_g$). Note that one may
equally well identify the $\{\vec{r}_{\vec{p}}\}$ with the lower left
corner of a mesh cell, if this is more practical.

We associate a shape function $S(\vec{x})$ with each particle,
normalized according to 
\begin{equation}
\int S(\vec{x})\, {\rm d}\vec{x} = 1.
\end{equation}
To each mesh-cell, we then assign the fraction $W_{\vec{p}}
(\vec{x}_i)$ of particle $i$'s mass that falls into the cell indexed
by $\vec{p}$. This is given by the overlap of the mesh cell with the
shape function, namely:
\begin{equation}
W_{\vec{p}} (\vec{x}_i) = \int_{\vec{x}_{\vec{p}} -
  \frac{h}{2}}^{\vec{x}_{\vec{p}} + \frac{h}{2}}  S(\vec{x}_i -
\vec{x}_{\vec{p}})\,{\rm d}\vec{x} .
\end{equation}
The integration extends here over the cubical cell $\vec{p}$. 
By introducing the top-hat function
\begin{equation}
\Pi(\vec{x}) = \left\{
\begin{array}{ll}
1 & \mbox{for $|\vec{x}| \le \frac{1}{2}$} ,\\
0 & \mbox{otherwise} ,
\end{array}
\right.
\end{equation}
we can extend the integration boundaries to all space and write instead:
\begin{equation}
W_{\vec{p}} (\vec{x}_i) = \int \Pi\left(\frac{\vec{x}-\vec{x}_{\vec{p}}}{h}\right)\,  S(\vec{x}_i - \vec{x}_{\vec{p}})\,{\rm d}\vec{x}.
\end{equation}
Note that this also shows that the assignment function $W$ is a
convolution of $\Pi$ with $S$.  The full density in grid cell
$\vec{p}$ is then given
\begin{equation}
\rho_{\vec{p}} = \frac{1}{h^3}\sum_{i=1}^N m_i W_{\vec{p}}(\vec{x}_i).
\end{equation}

These general formula evidently depend on the specific choice one
makes for the shape function $S(\vec{x})$. Below, we discuss a few of
the most commonly employed low-order assignment schemes.

\subsubsection{Nearest grid point (NGP) assignment}

\begin{figure}
\sidecaption
\resizebox{7.5cm}{!}{\includegraphics{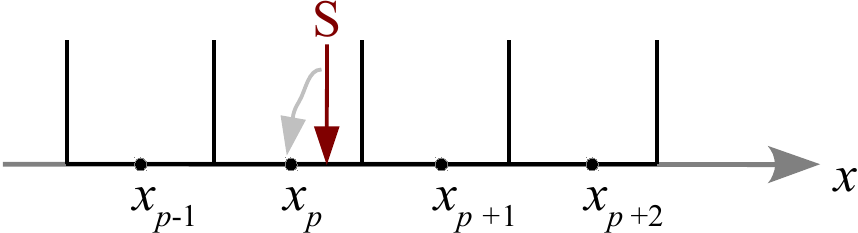}}
\caption{Sketch of the nearest grid point (NGP) assignment
  scheme. This simple binning scheme simply assigns the mass of a
  particle completely to the one mesh cell in which it falls.}
\label{fig_ngp}
\end{figure}

The simplest possible choice for $S$ is a Dirac $\delta$-function. One
then gets:
\begin{equation}
  W_{\vec{p}} (\vec{x}_i) = \int \Pi\left(\frac{\vec{x}-\vec{x}_{\vec{p}}}{h}\right)\,  \delta(\vec{x}_i - \vec{x}_{\vec{p}})\,{\rm d}\vec{x} = 
  \Pi\left(\frac{\vec{x}_i-\vec{x}_{\vec{p}}}{h}\right).
\end{equation}
In other words, this means that $W_{\vec{p}}$ is either 1 (if the
coordinate of particle $i$ lies inside the cell), or otherwise it is
zero. Consequently, the mass of particle $i$ is fully assigned to
exactly one cell -- the nearest grid point, as sketched in
Figure~\ref{fig_ngp}.

\subsubsection{Clouds-in-cell (CIC) assignment}

\begin{figure}
\sidecaption
\resizebox{7.5cm}{!}{\includegraphics{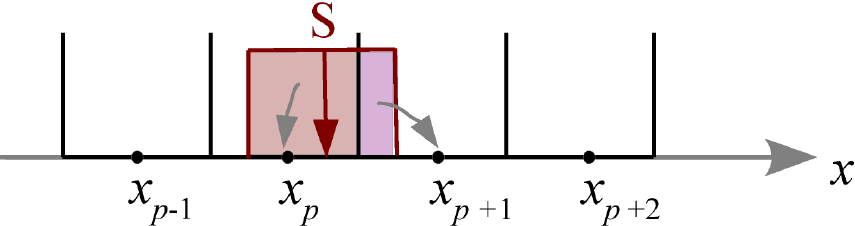}}
\caption{Sketch of the clouds-in-cell (CIC) assignment scheme. The
  fraction of mass assigned to a given cell is given by the fraction
  of the cubical cloud shape of the particle that overlaps with the
  cell.}
\label{fig_cic}
\end{figure}

Here one adopts as shape function
\begin{equation}
S(\vec{x}) = \frac{1}{h^3} \Pi\left(\frac{\vec{x}}{h}\right) ,
\end{equation}
which is the same cubical `cloud' shape as that of individual mesh
cells. The assignment function is
\begin{equation}
  W_{\vec{p}} (\vec{x}_i) = \int \Pi\left(\frac{\vec{x}-\vec{x}_{\vec{p}}}{h}\right)\,  \frac{1}{h^3}\Pi\left(\frac{\vec{x}_i - \vec{x}_{\vec{p}}}{h}\right)\,{\rm d}\vec{x},
\end{equation}
which only has a non-zero (and then constant) integrand if the cubes
centered on $\vec{x}_i$ and $\vec{x}_{\vec{p}}$ overlap. How can this
overlap be calculated? The 1D sketch of Fig.~\ref{fig_cic} can help to
make this clear.

Recall that for one of the dimensions we have $x_p = (p_x+1/2)h$, for
$p\in\{0$, $1$, $2$, $\ldots$, $N-1\}$. For a given particle
coordinate $x_i$ we may first calculate a `floating point index' by
inverting this relation, yielding $p_f = x_i/h - 1/2$. The index of
the left cell of the two cells with some overlap is then given by $p =
\lfloor p_f \rfloor$, where the brackets denote the integer floor,
i.e.~the largest integer not larger than $p_f$. We may then further
define $p^\star\equiv p_f - p$, which is a number between $0$ and
$1$. From the sketch, we see that the length of the overlap of the
particle's cloud with the cell $p$ is $h - h p^\star$, hence the
assignment function at cell $p$ takes on the value $W_p = 1 - m^\star$
for this location of the particle, whereas the assignment function for
the neighboring cell $p+1$ will take on the value $W_{p+1} = m^\star$.

\begin{figure}
\sidecaption
\resizebox{3.3cm}{!}{\includegraphics{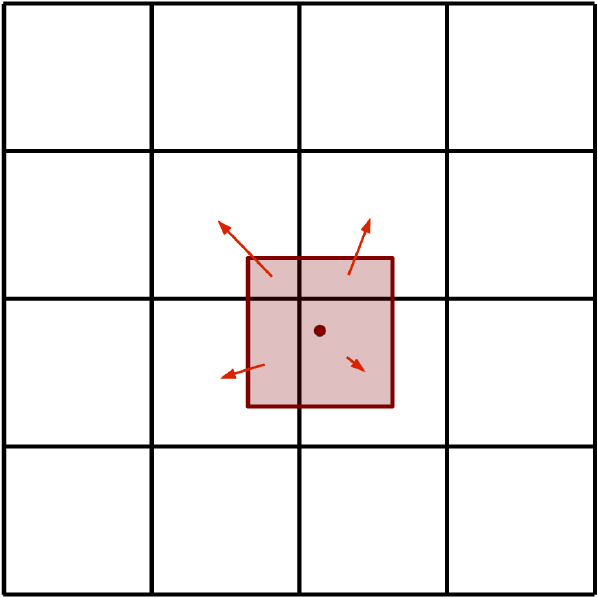}}
\caption{Sketch of CIC assignment of a particle to a two-dimensional
  mesh.}
\label{fig_cic2d}
\end{figure}

These considerations readily generalize to 2D and 3D. For example, in
2D (as sketched in Fig.~\ref{fig_cic2d}), we first assign to the
$y_i$-coordinate of point $i$ a `floating point index' $q_f = y_i/h -
1/2$. We can then use this to compute a cell index as the integer
floor $q = \lfloor q_f\rfloor$, and a fractional contribution $q^\star
= q_f - q$.  Finally, we obtain the following weights for the assignment
of a particle's mass to the four cells its `cloud' touches in 2D (as
sketched):
\begin{eqnarray}
W_{p,q} & = & (1-p^\star)(1-q^\star)\\
W_{p+1,q} & = & p^\star(1-q^\star)\\
W_{p,q+1} & = & (1-p^\star) q^\star\\
W_{p+1,q+1} & = & p^\star q^\star
\end{eqnarray}
In the corresponding 3D case, each particle contributes to the weight
functions of 8 cells, or in other words, it is spread over 8 cells.

\subsubsection{Triangular shaped clouds (TSC) assignment}

\begin{figure}
\sidecaption
\resizebox{7.5cm}{!}{\includegraphics{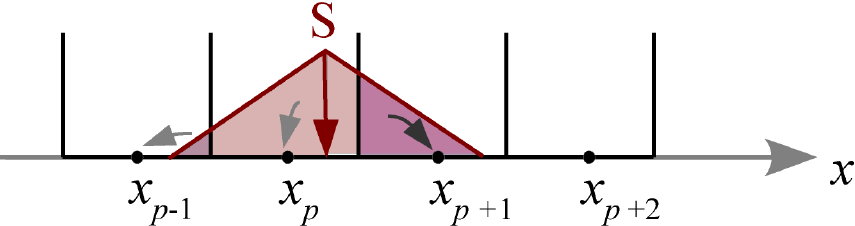}}
\caption{Sketch of triangular-shaped-clouds (TSC) assignment. Here a
  particle is spread to three cells in one dimension.}
\label{fig_tsc}
\end{figure}

One can construct a systematic sequence of ever higher-order shape
functions by adding more convolutions with the top-hat kernel. For
example, the next higher order (in 3D) is given by 
\begin{eqnarray}
W_{\vec{p}} (\vec{x}_i) & = & \int 
\Pi\left(\frac{\vec{x}-\vec{x}_{\vec{p}}}{h}\right)\,  
\frac{1}{h^3} \Pi\left(\frac{\vec{x}_i - \vec{x} - \vec{x}'}{h}\right)\,
\frac{1}{h^3} \Pi\left(\frac{\vec{x}'}{h}\right)\,
{\rm d}\vec{x}\,{\rm d}\vec{x}'\\
&=& \frac{1}{h^6} \int 
\Pi\left(\frac{\vec{x}-\vec{x}_{\vec{p}}}{h}\right)\,  
\Pi\left(\frac{\vec{x}_i - \vec{x}}{h}\right)\,
\Pi\left(\frac{\vec{x}'-\vec{x}}{h}\right)\,
{\rm d}\vec{x}\,{\rm d}\vec{x}'.
\end{eqnarray}
This still has a simple geometric interpretation. If one pictures the
kernel shape as a triangle with total base length $2h$, then the
fraction assigned to a certain cell is given by the area of overlap of
this triangle with the cell of interest (see Fig.~\ref{fig_tsc}). The
triangle will now in general touch 3 cells per dimension, making an
evaluation correspondingly more expensive. In 3D, 27 cells are touched
for every particle.

What's the advantage of using TSC over CIC, if any? Or should one
stick with the computationally cheap NGP? The assignment schemes
differ in the smoothness and differentiability of the reconstructed
density field. In particular, for NGP, the assigned density and hence
the resulting force jump discontinuously when a particle crosses a
cell boundary. The resulting force law will then at best be piece-wise
constant.

In contrast, the CIC scheme produces a force that is piece-wise linear
and continuous, but its first derivative jumps. Here the information
where a particle is inside a certain cell is not completely lost,
unlike in NGP.

Finally, TSC is yet smoother, and also the first derivative of the
force is continuous. See Table~\ref{tab_schemes} for a brief summary
of these assignment schemes. Which of these schemes is the preferred
choice is ultimately problem-dependent. In most cases, CIC and TSC are
quite good options, providing sufficient accuracy with still
reasonably small (and hence computationally efficient) assignment
kernels. The latter get invariably more extended for higher-order
assignment schemes, which not only is computationally ever more costly
but also invokes additional communication overheads in parallelization
schemes.

\begin{table}
\caption{Commonly used shape functions.} \label{tab_schemes}
\begin{tabular}{p{1.6cm}p{2.9cm}p{2cm}p{3.3cm}}
\hline\noalign{\smallskip}
Name & Cloud shape $S(x)$ & \# of cells used & assignment function shape\\
\noalign{\smallskip}\svhline\noalign{\smallskip}
NGP & $\delta(x)$ & $1^d$ & $\Pi$\\
CIC & $\frac{1}{h^d} \Pi \left( \frac{\vec{x}}{h} \right)$ & $2^d$ & $\Pi \star \Pi$\\
TSC & $\frac{1}{h^d} \Pi \left( \frac{\vec{x}}{h} \right) \star \frac{1}{h^d} \Pi \left( \frac{\vec{x}}{h} \right)$ & $3^d$ & $\Pi \star \Pi \star \Pi$\\
\noalign{\smallskip}\hline\noalign{\smallskip}
\end{tabular}
\end{table}

\subsubsection{Solving for the gravitational potential}

Once the density field is obtained, we would like to solve Poisson's
equation
\begin{equation}
\nabla^2 \Phi = 4\pi G \, \rho,
\end{equation}
and obtain the gravitational potential discretized on the same
mesh. There are primarily two methods that are in widespread use for
this.

First, there are Fourier-transform based methods which exploit the
fact that the potential can be viewed as a convolution of a Green's
function with the density field. In Fourier-space, one can then use
the convolution theorem and cast the computationally expensive
convolution into a cheap algebraic multiplication. Due to the
importance of this approach, we will discuss it extensively in
subsection~\ref{sec_fourier}.

Second, there are also iterative solvers for Poisson's equation which
yield a solution directly in real-space. Simple versions of such
iteration schemes use Jacobi or Gauss-Seidel iteration, more
complicated ones employ a sophisticated multi-grid approach to speed
up convergence. We shall discuss these methods in
subsection~\ref{sec_multigrid}.

\subsubsection{Calculation of the forces}

Let's assume for the moment that we already obtained the gravitational
potential $\Phi$ on the mesh, with one of the methods mentioned
above. We would then like to get the acceleration field from
\begin{equation}
\vec{a} = -\nabla \Phi.
\end{equation}
One can achieve this by calculating a numerical derivative of the
potential by {\em finite differencing}. For example, the simplest
estimate of the force in the $x$-direction would be
\begin{equation}
a_x^{(i,j,k)} = - \frac{\Phi^{(i+1,j,k)} - \Phi^{(i-1,j,k)}}{2 h} , 
\end{equation}
where $\vec{p}= (i,j,k)$ is a cell index.  The truncation error of
this expression is ${\cal O}(h^2)$, hence the estimate of the
derivative is second-order accurate.

Alternatively, one can use larger \emph{stencils} to obtain a more
accurate finite difference approximation of the derivative, at greater
computational cost. For example, the 4-point expression
\begin{equation}
  a_x^{(i,j,k)} = - \frac{1}{2h}\left\{ 
    \frac{4}{3} \left[ \Phi^{(i+1, j, k)} - \Phi^{(i-1,j,k)}\right] - \frac{1}{6}\left[\Phi^{(i+2,j,k)}- \Phi^{(i-2,j,k)}\right]
  \right\}
\end{equation}
can be used, which has a truncation error of ${\cal O}(h^4)$, as
verified through simple Taylor expansions.

For the $y$- and $z$-dimensions, corresponding formulae, where $j$ or
$k$ are varied and the other cell coordinates are held fixed, can be
used. Whether a second- or fourth-order discretization formula should
be used depends again on the question which compromise between
accuracy and speed is best for a given problem. In many collisionless
simulation set-ups, the residual truncation error of the second-order
finite difference approximation of the force will be negligible
compared to other errors inherent in the simulation methodology, hence
the second-order formula would then be expected to be sufficient. But
this cannot be generalized to all situations and simulation setups; if
in doubt, it is best to explicitly test for this source of error.

\subsubsection{Interpolating from the mesh to the particles}

Once we have the force field on a mesh, we are not yet fully done. We
actually desire the forces at the particle coordinates of the N-body
system, not at the coordinates of the mesh cells of our auxiliary
computational grid. We are hence left with the problem of
interpolating the forces from the mesh to the particle coordinates.

Recall that we defined the density field in terms of mass assignment
functions, of the form
\begin{equation}
\rho_{\vec{p}} = 
\frac{1}{h^3} \sum_i W_{\vec{p}}(\vec{x}_i)
=
\frac{1}{h^3} \sum_i W(\vec{x}_i - \vec{x}_{\vec{p}}).
\end{equation}
Here we introduced in the last expression an alternative notation for
the weight assignment function.

Assume that we have computed the acceleration field on the grid,
$\{\vec{a}_{\vec{p}}\}$. It turns out to be very important to {\em use
  the same} assignment kernel as used in the density construction also
for the force interpolation, i.e.~the force at coordinate $\vec{x}$
for a mass $m$ needs to be computed as
\begin{equation}
\vec{F}(\vec{x}) = m \sum_{\vec{p}} \vec{a}_{\vec{p}} W(\vec{x} - \vec{x}_{\vec{p}}),
\end{equation}
where $W$ denotes the assignment function used for computing the
density field on the mesh.  This requirement results from the desire
to have a vanishing {\em self-force}, as well as pairwise
antisymmetric forces between every particle pair. The self-force is
the force that a particle would feel if just it alone would be present
in the system.  If numerically this force would evaluate to a non-zero
value, the particle would accelerate all by itself, violating momentum
conservation. Likewise, for two particles, we require that the forces
they mutually exert on each other are equal in magnitude and opposite
in direction, such that momentum conservation is manifest.

We now show that using the same kernels for the mass assignment and
force interpolation protects against these numerical artefacts
\citep{Hockney1988}.  We start by noting that the acceleration field
at a mesh point $\vec{p}$ depends linearly on the mass at another mesh
point $\vec{p}'$, which is a manifestation of the superposition
principle (this can, for example, also be seen when Fourier techniques
are used to solve the Poisson equation). We can hence express the
field as
\begin{equation}
\vec{a}_{\vec{p}} = \sum_{\vec{p}'} \vec{d}(\vec{p},\vec{p}') \, h^3\rho_{\vec{p}'},
\end{equation}
 with a Green's function $\vec{d}(\vec{p}, \vec{p}')$. This
 vector-valued Green's function for the force is antisymmetric,
 i.e.~it changes sign when the two points in the arguments are swapped.
 Note that $h^3\rho_{\vec{p}'}$ is simply the mass contained in mesh
 cell $\vec{p}'$.

We can now calculate the self-force resulting from the density
assignment and interpolation steps:
\begin{eqnarray}
\vec{F}_{\rm self}(\vec{x}_i) & = & m_i \vec{a}_i(\vec{x}_i) = m_i\sum_{\vec{p}}  W(\vec{x}_i - \vec{x}_{\vec{p}}) \vec{a}_{\vec{p}} \\
& = &  m_i\sum_{\vec{p}}  W(\vec{x}_i - \vec{x}_{\vec{p}}) \sum_{\vec{p}'} \vec{d}(\vec{p},\vec{p}')  h^3 \rho_{\vec{p}'} \\
& = & m_i\sum_{\vec{p}}  W(\vec{x}_i - \vec{x}_{\vec{p}}) \sum_{\vec{p}'} \vec{d}(\vec{p},\vec{p}') m_i W(\vec{x}_i - \vec{x}_{\vec{p}'})  \\
& = & m_i^2\sum_{\vec{p}, \vec{p}'}  \vec{d}(\vec{p},\vec{p}') W(\vec{x}_i - \vec{x}_{\vec{p}})    W(\vec{x}_i - \vec{x}_{\vec{p}'})  \\
& = & 0.
\end{eqnarray}
Here we have started out with the interpolation from the mesh-based
acceleration field, and then inserted the expansion of the latter as
as convolution over the density field of the mesh. Finally, we put in
the density contribution created by the particle $i$ at a mesh cell
$\vec{p}'$. We then see that the double sum vanishes because of the
antisymmetry of ${\vec{d}}$ and the symmetry of the kernel product
under exchange of $\vec{p}$ and $\vec{p}'$.  Note that this however
only works because the kernels used for force interpolation and
density assignment are indeed equal -- it would have not worked out
if they would be different, which brings us back to the point
emphasized above.

Now let's turn to the force antisymmetry. The force exerted on a
particle 1 of mass $m_1$ at location $\vec{x}_1$ due to a particle 2
of mass $m_2$ at location $\vec{x}_2$ is given by
\begin{eqnarray}
\vec{F}_{12} & = & m_1 \vec{a}(\vec{x}_1) = m_1\sum_{\vec{p}} W(\vec{x}_1 - \vec{x}_{\vec{p}}) \vec{a}_{\vec{p}} \\
& = & 
 m_1\sum_{\vec{p}} W(\vec{x}_1 - \vec{x}_{\vec{p}}) \sum_{\vec{p}'}  \vec{d}(\vec{p},\vec{p}') h^3 \rho_{\vec{p}'} \\
& = & m_1\sum_{\vec{p}} W(\vec{x}_1 - \vec{x}_{\vec{p}}) \sum_{\vec{p}'}  \vec{d}(\vec{p},\vec{p}') m_2 W(\vec{x}_2 - \vec{x}_{\vec{p}'}) \\
& = & m_1 m_2 \sum_{\vec{p}, \vec{p}'}   \vec{d}(\vec{p},\vec{p}')  W(\vec{x}_1 - \vec{x}_{\vec{p}})  W(\vec{x}_2 - \vec{x}_{\vec{p}'}) .
\end{eqnarray}
Likewise, we obtain for the force experienced by particle 2 due to particle 1: 
\begin{eqnarray}
\vec{F}_{21} = m_1 m_2 \sum_{\vec{p}', \vec{p}}   \vec{d}(\vec{p},\vec{p}')  W(\vec{x}_2 - \vec{x}_{\vec{p}})  W(\vec{x}_1 - \vec{x}_{\vec{p}'}) .
\end{eqnarray}
We may swap the summation indices through relabeling and exploiting the
antisymmetry of $\vec{d}$, obtaining:
\begin{eqnarray}
\vec{F}_{21} = -m_1 m_2 \sum_{\vec{p}', \vec{p}}   \vec{d}(\vec{p},\vec{p}')  W(\vec{x}_1 - \vec{x}_{\vec{p}})  W(\vec{x}_2 - \vec{x}_{\vec{p}'})  .
\end{eqnarray}
Hence we have $\vec{F}_{12} + \vec{F}_{21} =0$, independent on where
the points are located on the mesh.

\subsection{Fourier techniques}  \label{sec_fourier}

Fourier transforms provide a powerful tool for solving certain partial
differential equations. In this subsection we shall consider the
particularly important example of using them to solve Poisson's
equation, but we note that the basic technique can be used in similar
form also for other systems of equations.

\subsubsection{Convolution problems}

Suppose we want to solve Poisson's equation,
\begin{equation}
\nabla^2 \Phi = 4\pi G \rho,
\end{equation}
for a given density distribution $\rho$. Actually, we can readily
write down a solution for a non-periodic space, since we know the
Newtonian potential of a point mass, and the equation is linear. The
potential is simply a linear superposition of contributions from
individual mass elements, which in the continuum can be written as the
integration:
\begin{equation}
\Phi(\vec{x}) =  - \int   G \, \frac{\rho(\vec{x}') \, {\rm d}\vec{x}' }{|\vec{x}-\vec{x}'|}.
\end{equation}
This is recognized to be a convolution integral of the form
\begin{equation}
\Phi(\vec{x}) =  \int   g(\vec{x}-\vec{x}') \, \rho(\vec{x}') \, {\rm d}\vec{x}', 
\end{equation}
where
\begin{equation}
g(\vec{x}) =  - \frac{G}{|\vec{x}|}
\end{equation}
is the {\em Green's function} of Newtonian gravity. The convolution
may also be formally written as:
\begin{equation}
\Phi =     g \star \rho.
\end{equation}

We now recall the {\em convolution theorem}, which says that the
Fourier transform of the convolution of two functions is equal to the
product of the individual Fourier transforms of the two functions,
i.e.
\begin{equation}
{\cal F}(  f \star g) =  {\cal F}(  f) \cdot {\cal F}( g), 
\end{equation}
where ${\cal F}$ denotes the Fourier transform and $f$ and $g$ are the
two functions. A convolution in real space can hence be transformed
to a much simpler, point-by-point multiplication in Fourier
space. 

There are many problems where this can be exploited to arrive at
efficient calculational schemes, for example in solving Poisson's
equation for a given density field. Here the central idea is to
compute the potential through
\begin{equation}
\Phi = {\cal F}^{-1} \left[ {\cal F}(  g) \cdot {\cal F}( \rho)\right], 
\end{equation}
i.e. in Fourier space, with $\hat \Phi(\vec{k}) \equiv {\cal F}(\Phi)
$, we have the simple equation
\begin{equation}
\hat\Phi(\vec{k}) = \hat g(\vec{k} ) \cdot \hat \rho(\vec{k}).
\label{eqnFT}
\end{equation}

\subsubsection{The continuous Fourier transform}

But how do we solve this in practice?  Let's first assume that we have
{\em periodic boundary conditions} with a box of size $L$ in each
dimension.  The continuous $\rho(\vec{x})$ can in this case be written
as a Fourier series of the form
\begin{equation}
\rho(\vec{x}) = \sum_{\vec{k}} \rho_{\vec{k}}\, {\rm e}^{i \vec{k}\vec{x}},
\end{equation}
where the sum over the $\vec{k}$-vectors extends over a discrete
spectrum of wave vectors, with
\begin{equation}
\vec{k} \in \frac{2\pi}{L} \left(
\begin{array}{c}
n_1\\
n_2\\
n_3
\end{array}
 \right),
\end{equation}
where $n_1, n_2, n_3$ are from the set of positive and negative
integer numbers.  The allowed modes in $\vec{k}$ hence form an
infinitely extended Cartesian grid with spacing $2\pi/L$. Because of
the periodicity condition, only these waves `fit' into the box.  For
a real field such as $\rho$, there is also a reality constraint of the
form $\rho_{\vec{k}} = \rho_{-\vec{k}}^\star$, hence the modes are not
all independent.  The Fourier coefficients can be calculated as
\begin{equation}
  \rho_{\vec{k}} = \frac{1}{L^3} \int_{V} \rho(\vec{x}) \,{\rm e}^{-i \vec{k}\vec{x}} {\rm d}\vec{x},
\label{eqnFTcoeff}
\end{equation}
where the integration is over one instance of the periodic box.

More generally, the periodic Fourier series features the following
orthogonality and closure relationships:
\begin{equation}
\frac{1}{L^3} \int {\rm d}\vec{x}\; {\rm e}^{i(\vec{k}-\vec{k}')} = \delta_{\vec{k},\vec{k}'},
\end{equation}
\begin{equation}
\frac{1}{L^3} \sum_{\vec{k}} {\rm e}^{i\vec{k}\vec{x}} = \delta(\vec{x}), 
\end{equation}
where the first relation gives a Kronecker delta, the second a Dirac
$\delta$-function.

Let's now look at the Poisson equation again and replace the potential
and the density field with their corresponding Fourier series:
\begin{equation}
\nabla^2 \left(
\sum_{\vec{k}} \Phi_{\vec{k}}\, {\rm e}^{ i\vec{k}\vec{x}}
\right)
 = 4\pi G \left(
\sum_{\vec{k}} \rho_{\vec{k}}\, {\rm e}^{ i\vec{k}\vec{x}}
\right).
\end{equation}
We see that we can easily carry out the spatial derivate on the left
hand side, yielding:
\begin{equation}
\sum_{\vec{k}}  \left( - \vec{k}^2
\Phi_{\vec{k}} \right)\, {\rm e}^{ i\vec{k}\vec{x}}
 = 4\pi G 
\sum_{\vec{k}} \rho_{\vec{k}}\, {\rm e}^{ i\vec{k}\vec{x}} .
\end{equation}
The equality must hold for each of the Fourier modes separately, hence
we infer
\begin{equation}
\Phi_{\vec{k}} 
 = - \frac{4\pi G}{
\vec{k}^2}  \rho_{\vec{k}}.
\end{equation}
Comparing with equation (\ref{eqnFT}), this means we have identified
the Green's function of the Poisson equation in a periodic space as
\begin{equation}
g_{\vec{k}} 
 = - \frac{4\pi G}{
\vec{k}^2}. 
\end{equation}

\subsubsection{The discrete Fourier transform (DFT)}

The above considerations were still for a continuos density field.  On
a computer, we will usually only have a discretized version of the
field $\rho(\vec{x})$, defined at a set of points. Assuming we have
$N$ equally spaced points per dimension, the $\vec{x}$ positions may
only take on the discrete positions
\begin{equation}
\vec{x}_{\vec{p}} = \frac{L}{N}
\left(
\begin{array}{c}
p_1\\
p_2\\
p_3
\end{array}
\right) \;\; \mbox{where $p_1, p_2, p_3 \in \{0, 1,\ldots, N-1\}$}.
\end{equation}
With the replacement ${\rm d^3}\vec{x} \to (L/N)^3$, we can cast the
Fourier integral (\ref{eqnFTcoeff}) into a discrete sum:
\begin{equation}
\rho_{\vec{k}} = \frac{1}{N^3} \sum_{\vec{p}} \rho_{\vec p} \, {\rm e}^{-i\vec{k}\vec{x}_{\vec{p}}}.
\end{equation}
Because of the periodicity and the finite number of density values
that is summed over, it turns out that this also restricts the number
of $\vec{k}$ values that give different answers -- shifting $\vec{k}$
in any of the dimensions by $N$ times the fundamental mode $2\pi/L$
gives again the same result. We may then for example select as primary
set of $\vec{k}$-modes the values
\begin{equation}
\vec{k}_{\vec{l}} = \frac{2\pi }{L}
\left(
\begin{array}{c}
l_1\\
l_2\\
l_3
\end{array}
\right) \;\; \mbox{where $l_1, l_2, l_3 \in \{0, 1,\ldots, N-1\}$},
\end{equation}
and the construction of $\rho$ through the Fourier series becomes a
finite sum over these $N^3$ modes.  We have now arrived at the {\em
  discrete Fourier transform} (DFT), which can equally well be written
as:
\begin{equation}
  \hat\rho_{\vec{l}} = \frac{1}{N^3} \sum_{\vec{p}} \rho_{\vec{p}} \, {\rm e}^{-i \frac{2\pi}{N} \vec{l}\vec{p}},
\end{equation}
\begin{equation}
  \rho_{\vec{p}} = \sum_{\vec{l}} \hat\rho_{\vec{l}} \, {\rm e}^{i \frac{2\pi}{N} \vec{l}\vec{p}}.
\end{equation}

Here are some notes about different aspects of the Fourier pair
defined by these relations:
\begin{itemize}
\item The two transformations are an invertible linear mapping of a
  set of $N^3$ (or $N$ in 1D) complex values $\rho_{\vec{p}}$ to $N^3$
  complex values $\hat\rho_{\vec{l}}$, and vice versa.
\item To label the frequency values, $\vec{k} = (2\pi/L) \cdot
  \vec{l}$, one often conventionally uses the set $l\in\{-N/2, \ldots,
  -1,0,1,\ldots, \frac{N}{2}-1\}$ instead of $l\in\{0, 1,\ldots,
  N-1\}$, which is always possible because shifting $l$ by multiples
  of $N$ does not change anything as this yields only a $2\pi$
  phase factor. With this convention, the occurrence of both negative and
  positive frequencies is made more explicit, and they are arranged
  quasi-symmetrically in a box in $\vec{k}$-space centered on
  $\vec{k}=(0,0,0)$. The box extends out to
\begin{equation}
k_{\rm max} = \frac{N}{2} \frac{2\pi}{L},
\end{equation}
which is the so-called Nyquist frequency
\citep[e.g.][]{Diniz2002}. Adding waves beyond the Nyquist frequency
in a reconstruction of $\rho$ on a given grid would add redundant
information that could not be unambiguously recovered from the
discretized density field. (Instead, the power in these waves would be
erroneously mapped to lower frequencies -- this is called {\em
  aliasing}, see also the so-called {\em sampling theorem}.)
\item Parseval's theorem relates the quadratic norms of the transform
  pair, namely
\begin{equation}
\sum_{\vec{p}} |\rho_{\vec{p}}|^2  = N^3  \sum_{\vec{l}} |\hat\rho_{\vec{l}}|^2. 
\end{equation}

\item The $1/N^3$ normalization factor could equally well be placed in
  front of the Fourier series instead of the Fourier transform, or one
  may split it symmetrically and introduce a factor $1/\sqrt{N^3}$ in
  front of both. This is just a matter of convention, and all of these
  alternative conventions are sometimes used.

\item In fact, many computer libraries for the DFT will omit the
  factor $N$ completely and leave it up to the user to introduce it
  where needed. Commonly, the DFT library functions define as forward
  transform of a set of $N$ complex numbers $x_j$, with
  $j\in\{0,\ldots,N-1\}$, the set of $N$ complex numbers:
\begin{equation}
y_k = \sum_{j=0}^{N-1} x_j \, {\rm e}^{-i \frac{2\pi}{N} j\cdot k}.
\end{equation}
The backwards transform is then defined as 
\begin{equation}
y_k = \sum_{j=0}^{N-1} x_j \, {\rm e}^{i \frac{2\pi}{N} j\cdot k}.
\end{equation}
This form of writing the Fourier transform is now nicely symmetric,
with the {\em only difference} between forward and backward transforms
being the sign in the exponential function. However, in this case we
have that ${\cal F}^{-1}({\cal F}(\vec{x})) = N \vec{x}$, i.e.~to get
back to the original input vector $\vec{x}$ one must eventually divide
by $N$. Note that the multi-dimensional transforms are simply
Cartesian products of one-dimensional transforms, i.e.~those are
obtained as straightforward generalizations of the one-dimensional
definition.

\item Computing the DFT of $N$ numbers has in principal a
  computational cost of order ${\cal O}(N^2)$. This is because for
  each of the $N$ numbers one has to calculate $N$ terms and sum them
  up. Fortunately, in 1965, the {\em Fast Fourier Transform} (FFT)
  algorithm \citep{Cooley1965} has been discovered
  \citep[interestingly, Gauss had already known something
  similar;][]{Gauss1866}. This method for calculating the DFT
  subdivides the problem recursively into smaller and smaller
  blocks. It turns out that this divide and conquer strategy can
  reduce the computational cost to ${\cal O}(N\log N)$, which is a
  very significant difference. The result of the FFT algorithm is
  mathematically identical to the DFT.  But actually, in practice, the
  FFT is even better than a direct computation of the DFT, because as
  an aside the FFT algorithm also reduces the numerical floating point
  round-off error that would otherwise be incurred. It is ultimately
  only because of the existence of the FFT algorithm that Fourier
  methods are so widely used in numerical calculations and applicable
  to even very large problem sizes.

\end{itemize}

\subsubsection{Storage conventions for the DFT}

\begin{figure}
\sidecaption
\resizebox{7.5cm}{!}{\includegraphics{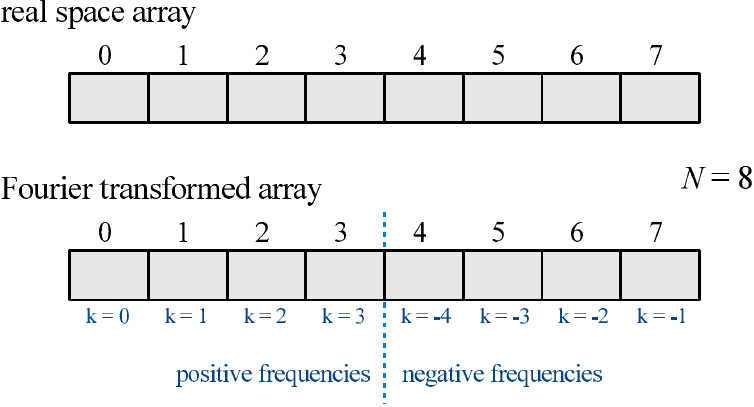}}
\caption{Commonly employed storage convention for DFTs. The positive
  frequencies are stored in the lower half of the array, the negative
  ones in the upper half.} \label{fig_dft_storage}
\end{figure}

Most numerical libraries for computing the FFT store both the original
field and its Fourier transform as simple arrays indexed by $k \in
\{0,\ldots, N-1\}$. The negative frequencies will then be stored in
the upper half of the array, consistent with what one obtains by
subtracting $N$ from the linear index. The example shown in
Figure~\ref{fig_dft_storage} for $N=8$ in 1D may help to make this
clear.

Correspondingly, in 2D, the grid of real-space values is mapped to a
grid of $k$-space values of the same dimensions. Again, negative
frequencies seem to be stored `backwards', with the smallest negative
frequency having the largest linear index, and the most negative
frequency appearing as first value past the middle of the mesh. But
note that this is consistent with the translational invariance in
$k$-space with respect to shifts of the indices by multiples of
$N$. 

\begin{figure}
\sidecaption
\resizebox{7.5cm}{!}{\includegraphics{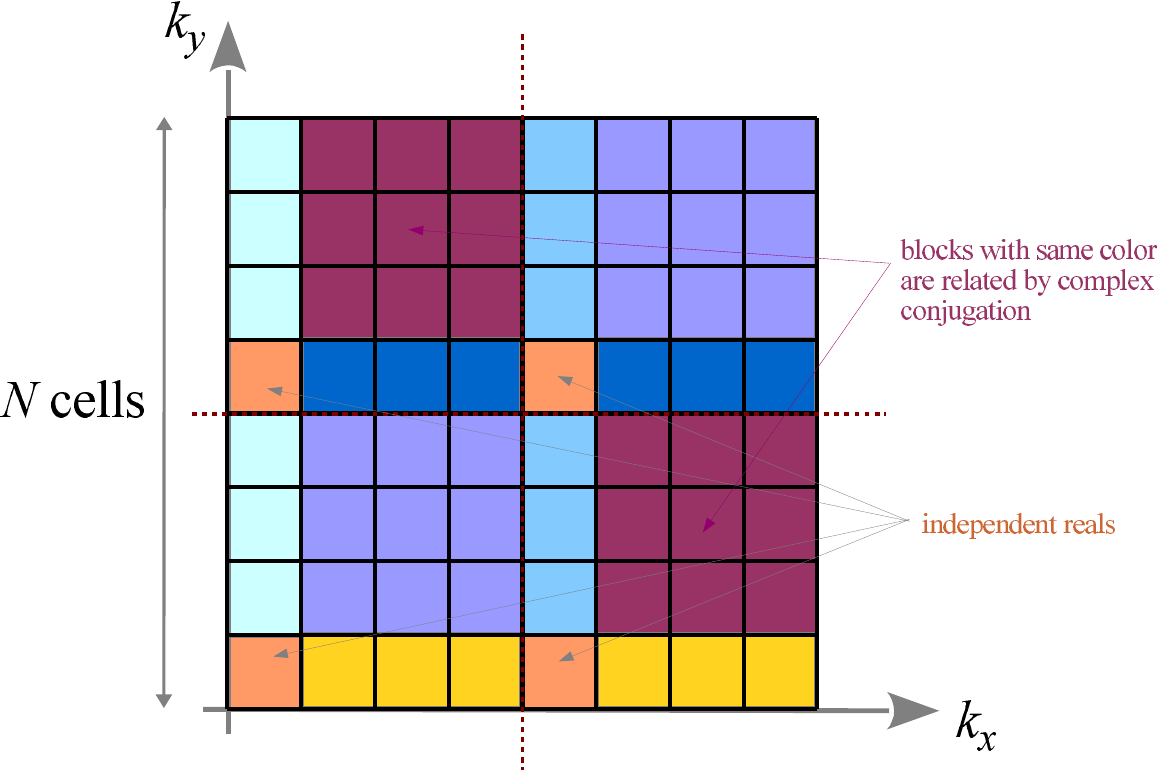}}
\caption{Sketch illustrating the implications of the reality
  constraint for the FFT of a field of reals in 2D. Different pairs of
  cells are related to each other as complex conjugate numbers
  (labeled as colored blocks), and some are aliased to themselves
  (orange) so that they end up being real. }
\label{fig_real_fft}
\end{figure}

Finally, when we have a real real-space field (such as the physical
density), the discrete Fourier transform fulfills a reality constraint
of the form $\hat\rho_{\vec{k}} = \hat\rho_{-\vec{k}}^\star$. This
implies a set of relations between the complex values that make up the
Fourier transform of $\rho$, reducing the number of values that can be
chosen arbitrarily. What does this imply in the discrete case?
Consider the sketch shown in Fig.~\ref{fig_real_fft}, in which regions
of like color are related to each other by the reality
constraint. Note that $k_x= N/2$ indices are aliased to themselves
under complex conjugation, i.e.~negating this gives $k_x={-N/2}$, but
since $N$ can be added, this mode really maps again to $k_x =
N/2$. Nevertheless, for the yellow regions there are always different
partner cells when one considers the corresponding $-\vec{k}$
cell. Only for the orange cells this is not the case; those are mapped
to themselves and are hence real due to the reality contraint.

If we now count how many independent numbers we have in the Fourier
transformed grid of a 2D real field, we find
\begin{equation}
  2\left(\frac{N}{2}-1\right)^2 \times 2 \;+\; 4 \left(\frac{N}{2}-1\right) \times 2 \; +\;  4 \times 1.
\label{eqsum}
\end{equation}
The first term accounts for the two square-shaped regions that have
different mirrored regions. Those contain $(\frac{N}{2}-1)^2$ complex
numbers, each with two independent real and imaginary values. Then
there are 4 different sections of rows and columns that are related to
each other by mirroring in $k$-space. Those contain $(\frac{N}{2}-1)$
complex numbers each. Finally, there are 4 independent cells that are
real and hence account for one independent value each. Reassuringly,
the sum of equation (\ref{eqsum}) works out to $N^2$, which is the
result we expect: the number of independent values in Fourier space
must be exactly equal to the $N^2$ real values we started out with,
otherwise we would not expect a strictly reversible transformation.

\subsubsection{Non-periodic problems with `zero padding'}

Can we use the FFT/DFT techniques discussed above also to calculate
non-periodic force fields? At first, this may seem impossible since
the DFT is intrinsically periodic. However, through the
so-called zero-padding trick one can circumvent this limitation.

\begin{figure}
\sidecaption
\resizebox{7.5cm}{!}{\includegraphics{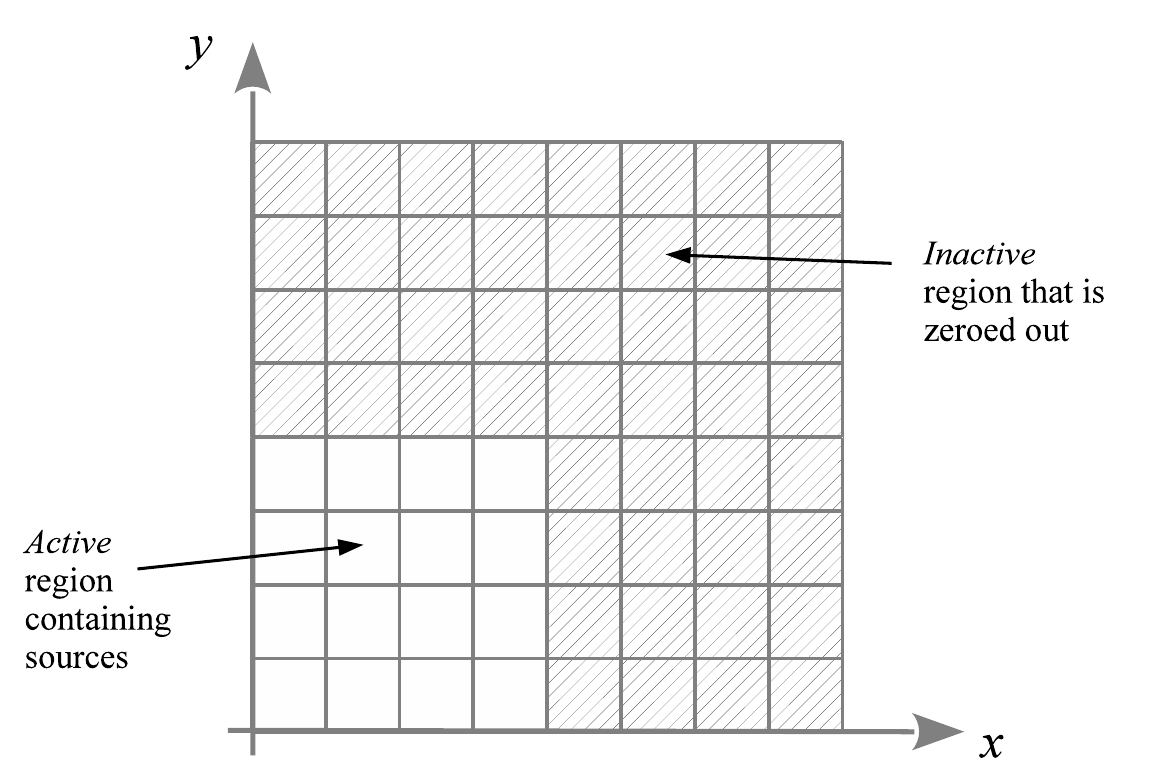}}
\caption{Sketch of zero padding used to treat non-periodic problems
  with the discrete Fourier transform.}
\label{fig_zeropadding}
\end{figure}

Let's discuss the procedure based on a 2D example (it works also in
1D or 3D, of course):
\begin{enumerate}

\item We need to arrange our mesh such that the source distribution
  lives only in one quarter of the mesh, the rest of the density field
  needs to be zeroed out. Schematically we hence have the situation
  depicted in Figure~\ref{fig_zeropadding}.

\item We now set up our desired real-space Green's function, i.e.~this
  is the response of a mass at the origin. The Green's function
  for the whole mesh is set-up as $g_{N-i,j} = g_{i, N-j}= g_{N-i,N-j}
  = g_{i,j}$ where $0\le i,j \le N/2$. This is equivalent to defining
  $g$ everywhere on the mesh, and using as relevant distance the
  distance to the \emph{nearest periodic image} of the origin. Note
  that by replicating $g$ with the condition of periodicity, the
  tessellated mesh then effectively yields a Green's function that is
  nicely symmetric around the origin.

\item We now want to carry out the real-space convolution
\begin{equation}
\phi = g \star \rho 
\end{equation}
by using the definition of the discrete, periodic convolution
\begin{equation}
\Phi_{\vec{p}} = \sum_{\vec{n}} g_{\vec{p}-\vec{n}}\, \rho_{\vec{n}},
\label{eqnzeropadding}
\end{equation}
where both $g$ and $\rho$ are treated as periodic fields for which
adding multiples of $N$ to the indices does not change anything.  We
see that this sum indeed yields the correct result for the
non-periodic potential in the quarter of the mesh that contains our
source distribution. This is because the Green's function `sees' only
one copy of the source distribution in this sector; the zero-padded
region is big enough to prevent any cross-talk from the (existing)
periodic images of the source distribution. This is different in the
other three quadrants of the mesh. Here we obtain incorrect potential
values that are basically useless and need to be discarded.

\item Given that equation (\ref{eqnzeropadding}) yields the correct
  result in the region of the mesh covered by the sources, we may now
  just as well use periodic FFTs in the usual way to carry out this
  convolution quickly! A downside of this procedure is that it
  features an enlarged cost in terms of CPU and memory usage. Because
  we have to effectively double the mesh-size compared to the
  corresponding periodic problem, the cost goes up by a factor of 4 in
  2D, and by a factor of 8 in 3D.

\item We note that \citet{James1977} proposed an ingenious trick based
  that allows a more efficient treatment of isolated source
  distributions. Through suitably determined correction masses on the
  boundaries, the memory and cpu cost can be reduced compared to the
  zero-padding approach described above.

\end{enumerate}

\subsection{Multigrid techniques}  \label{sec_multigrid}

Let's return once more to the problem of solving Poisson's equation,
\begin{equation}
\nabla^2\Phi = 4\pi G \rho,
\end{equation}
and consider first the one-dimensional problem, i.e.
\begin{equation}
\frac{\partial^2\Phi}{\partial x^2} = 4\pi G \rho(x).
\end{equation}
The spatial derivative on the left hand-side can be approximated as
\begin{equation}
\left(\frac{\partial^2\Phi}{\partial x^2}\right)_i \simeq
 \frac{\Phi_{i+1} - 2 \Phi_i +\Phi_{i-1}}{h^2},
\label{eqn2ndder}
\end{equation}
where we have assumed that $\Phi$ is discretized with $N$ points on a
regular mesh with spacing $h$, and $i$ is the cell index. This means
that we have the equations
\begin{equation}
 \frac{\Phi_{i+1} - 2 \Phi_i +\Phi_{i-1}}{h^2} = 4 \pi G \rho_i.
\end{equation}
There are $N$ of these equations, for the $N$ unknowns $\Phi_i$, with
$i \in \{ 0, 1,\ldots, N-1\}$. This means we should in principle be
able to solve this algebraically! In other words, the system of
equations can be rewritten as a standard linear set of equations, in
the form
\begin{equation}
\vec{A} \vec{x} = \vec{b},
\label{eqnDirect}
\end{equation}
with a vector of unknowns, $\vec{x} = (\Phi_i)$, and a right hand side
$\vec{b} = \frac{4\pi G}{h^2} \vec{\rho}$. In the 1D case, the matrix
$\vec{A}$ (assuming periodic boundary conditions) is explicitly given
as
\begin{equation}
\vec{A}  = \left(
\begin{array}{rrrrrr}
-2 &  1  &    &   & & 1  \\
1  & -2  &  1 &   &  & \\
   &  1  & -2 & 1 &  & \\
   &    & \ldots &  &  &  \\
   &   &   & 1  & -2 & 1  \\
1  &     &  &    & 1 & -2  \\
\end{array}
\right).
\end{equation}
Solving equation (\ref{eqnDirect}) directly constitutes a matrix
inversion that can in principle be carried out by LU-decomposition or
Gauss elimination with pivoting \citep[e.g.][]{Press1992}. However,
the computational cost of these procedures is of order ${\cal
  O}(N^3)$, meaning that it becomes extremely costly with growing $N$,
and rather sooner than later infeasible, already for problems of small
to moderate size.

\subsubsection{Jacobi iteration}

However, if we are satisfied with an approximate solution, then we can
turn to iterative solvers that are much faster.  Suppose we decompose
the matrix $\vec{A}$ as
\begin{equation}
\vec{A} = \vec{D} - (\vec{L}+\vec{U}),
\end{equation}
where $\vec{D}$ is the diagonal part, $\vec{L}$ is the (negative)
lower diagonal part and $\vec{U}$ is the upper diagonal part. Then we
have
\begin{equation}
\left[\vec{D} - (\vec{L}+\vec{U})\right] \vec{x} = \vec{b},
\end{equation}
and from this
\begin{equation}
\vec{x} = \vec{D}^{-1}\vec{b} + \vec{D}^{-1}(\vec{L}+\vec{U}) \vec{x}.
\end{equation}
We can use this to define an iterative sequence of vectors $\vec{x}^{n}$:
\begin{equation}
\vec{x}^{(n+1)} = \vec{D}^{-1}\vec{b} + \vec{D}^{-1}(\vec{L}+\vec{U}) \vec{x}^{(n)}.
\end{equation}
This is called Jacobi iteration \citep[e.g.][]{saad2003}. Note that
$\vec{D}^{-1}$ is trivially obtained because $\vec{D}$ is
diagonal. i.e.~here $(\vec{D}^{-1})_{ii} = 1 / \vec{A}_{ii}$.

The scheme converges if and only if the so-called convergence matrix
\begin{equation}
\vec{M} = \vec{D}^{-1}(\vec{L}+\vec{U})
\end{equation}
has only eigenvalues that are less than 1, or in other words, that the
spectral radius $\rho_s(\vec{M})$ fullfils
\begin{equation}
\rho_s(\vec{M}) \equiv \max_i |\lambda_i| < 1.
\end{equation}
We can easily derive this condition by considering the error vector
of the iteration. At step $n$ it is defined as
\begin{equation}
\vec{e}^{(n)} \equiv \vec{x}_{\rm exact} - \vec{x}^{(n)},
\end{equation}
where $\vec{x}_{\rm exact}$ is the exact solution. We can use this to
write the error at step $n+1$ of the iteration as
\begin{equation}
\vec{e}^{(n+1)} = \vec{x}_{\rm exact} - \vec{x}^{(n+1)} = 
\vec{x}_{\rm exact}  - 
\vec{D}^{-1}\vec{b} - \vec{D}^{-1}(\vec{L}+\vec{U}) \vec{x}^{(n)}
= 
\vec{M} \vec{x}_{\rm exact} - \vec{M}\vec{x}^{(n)} = \vec{M} \vec{e}^{(n)}  
\end{equation}
Hence we find
\begin{equation}
\vec{e}^{(n)} = \vec{M}^n \vec{e}^{(0)}.
\end{equation} 
This implies $|\vec{e}^{(n)}| \le [\rho_s(\vec{M})]^n |\vec{e}^{(0)}|$, and hence
convergence if the spectral radius is smaller than 1.

For completeness, we state the Jacobi iteration rule for the Poisson
equation in 3D when a simple 2-point stencil is used in each dimension
for estimating the corresponding derivatives:
\begin{eqnarray}
\Phi_{i,j,k}^{(n+1)} & = & 
\frac{1}{6}\left(
\Phi_{i+1,j,k} + \Phi_{i-1,j,k} + 
\Phi_{i,j+1,k} + \Phi_{i,j-1,k} + 
\Phi_{i,j,k+1} + \Phi_{i,j,k-1} \right. \nonumber \\
& & \left. - 4\pi G h^2 \rho_{i,j,k}
\right) .
\end{eqnarray}

\subsubsection{Gauss-Seidel iteration}

The central idea of Gauss-Seidel iteration is to use the updated
values as soon as they become available for computing further
updated values. We can formalize this as follows. Adopting the same
decomposition of $\vec{A}$ as before, we can write
\begin{equation}
(\vec{D}-\vec{L})\vec{x} = \vec{U}\vec{x} + \vec{b},
\end{equation}
from which we obtain
\begin{equation}
\vec{x} = (\vec{D}-\vec{L})^{-1}\vec{U}\vec{x} + (\vec{D}-\vec{L})^{-1}\vec{b},
\end{equation}
suggesting the iteration rule
\begin{equation}
  \vec{x}^{(n+1)} = (\vec{D}-\vec{L})^{-1}\vec{U}\vec{x}^{(n)} + (\vec{D}-\vec{L})^{-1}\vec{b}.
\end{equation}
This seems at first problematic, because we can't easily compute
$(\vec{D}-\vec{L})^{-1}$. But we can modify the last equation as
follows:
\begin{equation}
\vec{D} \vec{x}^{(n+1)} = \vec{U}\vec{x}^{(n)} + \vec{L}\vec{x}^{(n+1)} + \vec{b}.
\end{equation}
From which we get the alternative form:
\begin{equation}
  \vec{x}^{(n+1)} = \vec{D}^{-1} \vec{U}\vec{x}^{(n)} + \vec{D}^{-1}\vec{L}\vec{x}^{(n+1)} + \vec{D}^{-1}\vec{b}.
\end{equation}
Again, this may seem of little help because it looks like
$\vec{x}^{(n+1)}$ would only be implicitly given. However, if we start
computing the new elements in the first row $i=1$ of this matrix
equation, we see that no values of $\vec{x}^{(n+1)}$ are actually
needed, because $\vec{L}$ has only elements below the diagonal. For
the same reason, if we then proceed with the second row $i=2$, then
with $i=3$, etc., only elements of $\vec{x}^{(n+1)}$ from rows above
the current one are needed. So we can calculate things in this order
without problem and make use of the already updated values. It turns
out that this speeds up the convergence quite a bit, with one
Gauss-Seidel step often being close to two Jacobi steps.

\subsubsection{Red black ordering}

A problematic point about Gauss-Seidel is that the equations have to
be solved in a specific sequential order, meaning that this part
cannot be parallelized. Also, the result will in general depend on
which element is selected to be the first. To overcome this problem,
one can sometimes use so-called red-black ordering, which effectively
is a compromise between Jacobi and Gauss-Seidel.

\begin{figure}
  \sidecaption
  \resizebox{7.4cm}{!}{\includegraphics{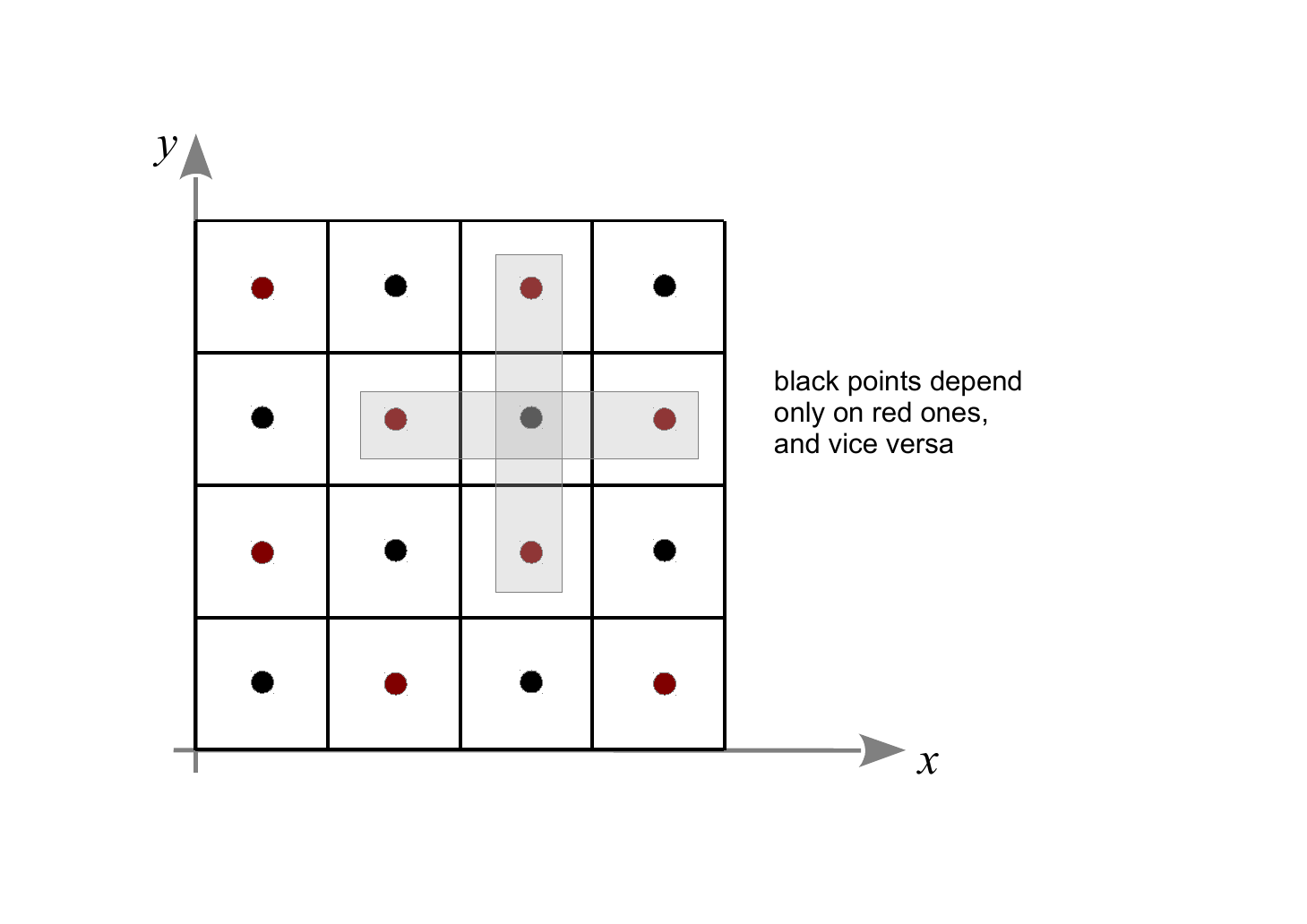}}
  \caption{Red-black ordering in which two interleaved chessboard-like
    patterns are formed that can be independently processed with
    immediate updating.}
\label{fig_redblack}
\end{figure}

Certain update rules, such as that for the Poisson equation, allow a
decomposition of the cells into disjoint sets whose update rules
depend only on cells from other sets, as shown in
Fig.~\ref{fig_redblack}. For example, for the Poisson equation, this
is the case for a chess-board like pattern of `red' and `black' cells.

One can then first update all the black points (which rely only on the
red points), followed by an update of all the red points (which rely
only on the black ones). In the second of these two half-steps, one
can then use the updated values from the first half-step, making it
intuitively clear why such a scheme can almost double the convergence
rate relative to Jacobi.

\subsubsection{The multigrid technique}

Iterative solvers like Jacobi or Gauss-Seidel often converge quite
slowly, in fact, the convergence seems to ``stall'' after a few steps
and proceeds only anemically. One also observes that high-frequency
errors in the solution are damped out quickly by the iterations, but
long-wavelength errors die out much more slowly. Intuitively this is
not unexpected: In every iteration, only neighboring points
communicate, so the information ``travels'' only by one cell (or more
generally, one stencil length) per iteration. And for convergence, it
has to propagate back and forth over the whole domain a few times.

\runinhead{Idea} By going to a coarser mesh, we may be able to compute
an improved initial guess which may help to speed up the convergence
on the fine grid \citep{Brandt1977}. Note that on the coarser mesh,
the relaxation will be computationally cheaper (since there are only
$1/8$ as many points in 3D, or $1/4$ in 2D), and the convergence rate
should be faster, too, because the perturbation is there less smooth
and effectively on a smaller scale relative to the coarser grid.

So schematically, we, for example, might imagine an iteration scheme
where we first iterate the problem $\vec{A}\vec{x} = \vec{b}$ on a
mesh with cells $4h$, i.e.~for times coarser than the fine mesh. Once
we have a solution there, we continue to iterate it on a mesh
coarsened with cell size $2h$, and only finally we iterate to solution
on the fine mesh with cell size $h$.

A couple of questions immediately come up when we want to work out the
details of this basic idea:
\begin{enumerate}
\item How do we get from a coarse solution to a guess on a finer grid?
\item How should we solve $\vec{A}\vec{x} = \vec{b}$ on the coarsened mesh?
\item What if there is still an error left with long wavelength on the
  fine grid?
\end{enumerate}

In order to make things work, we clearly need mappings from a finer
grid to a coarser one, and vice versa. This is the most important
issue to solve.

\subsubsection{Prolongation and restriction operations}

\runinhead{Coarse-to-fine} This transition is an interpolation step,
or in the language of multigrid methods \citep{Briggs2000}, it is
called \emph{prolongation}. Let $\vec{x}^{(h)}$ be a vector defined on
a mesh $\Omega^{(h)}$ with $N$ cells and spacing $h$, covering our
computational domain. Similarly, let $\vec{x}^{(2h)}$ be a vector
living on a coarser mesh $\Omega^{(2h)}$ with twice the spacing and
half as many points per dimension. We now define a linear
interpolation operator $\vec{I}_{2h}^{h}$ that maps points from the
coarser to the fine mesh, as follows:
\begin{equation}
\vec{I}_{2h}^{h} \,\vec{x}^{(2h)} = \vec{x}^{(h)}.
\end{equation}
A simple realization of this operator in 2D would be the following:
\begin{equation}
\vec{I}_{2h}^{h}: \;\;
\left.
\begin{array}{ll}
x^{(h)}_{2i} = x^{(2h)}_{i} & \\
               & \mbox{for $0\le i < \frac{N}{2}$}. \\
x^{(h)}_{2i+1} = \frac{1}{2}( x^{(2h)}_{i} + x^{(2h)}_{i+1}) & \\
\end{array}
\right.
\end{equation}
Here, every second point is simply injected from the coarse to the
fine mesh, and the intermediate points are linearly interpolated from
the neighboring points, which here boils down to a simple arithmetic
average.

\runinhead{Fine-to-coarse} The converse mapping represents a smoothing
operation, or a \emph{restriction} in multigrid-language. We can
define the restriction operator as
\begin{equation}
\vec{I}_{h}^{2h} \, \vec{x}^{(h)} = \vec{x}^{(2h)},
\end{equation}
which hence takes a vector defined on the fine grid $\Omega^{(h)}$ to
one that lives on the coarse grid $\Omega^{(2h)}$.  Again, lets give a
simple realization example in 2D:
\begin{equation}
\vec{I}_{h}^{2h}: \;\;
x^{(2h)}_{i} = \frac{x^{(h)}_{2i-1} + 2 x^{(h)}_{2i} + x^{(h)}_{2i+1}}{4} 
\;\;
\mbox{for $0\le i < \frac{N}{2}$} .
\end{equation}
Evidently, this is a smoothing operation with a simple 3-point stencil.

One usually chooses these two operators such that the transpose of one
is proportional to the other, i.e.~they are related as follows:
\begin{equation}
\vec{I}_{h}^{2h} = c\, [\vec{I}_{2h}^{h}]^{\rm T},
\end{equation}
where $c$ is a real number.

In a shorter notation, the above prolongation 
operator can be written as
\begin{equation}
\mbox{1D-prolongation,}  \;\; \vec{I}_{2h}^{h}: \;\; \left] \; \frac{1}{2} \;\; 1 \;\; \frac{1}{2} \;\right[,
\end{equation}
which means that every coarse point is added with these weights to
three points of the fine grid. The fine-grid points accessed with
weight $1/2$ will get contributions from two coarse grid points.
Similarly, the restriction operator can be written with the short-hand
notation
\begin{equation}
\mbox{1D-restriction,}  \;\; \vec{I}_{h}^{2h}: \;\; \left[ \; \frac{1}{4} \;\; \frac{1}{2} \;\; \frac{1}{4} \;\right].
\end{equation}
This expresses that every coarse grid point is a weighted sum of three
fine grid points.

For reference, we also state the corresponding low-order prolongation
and restriction operators in 2D:
\begin{eqnarray}
\mbox{2D-prolongation,}  \;\; \vec{I}_{2h}^{h}: & \;\; \left]
\begin{array}{ccc}
\frac{1}{4} & \frac{1}{2} & \frac{1}{4} \\
\\
\frac{1}{2} &       1     & \frac{1}{2} \\
\\
\frac{1}{4} & \frac{1}{2} & \frac{1}{4} \\
\end{array}
\right[
\\
\ \nonumber
\\
\ \nonumber
\\
\mbox{2D-restriction,}  \;\; \vec{I}_{h}^{2h}: & \;\; \left[
\begin{array}{ccc}
\frac{1}{16} & \frac{1}{8} & \frac{1}{16} \\
\\
\frac{1}{8} &       \frac{1}{4}     & \frac{1}{8} \\
\\
\frac{1}{16} & \frac{1}{8} & \frac{1}{16} \\
\end{array}
\right]
\end{eqnarray}
Corresponding extensions to 3D can be readily derived.

\subsubsection{The multigrid V-cycle}

An important role in the multigrid approach plays the error vector,
defined as
\begin{equation}
\vec{e} \equiv \vec{x}_{\rm exact} - \tilde{\vec{x}},
\end{equation}
where $\vec{x}_{\rm exact}$ is the exact solution, and
$\tilde{\vec{x}}$ the (current) approximate solution.  Another
important concept is the {\em residual}, defined as
\begin{equation}
\vec{r} \equiv \vec{b} - \vec{A}\tilde{\vec{x}}.
\end{equation}
Note that the pair of error and residual are solutions of the original
linear system, i.e.~we have
\begin{equation}
\vec{A}\vec{e} = \vec{r}.
\end{equation}

\runinhead{Coarse-grid correction scheme} We now define a function
that is supposed to return an improved solution
$\tilde{\vec{x}}'^{(h)}$ for the problem
$\vec{A}^{(h)}\vec{x}^{(h)}=\vec{b}^{(h)}$ on grid level $h$, based on
some starting guess $\tilde{\vec{x}}^{(h)}$ and a right hand side
$\vec{b}^{(h)}$. This so-called {\em coarse grid correction},
\begin{equation}
\tilde{\vec{x}}'^{(h)}  = {\rm CG}(\tilde{\vec{x}}^{(h)}, \vec{b}^{(h)}),
\end{equation}
proceeds along the following steps:
\begin{enumerate}
\item Carry out a relaxation step on $h$ (for example by using one
  Gauss-Seidel or one Jacobi iteration).
\item Compute the residual: 
$\vec{r}^{(h)} = \vec{b}^{(h)} - \vec{A}^{(h)}\tilde{\vec{x}}^{(h)}$.
\item Restrict the residual to a coarser mesh:
$\vec{r}^{(2h)} = \vec{I}_{h}^{2h}\, \vec{r}^{(h)}$.
\item Solve $\vec{A}^{(2h)}\vec{e}^{(2h)}=\vec{r}^{(2h)}$ on the
  coarsened mesh, with $\tilde{\vec{e}}^{(2h)}= 0$ as initial guess.
\item Prolong the obtained error $\vec{e}^{(2h)}$ to the finer
  mesh, $\vec{e}^{(h)} = \vec{I}_{2h}^h\, \vec{e}^{(2h)}$, and use it to
correct the current solution on the fine grid: 
$\tilde{\vec{x}}'^{(h)} = \tilde{\vec{x}}^{(h)} + \vec{e}^{(h)}$.
\item Carry out a further relaxation step on the fine mesh $h$.
\end{enumerate}

How do we carry out step 4 in this scheme? We can use recursion!
Because what we have to do in step 4 is exactly the job description of
the function ${\rm CG}(.,.)$. However, we also need a stopping
condition for the recursion, which is simply a prescription that tells
us under which conditions we should skip steps 2 to 5 in the above
scheme. We can do this by simply saying that further coarsening of the
problem should stop once we have reached a minimum number of cells
$N$. At this point we either just do the relaxation steps, or we solve
the remaining problem exactly.

\begin{figure}
\sidecaption
\resizebox{7.5cm}{!}{\includegraphics{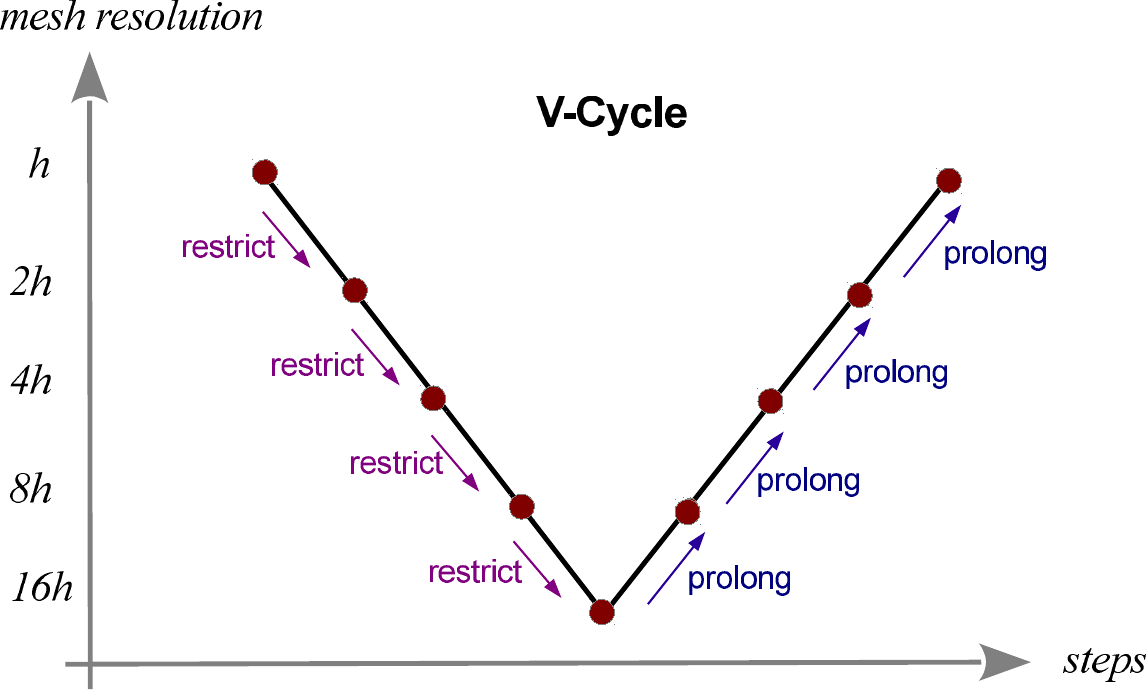}}
\caption{The typical V-cycle of a multigrid iteration scheme. The
  current solution on a fine mesh is recursively restricted to coarser
  meshes. Coarse-grid corrections are then prolonged back up to the
  finer meshes, interleaved with one Gauss-Seidl or Jacobi iteration
  at the corresponding mesh level.}
\label{fig_vcycle}
\end{figure}

\runinhead{V-Cycle} When the coarse grid correction scheme is
recursively called, we arrive at the schematic diagram shown in
Fig.~\ref{fig_vcycle} for how the iteration progresses, which is
called a V-cycle. It turns out that the V-cycle rather dramatically
speeds up the convergence rate of simple iterative solvers for linear
systems of equations. It is easy to show that the computational cost
of one V-cycle is of order ${\cal O}(N_{\rm grid})$, where $N_{\rm
  grid}$ is the number of grid cells on the fine mesh. A convergence
to truncation error (i.e.~machine precision) requires several V-cycles
and involves a computational cost of order ${\cal O}(N_{\rm grid}\log
N_{\rm grid})$. For the Poisson equation, this is the same cost
scaling as one gets with FFT-based methods. In practice, good
implementations of the two schemes should roughly be equally fast. In
cosmology, a multigrid solver is for example used by the {\small
  MLAPM} \citep{Knebe2001} and {\small RAMSES} codes
\citep{Teyssier2002}. An interesting advantage of multigrid is that it
requires less data communication when parallelized on distributed
memory machines.

One problem we haven't addressed yet is how one finds the operator
$\vec{A}^{(2h)}$ required on the coarse mesh. The two most commonly
used options for this are:
\begin{itemize}
\item Direct coarse grid approximation: Here one simply uses the same
  discrete equations on the coarse grid as on the fine grid, just
  scaled by the grid resolution $h$ as needed.  In this case, the
  stencil of the matrix does not change.
\item Galerkin coarse grid approximation: Here one defines the coarse
  operator as
\begin{equation}
\vec{A}^{(2h)} = \vec{I}_h^{2h}\, \vec{A}^{(h)} \, \vec{I}_{2h}^{h} ,
\end{equation}
which is formally the most consistent way of defining
$\vec{A}^{(2h)}$, and in this sense optimal. However, computing the
matrix in this way can be a bit cumbersome, and it may involve a
growing size of the stencil, which then leads to an enlarged
computational cost.
\end{itemize}

\subsubsection{The full multigrid method}

The V-cycle scheme discussed thus far still relies on an initial guess
for the solution, and if this guess is bad, one has to do more
V-cycles to reach satisfactory convergence.  This raises the question
on how one may get a good guess. If one is dealing with the task of
repeatedly having to solve the same problem over and over again with
only small changes from solution to solution (as will often be the
case in dynamical simulation problems) one may be able to simply use
the solution from the previous timestep as a guess. In all other
cases, one can allude to the following idea: Let's get a good guess by
solving the problem on a coarser grid first, and then interpolate the
coarse solution to the fine grid as a starting guess.

\begin{figure}
\sidecaption
\resizebox{7.5cm}{!}{\includegraphics{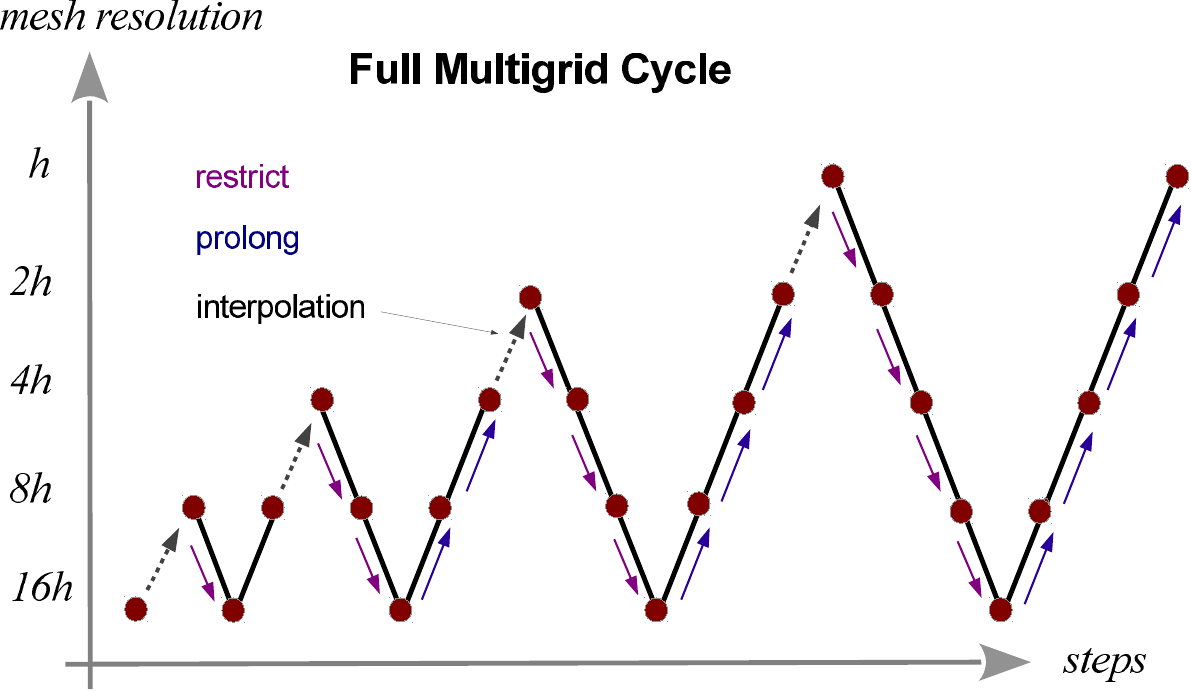}}
\caption{The full multigrid cycle in which also the problem of finding
  an adequate starting guess is addressed.}
\label{fig_fullmultigrid}
\end{figure}

But at the coarser grid, one is then again confronted with the task to
solve the problem without a starting guess. Well, we can simply
recursively apply the idea again, and delegate the finding of a good
guess to a yet coarser grid, etc. This then yields the \emph{full
  multigrid cycle}, as depicted in Fig.~\ref{fig_fullmultigrid}. It
involves the following steps:
\begin{enumerate}
\item Initialize the right hand side on all grid levels,
  $\vec{b}^{(h)}$, $\vec{b}^{(2h)}$, $\vec{b}^{(4h)}$, $\ldots$,
  $\vec{b}^{(H)}$, down to some coarsest level $H$.
\item Solve the problem (exactly) on the coarsest level $H$.
\item Given a solution on level $i$ with spacing $2h$, map it to the
  next level $i+1$ with spacing $h$ and obtain the initial guess
  $\tilde{\vec{x}}^{(h)} = I_{2h}^h\,\vec{x}^{(2h)}$.
\item Use this starting guess to solve the problem on the 
level $i+1$ with one V-cycle.
\item Repeat Step 3 until the finest level is reached.
\end{enumerate}

The computational cost of such a full multigrid cycle is still of
order the number of mesh cells, as before.

\subsection{Hierarchical multipole methods (``tree codes'')}

Another approach for a real-space evaluation of the gravitational
field are so-called tree codes \citep{Barnes1986}.  In cosmology, they
are for example used in the {\small PKDGRAV/GASOLINE}
\citep{Wadsley2004} and {\small GADGET} \citep{Springel2001gadget,
  Springel2005} codes.

\subsubsection{Multipole expansion}

The central idea is here to use the multipole expansion of a distant
group of particle to describe its gravity, instead of summing up the
forces from all individual particles.

\begin{figure}
\sidecaption
\resizebox{5.5cm}{!}{\includegraphics{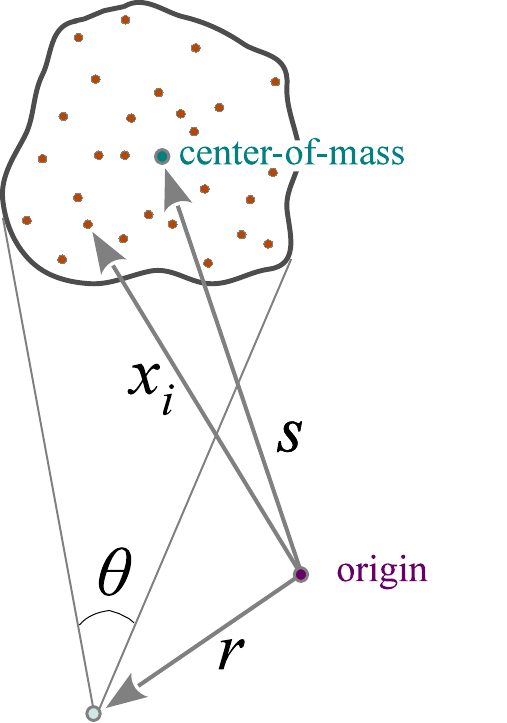}}
\caption{Multipole expansion for a group of distant
  particles. Provided the reference point $\vec{r}$ is sufficiently
  far away, the particles are seen under a small opening angle
  $\theta$, and the field created by the particle group can be
  approximated by the monopole term at its center of mass, augmented
  with higher order multipole corrections if desired.}
\label{fig_multipole}
\end{figure}

\noindent The potential of the group is given by
\begin{equation}
\Phi(\vec{r}) = -G \sum_i \frac{m_i}{|\vec{r} - \vec{x}_i|},
\end{equation}
which we can re-write as
\begin{equation}
\Phi(\vec{r}) = -G \sum_i \frac{m_i}{|\vec{r} - \vec{s} + \vec{s} - \vec{x}_i|}.
\end{equation}
Now we expand the denominator assuming $|\vec{x}_i - \vec{s}| \ll
|\vec{r} - \vec{s}|$, which will be the case provided the {\em opening
  angle} $\theta$ under which the group is seen is sufficiently small,
as sketched in Figure~\ref{fig_multipole}.  We can then use the Taylor
expansion
\begin{equation}
  \frac{1}{|\vec{y} +  \vec{s} - \vec{x}_i|} = \frac{1}{|\vec{y}|}
  - \frac{\vec{y}\cdot (\vec{s}-\vec{x}_i)}{|\vec{y}|^3} 
  +\frac{1}{2}\frac{\vec{y}^T\left[ 3 (\vec{s}-\vec{x}_i) (\vec{s}-\vec{x}_i)^T - (\vec{s}-\vec{x}_i)^2\right] \vec{y}}{|\vec{y}|^5} + \ldots  ,
\end{equation}
where we introduced $\vec{y}\equiv \vec{r} - \vec{s}$ as a
short-cut. The first term on the right hand side gives rise to the
monopole moment, the second to the dipole moment, and the third to the
quadrupole moment. If desired, one can continue the expansion to ever
higher order terms.

These multipole moments then become properties of the group of particles:
\begin{equation}
\mbox{monopole:} \;\;\; M = \sum_i m_i ,
\end{equation}
\begin{equation}
\mbox{quadrupole:} \;\;\; Q_{ij} = \sum_k m_k \left[ 3(\vec{s}-\vec{x}_k)_i (\vec{s}-\vec{x}_k)_j - \delta_{ij}(\vec{s}-\vec{x}_k)^2\right].
\end{equation}
The dipole vanishes, because we carried out the expansion relative to
the center-of-mass, defined as
\begin{equation}
\vec{s} = \frac{1}{M}\sum_i m_i \vec{x}_i.
\end{equation}
If we restrict ourselves to terms of up to quadrupole order, we hence arrive at the expansion
\begin{equation}
\Phi(\vec{r}) = -G\left(\frac{M}{|\vec{y}|} + \frac{1}{2} \frac{\vec{y}^T\vec{Q}\vec{y}}{|\vec{y}|^5} \right) , \;\;\;\;\;\; \vec{y} =\vec{r}-\vec{s}, 
\end{equation}
from which also the force can be readily obtained through differentiation. Recall that we expect the expansion to be accurate if
\begin{equation}
\theta \simeq \frac{ \left<|\vec{x}_i - \vec{s}|\right>}{|\vec{y}|} \simeq \frac{l}{y} \ll 1,
\end{equation}
where $l$ is the radius of the group.

\subsubsection{Hierarchical grouping}

Tree algorithms are based on a hierarchical grouping of the particles,
and for each group, one then pre-computes the multipole moments for
later use in approximations of the force due to distant
groups. Usually, the hierarchy of groups is organized with the help of
a tree-like data structure, hence the name ``tree algorithms''.

\begin{figure}
\sidecaption[b]
\resizebox{5.0cm}{!}{\includegraphics{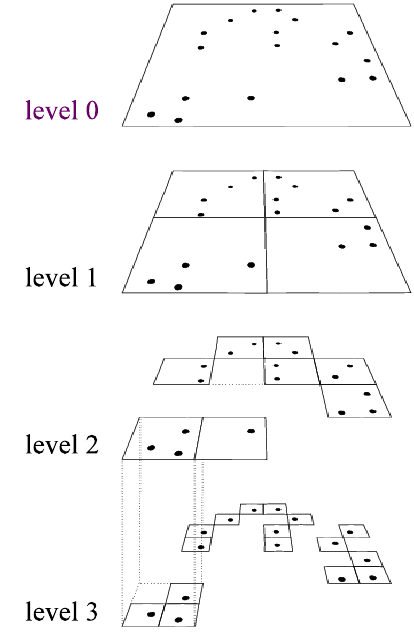}}
\caption{Organization of the \citet{Barnes1986} tree in two dimensions
(quad tree). All particles are enclosed in a square-shaped box. This
is then hierarchically subdivided until each particle finds itself in
a node on its own. Empty cells do not need to be stored.}
\label{fig_tree2d}
\end{figure}

There are different strategies for defining the groups. In the popular
\citet{Barnes1986} oct-tree, one starts out with a cube that contains
all the particles. This cube is then subdivided into 8 sub-cubes of
half the size in each spatial dimension. One continues with this
refinement recursively until each subnode contains only a single
particle. Empty nodes (sub-cubes) need not be
stored. Figure~\ref{fig_tree2d} shows a schematic sketch how this can
look like in two dimensions.

\begin{itemize}
\item We note that the oct-tree is not the only possible grouping
  strategy. Sometimes kd-trees \citep{Stadel2001}, or binary trees
  where subdivisions are done along alternating spatial axes are used.
\item An important property of such hierarchical, tree-based groupings
  is that they are geometrically highly flexible and adjust to any
  clustering state the particles may have. They are hence
  automatically adaptive.
\item Also, there is no significant slow-down when severe clustering
  starts.
\item The simplest way to construct the hierarchical grouping is to
  sequentially insert particles into the tree, and then to compute the
  multipole moments recursively.
\end{itemize}

\subsubsection{Tree walk}

The force calculation with the tree then proceeds by {\em walking the
  tree}. Starting at the root node, one checks for every node whether
the opening angle under which it is seen is smaller than a prescribed
tolerance angle $\theta_c$. If this is the case, the multipole
expansion of the node can be accepted, and the corresponding partial
force is evaluated and added to an accumulation of the total
force. The tree walk along this branch of the tree can then be
stopped. Otherwise, one must open the tree node and consider all its
sub-nodes in turn.

The resulting force is {\em approximate} by construction, but the
overall size of the error can be conveniently controlled by the
tolerance opening angle $\theta_c$ \citep[see also][]{Salmon1994}. If
one makes this smaller, more nodes will have to be opened. This will
make the residual force errors smaller, but at the price of a higher
computational cost. In the limit of $\theta_c \to 0$ one gets back to
the expensive direct summation force.

An interesting variant of this approach to walk the tree is obtained
by not only expanding the potential on the source side into a
multipole expansion, but also around the target coordinate. This can
yield a substantial additional acceleration and results in so-called
fast multipole methods (FFM). The {\small FALCON} code of
\citet{Dehnen2000,Dehnen2002} employs this approach. A further
advantage of the FFM formulation is that force anti-symmetry is
manifest, so that momentum conservation to machine precision can be
achieved. Unfortunately, the speed advantages of FFM compared to
ordinary tree codes are significantly alleviated once individual
time-step schemes are considered. Also, FFM is more difficult to
parallelize efficiently on distributed memory machines.

\subsubsection{Cost of the tree-based force computations}

How do we expect the total cost of the tree algorithm to scale with
particle number $N$?  For simplicity, let's consider a sphere of size
$R$ containing $N$ particles that are approximately
homogeneously distributed.
The mean particle spacing of these
particles will then be
\begin{equation}
d = \left[\frac{(4\pi/3) R^3}{N}\right]^{1/3}.
\end{equation}
We now want to estimate the number of nodes that we need for
calculating the force on a central particle in the middle of the
sphere. We can identify the computational cost with the number of
interaction terms that are needed. Since the used nodes must
tessellate the sphere, their number can be estimated as
\begin{equation}
N_{\rm nodes} = \int_d^{R} \frac{4\pi r^2 \,{\rm d}r}{l^3(r)},
\end{equation}
where $l(r)$ is the expected node size at distance $r$, and $d$ is the
characteristic distance of the nearest particle. Since we expect the
nodes to be close to their maximum allowed size, we can set $l \simeq
\theta_c r$. We then obtain
\begin{equation}
  N_{\rm nodes} = \frac{4\pi}{\theta_c^3} \ln \frac{R}{d} \propto \frac{\ln N}{\theta_c^3}.
\end{equation}
The total computational cost for a calculation of the forces for all
particles is therefore expected to scale as ${\cal O}(N\ln N)$. This
is a very significant improvement compared with the $N^2$-scaling of
direct summation.

We may also try to estimate the expected typical force errors. If we
keep only monopoles, the error in the force per unit mass from one
node should roughly be of order the truncation error, i.e.~about
\begin{equation}
\Delta F_{\rm node} \sim \frac{G
  M_{\rm node}}{r^2} \theta^2.
\end{equation} 
The errors from multipole nodes will
add up in quadrature, hence
\begin{equation}
(\Delta F_{\rm
  tot})^2  \sim N_{\rm node} 
(\Delta F_{\rm
  node})^2 = N_{\rm node} \left( \frac{G M_{\rm node}}{r^2} \theta^2\right)^2 \propto \frac{\theta^4}{N_{\rm node}} \propto \theta^7.
\end{equation}
The force error for a monopoles-only scheme therefore scales as
$(\Delta F_{\rm tot}) \propto \theta^{3.5}$, roughly inversely as the
invested computational cost. A much more detailed analysis of the
performance characteristics of tree codes can be found, for example,
in \citet{Hernquist1987}.

\subsection{TreePM schemes}

While the high adaptivity of tree algorithms is particularly ideal for
strongly clustered particle distributions and when a high spatial
force accuracy is desired, the mesh-based approaches are usually
faster when only a coarsely resolved gravitational field on large
scales is required. In particular, the particle-mesh (PM)
approach based on Fourier techniques is probably the fastest method to
calculate the gravitational field on a homogenous mesh. The obvious
limitation of this method is however that the force resolution cannot
be better than the size of one mesh cell, and the latter cannot be easily
made small enough to resolve all the scales of interest in
cosmological simulations.

One interesting idea is to try to combine both approaches into a
unified scheme, where the gravitational field on large scales is
calculated with a PM algorithm, while the short-range forces are
delivered by a hierarchical tree method. Such TreePM schemes have
first been proposed by \citet{Xu1995} and \citet{Bagla2002}, and a
version similar to that of \citet{Bagla2002} is implemented in the
{\small GADGET2} code \citep{Springel2005}.

In order to achieve a clean separation of scales, one can consider the
potential in Fourier space. The individual modes $\Phi_{\vec{k}}$ can
be decomposed into a long-range and a short-range part, as follows:
\begin{equation}
\Phi_{\vec{k}} =
\Phi^{{\rm long}}_{\vec{k}} + \Phi^{{\rm short}}_{\vec{k}},
\end{equation}
where
\begin{equation}
\Phi^{{\rm long}}_{\vec{k}} = \Phi_{\vec{k}} \exp(-\vec{k}^2 r_s^2),
\end{equation}
and
\begin{equation}
\Phi^{{\rm short}}_{\vec{k}} = \Phi_{\vec{k}} [1-\exp(-\vec{k}^2 r_s^2)],
\label{eqnshortrangepotential}
\end{equation}
with $r_s$ describing the spatial scale of the force-split.  Due to
the exponential cut-off of the Fourier-spectrum of the long-range
force, a PM grid of finite size can be used to fully resolve this
force component (this is achieved once the cell size is a few times
smaller than $r_s$). Compared to the ordinary PM-scheme, the only
change is that the Greens function in Fourier-space gets an additional
exponential smoothing factor. Thanks to this force-shaping factor,
inaccuracies such as force anisotropies from the mesh geometry can be
made arbitrarily small, so that the long-range force in the transition
region between the force components is accurately
computed by the PM scheme.

To calculate the short-range force, one transforms
equation~(\ref{eqnshortrangepotential}) back to real space. Assuming a
single point mass $m$ somewhere in a periodic box of size $L$, this
becomes for
$r_s \ll L$: 
\begin{equation} 
\Phi^{{\rm short}}(\vec{x}) = - G 
\frac{m}{r} {\rm erfc}\left(\frac{r} {2 r_s}\right),
\label{eqnshortrange}
\end{equation}
where $r = \min(|\vec{x}-\vec{r} - \vec{n}L|)$ is defined as the
smallest distance of any of the periodic images ($\vec{n}$ is an
arbitrary integer triplet) of the point mass at $\vec{r}$ relative to
the point~$\vec{x}$.  Now, this is recognized as the ordinary
Newtonian potential, modified with a truncation factor that rapidly
turns off the force at a finite distance of order $r_s$. In fact, the
force drops to about 1\% of its Newtonian value for $r \simeq 4.5
r_s$, and quickly becomes completely negligible at still larger
separations.

The potential (\ref{eqnshortrange}) can still be treated with a
hierarchical tree algorithm, except for the simplification that any
tree node more distant than a finite cut-off range (of order $\sim 5
r_s$) can be immediately discarded in the tree walk. This can yield a
significant speed-up relative to a plain tree code, because the
tree-walk can now be restricted to a small region around the target
particle as opposed to having to be carried out for the full
volume. Also, periodic boundary conditions do not have to be included
explicitly through Ewald summation \citep{Hernquist1991} any more,
rather they are absorbed in the periodic PM force. Another advantage
is that for close to homogeneous particle distributions, the PM method
used for long-range forces delivers a precise force quickly, whereas a
pure tree code struggles in this regime to reach the required force
accuracy, simply because here large forces in all directions, which
almost completely compensate in the end, need to be evaluated with
high relative accuracy, otherwise they do not cancel out
properly. Finally, the hybrid TreePM scheme also offers the
possibility to split the time integration into a less frequent
evaluation of the long-range force, and a more frequent evaluation of
the short-range tree force, because the former is associated with
longer dynamical time scales than the latter. This can be exploited to
realize additional efficiency gains, and can in principle even be done
in a symplectic fashion \citep{Saha1992, Springel2005}.

\section{Basic gas dynamics}

Gravity is the dominant driver behind cosmic structure formation
\citep[e.g.][]{Mo2010}, but at small scales hydrodynamics in the
baryonic components becomes very important, too. In this section we
very briefly review the basic equations and some prominent phenomena
related to gas dynamics in order to make the discussion of the
numerical fluid solvers used in galaxy evolution more accessible. For
a detailed introduction to hydrodynamics, the reader is referred to
the standard textbooks on this subject \citep[e.g.][]{Landau1959,
  Shu1992}.

\subsection{Euler and Navier-Stokes equations}

The gas flows in astrophysics are often of extremely low density, making
internal friction in the gas extremely small. In the limit of assuming
internal friction to be completely absent, we arrive at the so-called
ideal gas dynamics as described by the Euler equations. Most
calculations in cosmology and galaxy formation are carried out under
this assumption. However, in certain regimes, viscosity may still
become important (for example in the very hot plasma of rich galaxy
clusters), hence we shall also briefly discuss the hydrodynamical
equations in the presence of physical viscosity, the Navier-Stokes
equations, which in a sense describe \emph{real} fluids as opposed to
ideal ones. Phenomena such as fluid instabilities or turbulence are
also best understood if one does not neglect viscosity completely.

\subsubsection{Euler equations} \label{SecEulerEqns}

If internal friction in a gas flow can be neglected, the dynamics of
the fluid is governed by the Euler equations:
\begin{equation}
\frac{\partial \rho}{\partial t}  + \nabla (\rho\vec{v}) = 0,
\end{equation}
\begin{equation}
\frac{\partial}{\partial t}(\rho \vec{v})  + 
\nabla (\rho\vec{v}\vec{v}^{T} + P) = 0,
\end{equation}
\begin{equation}
\frac{\partial}{\partial t}(\rho e)  + 
\nabla[ (\rho e + P)\vec{v}] = 0,
\end{equation}
where $e = u +\vec{v}^2/2$ is the total energy per unit mass, and $u$
is the thermal energy per unit mass.  Each of these equations is a
continuity law, one for the mass, one for the momentum, and one for
the total energy. The equations hence form a set of hyperbolic
conservation laws. In the form given above, they are not yet complete,
however. One still needs a further expression that gives the pressure
in terms of the other thermodynamic variables.  For an ideal gas, the
pressure law is
\begin{equation}
P = (\gamma -1 ) \rho u,
\end{equation}
where $\gamma = c_p/c_v$ is the ratio of specific heats. For a
monoatomic gas, we have $\gamma = 5/3$.

\subsubsection{Navier-Stokes equations}

Real fluids have internal stresses, due to {\em
  viscosity}. The effect of viscosity is to dissipate relative motions
of the fluid into heat.  The Navier-Stokes equations are then given by
\begin{equation}
\frac{\partial \rho}{\partial t}  + \nabla (\rho\vec{v}) = 0 ,
\end{equation}
\begin{equation}
\frac{\partial}{\partial t}(\rho \vec{v})  + 
\nabla (\rho\vec{v}\vec{v}^{T} + P) = \nabla\, \vec{\rm\Pi} ,
\label{eqnviscshearforce}
\end{equation}
\begin{equation}
\frac{\partial}{\partial t}(\rho e)  + 
\nabla[ (\rho e + P)\vec{v}] = \nabla (\vec{\rm\Pi}\vec{v}) .
\end{equation}
Here $\vec{\rm\Pi}$ is the so-called viscous stress tensor, which is a
material property. For $\vec{\rm\Pi}=0$, the Euler equations are
recovered.  To first order, the viscous stress tensor must be a linear
function of the velocity derivatives \citep{Landau1959}. The most
general tensor of rank-2 of this type can be written as
\begin{equation}
\vec{\rm\Pi} = \eta \left[\nabla \vec{v} + (\nabla \vec{v})^T -
  \frac{2}{3}(\nabla\cdot \vec{v})\vec{1}\right] + \xi (\nabla\cdot
\vec{v}) \vec{1} ,
\end{equation}
where $\vec{1}$ is the unit matrix.
Here $\eta$ scales the traceless part of the tensor and describes the
shear viscosity. $\xi$ gives the strength of the diagonal part, and is
the so-called bulk viscosity. Note that $\eta$ and $\xi$ can in
principle be
functions of local fluid properties, such as $\rho$, $T$,
etc. 

\runinhead{Incompressible fluids}
In the following we shall assume constant viscosity
coefficients. Also, we specialize to incompressible fluids with 
$\nabla\cdot \vec{v} =0$, which is a particularly important case in
practice. Let's see how the Navier-Stokes equations simplify in this
case. Obviously, $\xi$ is then unimportant and we only need to deal
with shear viscosity. Now, let us consider one of the components of
the viscous shear force described by equation~(\ref{eqnviscshearforce}):
\begin{eqnarray}
\frac{1}{\eta}(\nabla\,\vec{\rm\Pi})_x & = &
\frac{\partial}{\partial x}\left(2 \frac{\partial v_x}{\partial
      x}\right)
+
\frac{\partial}{\partial y} \left(\frac{\partial v_x}{\partial
      y} + \frac{\partial v_y}{\partial
      x}   \right)
+
\frac{\partial}{\partial z} \left(\frac{\partial v_x}{\partial
      z} + \frac{\partial v_z}{\partial
      x}   \right)  \nonumber \\
& = & \left(  \frac{\partial^2}{\partial x^2} 
+ \frac{\partial^2}{\partial y^2}
+ \frac{\partial^2}{\partial z^2}\right) v_x = \nabla^2 v_x ,
\end{eqnarray}
where we made use of the $\nabla\cdot \vec{v} =0$ constraint. If we
furthermore
introduce the {\em kinematic viscosity} $\nu$ as
\begin{equation}
\nu \equiv \frac{\eta}{\rho},
\end{equation}
we can write the equivalent of equation~(\ref{eqnviscshearforce}) in the compact form 
\begin{equation}
\frac{{\rm D}\,\vec{v}}{{\rm D}\,t}  =  - \frac{\nabla P}{\rho} + \nu \nabla^2
\vec{v},
\label{eqnNS}
\end{equation}
where the derivative on the left-hand side is the Lagrangian
derivative,
\begin{equation}
\frac{{\rm D}}{{\rm D}\,t} = \frac{\partial}{\partial t} + \vec{v}\cdot\nabla.
\end{equation}
We hence see that the motion of individual fluid elements responds to
pressure gradients and to viscous forces. The form (\ref{eqnNS}) of
the equation is also often simply referred to as the Navier-Stokes
equation.

\subsubsection{Scaling properties of viscous flows}
 
Consider the Navier-Stokes equations for some flow problem that is
characterized by some characteristic length $L_0$, velocity $V_0$, and
density scale $\rho_0$. We can then define dimensionless fluid
variables of the form
\begin{equation}
\hat{\vec{v}}  =  \frac{\vec{v}}{V_0},  \;\;\;\;\;\;\;\;\;\;\;\;
\hat{\vec{x}}  =  \frac{\vec{x}}{L_0},  \;\;\;\;\;\;\;\;\;\;\;\;
\hat P  =  \frac{P}{\rho_0 V_0^2} .
\end{equation}
Similarly, we define a dimensionless time, a dimensionless density, 
and a dimensionless Nabla
operator:
\begin{equation}
\hat{t}  =  \frac{t}{L_0/V_0},  \;\;\;\;\;\;\;\;\;\;\;\;
\hat{\rho}  =  \frac{\rho}{\rho_0},  \;\;\;\;\;\;\;\;\;\;\;\;
\hat{\nabla}  =  L_0 \nabla  .  \;\;\;\;\;\;\;\;\;\;\;\; 
\end{equation}
Inserting these definitions into the Navier-Stokes equation
(\ref{eqnNS}), we obtain the dimensionless equation
\begin{equation}
\frac{D\hat{\vec{v}}}{D\hat{t}}  =  - \frac{\hat{\nabla}
  \hat{P}}{\hat{\rho}} + \frac{\nu}{L_0 V_0}\, \hat{\nabla}^2
\hat{\vec{v}}.
\end{equation}
Interestingly, this equation involves one number,
\begin{equation}
{\rm Re} \equiv \frac{L_0 V_0}{\nu},
\end{equation}
which characterizes the flow and determines the structure of the
possible solutions of the equation. This is the so-called Reynolds
number. Problems which have similar Reynolds number are expected to
exhibit very similar fluid behavior. One then has {\em
  Reynolds-number similarity}. 
In contrast, the Euler equations (${\rm Re} \to \infty$) exhibit always
scale similarity because they are invariant under scale
transformations.

One intuitive interpretation one can give the Reynolds number is that
it measures the importance of inertia relative to viscous
forces. Hence:
\begin{equation}
{\rm Re} \approx \frac{\mbox{inertial forces}}{\mbox{viscous forces}}
\approx \frac{D\vec{v} / D t}{\nu\nabla^2 \vec{v}} \approx
\frac{V_0 / (L_0/V_0)}{ \nu V_0 / L_0^2} = \frac{L_0 V_0}{\nu} .
\end{equation}
If we have ${\rm Re}\sim 1$, we are completely dominated by viscosity.
On the other hand, for ${\rm Re} \to \infty$ viscosity becomes
unimportant and we approach an ideal gas.

\subsection{Shocks}

An important feature of hydrodynamical flows is that they can develop
shock waves in which the density, velocity, temperature and specific
entropy jump by finite amounts \citep[e.g.][]{Toro1997}. In the case
of the Euler equations, such shocks are true mathematical
discontinuities. Interestingly, shocks can occur even from perfectly
smooth initial conditions, which is a typical feature of hyperbolic
partial differential equations. In fact, acoustic waves with
sufficiently large amplitude will suffer from wave-steeping (because
the slightly hotter wave crests travel faster than the colder
troughs), leading eventually to shocks. Of larger practical importance
in astrophysics are however the shocks that occur when flows collide
supersonically; here kinetic energy is irreversibly transferred into
thermal energy, a process that also manifests itself with an increase
in entropy.

In the limit of vanishing viscosity (i.e.~for the Euler equations),
the differential form of the fluid equations breaks down at the
discontinuity of a shock, but the integral form (the {\em weak
  formulation}) remains valid. In other words this means that the flux
of mass, momentum and energy must remain continuous at a shock
front. Assuming that the shock connects two piecewise constant states,
this leads to the Rankine-Hugoniot jump conditions
\citep{Rankine1870}. If we select a frame of reference where the shock
is stationary ($v_s=0$) and denote the pre-shock state with $(v_1,
P_1, \rho_1)$, and the post-shock state as $(v_2, P_2, \rho_2)$ (hence
$v_1,v_2 > 0$), we have
\begin{equation}
\rho_1 v_2 = \rho_2 v_2,
\label{eqnRK1}
\end{equation}
\begin{equation}
\rho_1 v_1^2 + P_1 = \rho_2 v_2^2 + P_2,
\end{equation}
\begin{equation}
(\rho_1 e_1 + P_1) v_1 = (\rho_2 e_2 + P_2)  v_2.
\label{eqnRK3}
\end{equation}
For an ideal gas, the presence of a shock requires that the pre-shock
gas streams supersonically into the discontinuity, i.e.~$v_1 > c_1$,
where $c_1^2 = \gamma\, P_1/\rho_1$ is the sound speed in the
pre-shock phase. The Mach number
\begin{equation}
{\cal M} = \frac{v_1}{c_1}  
\end{equation}
measures the strength of the shock (${\cal M} > 1$). The shock itself
decelerates the fluid and compresses it, so that we have $v_2 < v_1$
and $\rho_2 > \rho_1$. It also heats it up, so that $T_2 > T_1$, and
makes the postshock flow subsonic, with $v_2 / c_2 < 1$.  Manipulating
equations (\ref{eqnRK1}) to (\ref{eqnRK3}), we can express the
relative jumps in the thermodynamic quantities (density, temperature,
entropy, etc.)  through the Mach number alone, for example:
\begin{equation}
\frac{\rho_2}{\rho_1} =  \frac{(\gamma+1) {\cal M}^2}{(\gamma -1){\cal
    M}^2 +2}.
\end{equation}

\subsection{Fluid instabilities}

In many situations, gaseous flows can be subject to fluid
instabilities in which small perturbations can rapidly grow, thereby
tapping a source of free energy. An important example of this are
Kelvin-Helmholtz and Rayleigh-Taylor instabilities, which we briefly
discuss in this subsection.

\begin{figure}
\sidecaption
\resizebox{6.5cm}{!}{\includegraphics{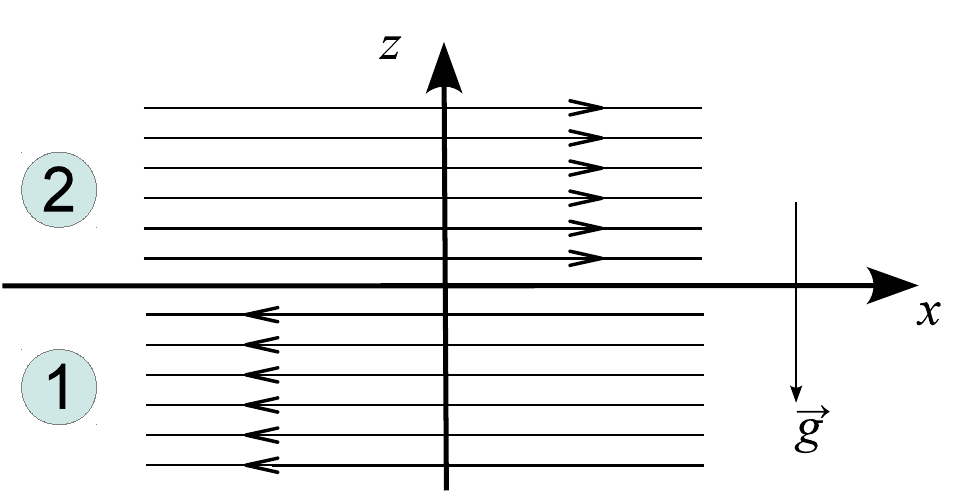}}
\caption{Geometry of a generic shear flow.}
\label{fig_shearflow}
\end{figure}

\runinhead{Stability of a shear flow} We consider a flow in the
$x$-direction, which in the lower half-space $z<0$ has velocity $U_1$
and density $\rho_1$, whereas in the upper half-space the gas streams
with $U_2$ and has density $\rho_2$. In addition there can be a
homogeneous gravitational field $\vec{g}$ pointing into the negative
$z$-direction, as sketched in Figure~\ref{fig_shearflow}.

The stability of the flow can be analysed through perturbation theory.
To this end, one can for example treat the flow as an incompressible
potential flow, and carry out an Eigenmode analysis in Fourier
space. With the help of Bernoulli's theorem one can then derive an
equation for a function $\xi(x, t) = z$ that describes the
$z$-location of the interface between the two phases of the
fluid. Details of this calculation can for example be found in
\citet{Pringle2007}.  For a single perturbative Fourier mode
\begin{eqnarray}
\xi & = & \hat\xi \exp[i(kx - \omega t)], 
\end{eqnarray}
one then finds that non-trivial solutions with $\hat\xi \ne 0$ are
possible for
\begin{equation}
\omega^2(\rho_1+\rho2) - 2\omega k(\rho_1 U_1 + \rho_2  U_2) +
k^2(\rho_1 U_1^2 + \rho_2 U_2^2) + (\rho_2-\rho1) k g  = 0,
\end{equation}
which is the {\em dispersion relation}. Unstable, exponentially
growing mode solutions appear if there are solutions for $\omega$ with
positive imaginary part. Below, we examine the dispersion relation for
a few special cases.

\begin{figure}
\hfill\resizebox{11cm}{!}{\includegraphics{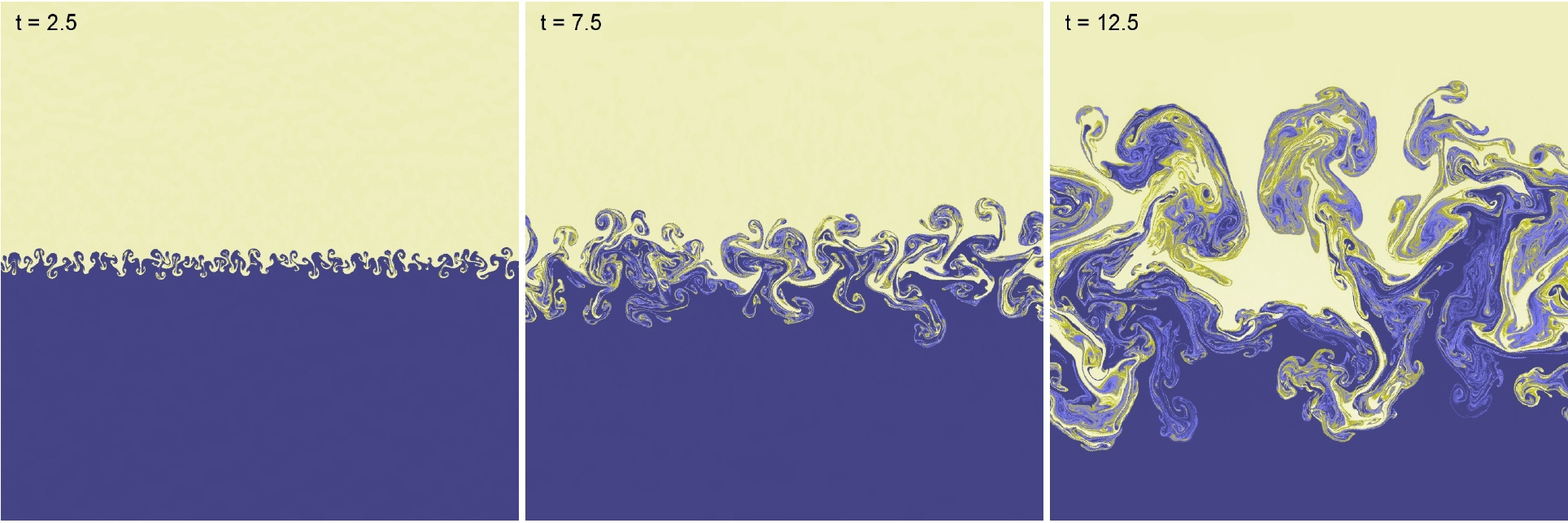}}
\caption{A growing Rayleigh-Taylor instability in which a lighter
  fluid (blue) is covered by a heaver fluid (yellow).}
\label{fig_rt}
\end{figure}

\runinhead{Rayleigh-Taylor instability} Let us consider the case of a
fluid at rest, $U_1 = U_2 = 0$. The dispersion relation simplifies to
\begin{equation}
\omega^2 = \frac{(\rho_1-\rho_2) k g}{\rho_1 + \rho_2}.
\end{equation}
We see that for $\rho_2> \rho_1$, i.e.~the denser fluid lies on top,
unstable solutions with $\omega^2 < 0$ exist. This is the so-called
Rayleigh-Taylor instability. It is in essence buoyancy driven and
leads to the rise of lighter material underneath heavier fluid in a
stratified atmosphere, as illustrated in the simulation shown in
Figure~\ref{fig_rt}. The free energy that is tapped here is the
potential energy in the gravitational field. Also notice that for an
ideal gas, arbitrary small wavelengths are unstable, and those modes
will grow fastest. If on the other hand we have $\rho_1> \rho_2$, then
the interface is stable and will only oscillate when perturbed.

\begin{figure}
\sidecaption
\resizebox{7.2cm}{!}{\includegraphics{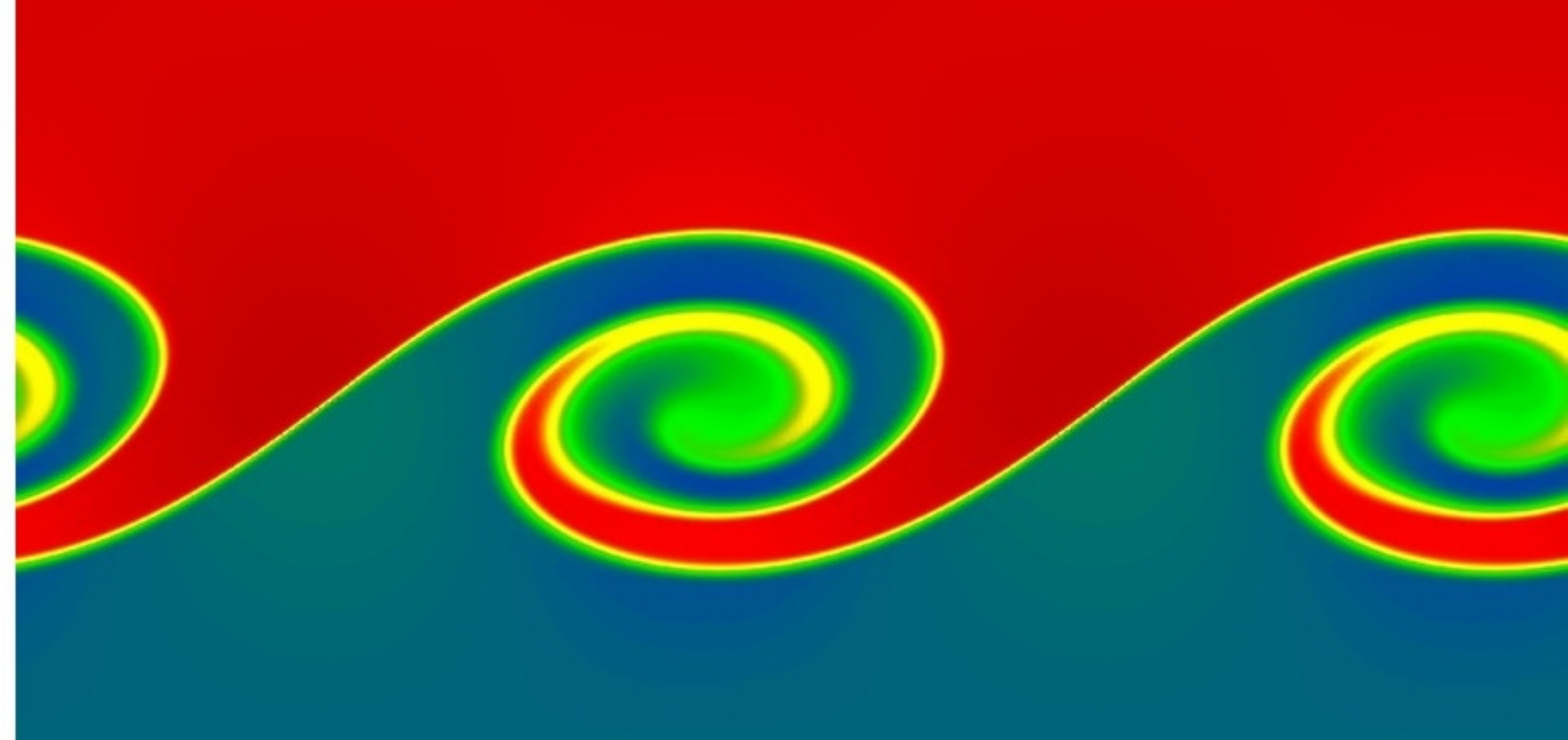}}
\caption{Characteristic Kelvin-Helmholtz billows arising in a shear flow.}
\label{fig_kh}
\end{figure}

\runinhead{Kelvin-Helmholtz instability}
If we set the gravitational field to zero, $g=0$, we have the
situation of a pure shear flow. In this case, the solutions of the
dispersion relation are given by
\begin{equation}
\omega_{1/2} = \frac{k (\rho_1 U_1 + \rho_2 U_2)}{\rho_1 + \rho_2} \pm
i k \frac{\sqrt{\rho_1 \rho_2}}{\rho_1 + \rho_2} |U_1 - U_2|.
\end{equation}
Interestingly, in an ideal gas there is an imaginary growing mode
component for every $|U_1 - U_2|>0$! This means that a small wave-like
perturbation at an interface will grow rapidly into large waves that
take the form of characteristic Kelvin-Helmholtz ``billows''. In the
non-linear regime reached during the subsequent evolution of this
instability the waves are rolled up, leading to the creation of vortex
like structures, as seen in Figure~\ref{fig_kh}. As the instability
grows fastest for small scales (high $k$), the billows tend
to get larger and larger with time.

Because the Kelvin-Helmholtz instability basically means that any
sharp velocity gradient in a shear flow is unstable in a freely
streaming fluid, this instability is particularly important for the
creation of fluid turbulence. Under certain conditions, some modes can
however be stabilized against the instability. This happens for
example if we consider shearing with $U_1 \ne U_2$ in a gravitational
field $g > 0$. Then the dispersion relation has the solutions
\begin{equation}
\omega = \frac{k(\rho_1 U_1 + \rho_2 U_2)}{\rho_1 +\rho_2}
\pm \frac{\sqrt{-k^2 \rho_1\rho_2(U_1-U_2)^2 - (\rho_1 +
    \rho_2)(\rho_2 - \rho_1) k g}}{\rho_1+\rho_2} .
\end{equation}
Stability is possible if two conditions are met. First, we need
$\rho_1 > \rho_2$, i.e.~the lighter fluid needs to be on top
(otherwise we would have in any case a Rayleigh-Taylor
instability). Second, the condition
\begin{equation}
(U_1-U_2)^2 < \frac{(\rho_1+\rho_2)(\rho_1-\rho_2) g }{k \rho_1 \rho_2}
\end{equation}
must be fulfilled. Compared to the ordinary Kelvin-Holmholtz
instability without a gravitational field, we hence see that
sufficiently small wavelengths are stabilized below a threshold
wavelength. The larger the shear becomes, the further this threshold
moves to small scales.

The Rayleigh-Taylor and Kelvin-Helmholtz instabilities are by no means
the only fluid instabilities that can occur in an ideal gas
\citep{Pringle2007}. For example, there is also the Richtmyer-Meshov
instability, which can occur when an interface is suddenly
accelerated, for example due to the passage of a shock wave. In
self-gravitating gases, there is the Jeans instability, which occurs
when the internal gas pressure is not strong enough to prevent a
positive density perturbation from growing and collapsing under its
own gravitational attraction. This type of instability is particularly
important in cosmic structure growth and star formation. If the gas
dynamics is coupled to external sources of heat (e.g.~through a
radiation field), a number of further instabilities are possible. For
example, a thermal instability \citep{Field1965} can occur when a
radiative cooling function has a negative dependence on
temperature. If the temperature drops somewhere a bit more through
cooling than elsewhere, the cooling rate of this cooler patch will
increase such that it is cooling even faster. In this way, cool clouds
can drop out of the background gas.

\subsection{Turbulence}

Fluid flow which is unsteady, irregular, seemingly random, and chaotic
is called {\em turbulent} \citep{Pope2000}. Familiar examples of such
situations include the smoke from a chimney, a waterfall, or the wind
field behind a fast car or airplane. The characteristic feature of
turbulence is that the fluid velocity varies significantly and
irregularly both in position and time. As a result, turbulence is a
statistical phenomenon and is best described with statistical
techniques.

If the turbulent motions are subsonic, the flow can often be
approximately treated as being incompressible, even for an equation of
state that is not particularly stiff. Then only solenoidal motions
that are divergence free can occur, or in other words, only shear
flows are present. We have already seen that such flows are subject to
fluid instabilities such as the Kelvin-Helmholtz instability, which
can easily produce swirling motions on many different scales. Such
vortex-like motions, also called {\em eddies}, are the conceptual
building blocks of Kolmogorov's theory of incompressible turbulence
\citep{Kolmogorov1941}, which yields a surprisingly accurate
description of the basic phenomenology of turbulence, even though many
aspects of turbulence are still not fully understood.

\subsubsection{Kolmogorov's theory of incompressible turbulence}
 
We consider a fully turbulent flow with characteristic velocity $U_0$
and length scale $L_0$. We assume that a quasi-stationary state for
the turbulence is achieved by some kind of driving process on large
scales, which in a time-averaged way injects an energy $\epsilon$ per
unit mass.  We shall also assume that the Reynolds number ${\rm Re}$
is large. We further imagine that the turbulent flow can be considered
to be composed of eddies of different size $l$, with characteristic
velocity $u(l)$, and associated timescale $\tau(l) = l/u(l)$.

For the largest eddies, $l\sim L_0$ and $u(l) \sim U_0$, hence
viscosity is unimportant for them. But large eddies are unstable und
break up, transferring their energy to somewhat smaller eddies. This
continues to yet smaller scales, until
\begin{equation}
{\rm Re}(l) = \frac{l u(l)}{\nu}
\end{equation}
reaches of order unity, where $\nu$ is the kineamtic viscosity. For
these eddies, viscosity will be very important so that their kinetic
energy is dissipated away. We will see that this transfer of energy to
smaller scales gives rise to the {\em energy cascade} of
turbulence. But several important questions are still unanswered:
\begin{enumerate}
\item What is the actual size of the smallest eddies that dissipate
  the energy?
\item How do the velocities $u(l)$ of the eddies vary with $l$ when
  the eddies become smaller?
\end{enumerate}

\runinhead{Kolmogorov's hypotheses} Kolmogorov conjectured a number of
hypotheses that can answer these questions. In particular, he
proposed:
\begin{itemize}
\item For high Reynolds number, the small-scale turbulent motions ($l
  \ll L_0$) become statistically isotropic. Any memory of large-scale
  boundary conditions and the original creation of the turbulence on
  large scales is lost.
\item  For high Reynolds number, the statistics of small-scale
  turbulent motions has a universal form and is only determined by
  $\nu$ and the energy injection rate per unit mass, $\epsilon$. 
\end{itemize}

From $\nu$ and $\epsilon$, one can construct characteristic Kolmogorov
length, velocity and timescales. Of particular importance is the {\em
  Kolmogorov length}:
\begin{equation}
\eta \equiv \left(\frac{\nu^3}{\epsilon}\right)^{1/4}.
\end{equation}
Velocity and timescales are given by 
\begin{equation}
u_\eta = (\epsilon \nu) ^{1/4}, \;\;\;\;\;\;\; \tau_\eta = \left(\frac{\nu}{\epsilon}\right)^{1/2}.
\end{equation}
We then see that the Reynolds number at the Kolmogorov scales is 
\begin{equation}
{\rm Re}(\eta) = \frac{\eta u_\eta}{\nu} = 1,
\end{equation} 
showing that they describe the dissipation range.
Kolmogorov has furthermore made a second similarity hypothesis, as
follows:
\begin{itemize}
\item For high Reynolds number, there is a range of scales $L_0 \gg l
  \gg \eta$ over which the statistics of the motions on scale $l$ take
  a universal form, and this form is {\em only} determined by
  $\epsilon$, {\em independent} of $\nu$.
\end{itemize}
In other words, this also means that viscous effects are unimportant
over this range of scales, which is called the {\em inertial range}.
Given an eddy size $l$ in the inertial range, one can construct its
characteristic velocity and timescale just from $l$ and $\epsilon$:
\begin{equation}
u(l) = (\epsilon l )^{1/3}, \hspace*{1cm}
\tau(l) = \left(\frac{l^2}{\epsilon}\right)^{1/3} .
\end{equation}
One further consequence of the existence of the inertial range is that
here the energy transfer rate
\begin{equation}
T(l) \sim \frac{u^2(l)}{\tau(l)}
\end{equation}
of eddies to smaller scales is expected to be scale-invariant. Indeed,
putting in the expected characteristic scale dependence we get $T(l)
\sim \epsilon$, i.e.~$T(l)$ is equal to the energy injection
rate. This also implies that we have
\begin{equation}
\epsilon \sim \frac{U_0^3}{L_0}.
\end{equation}
With this result we can also work out what we expect for the ratio
between the characteristic quantities of the
largest and smallest scales:
\begin{equation}
\frac{\eta}{L_0} \sim \left(\frac{\nu^3}{\epsilon L_0^4}\right)^{1/4}
= \left(\frac{\nu^3}{U_0^3 L_0^3}\right)^{1/4}
= {\rm Re}^{-\frac{3}{4}} ,
\end{equation}
\begin{equation}
\frac{u_\eta}{U_0} \sim \left(\frac{\epsilon \nu}{U_0^4}\right)^{1/4}
= \left(\frac{U_0^3 \nu}{L_0 U_0^4}\right)^{1/4}
= {\rm Re}^{-\frac{1}{4}} ,
\end{equation}
\begin{equation}
\frac{\tau_\eta}{\tau} \sim \left(\frac{\nu U_0^2}{\epsilon L_0^2}\right)^{1/2}
= \left(\frac{\nu U_0^2 L_0}{U_0^3 L_0^2}\right)^{1/2}
= {\rm Re}^{-\frac{1}{2}} .
\end{equation}
This shows that the Reynolds number directly sets the dynamic range of
the inertial range.

\subsubsection{Energy spectrum of Kolmogorov  turbulence}

Eddy motions on a length-scale $l$ correspond to wavenumber $k =
2\pi/l$. The kinetic energy $\Delta E$ contained between two
wave numbers $k_1$ and $k_2$ can be described by
\begin{equation}
\Delta E = \int_{k_1}^{k_2} E(k)\,{\rm d} k,
\end{equation}
where $E(k)$ is the so-called energy spectrum. For the inertial range
in Kolmogorov's theory, we know that $E(k)$ is a universal function
that only depends on $\epsilon$ and $k$. Hence $E(k)$ must be of the
form
\begin{equation}
E(k)  = C \, \epsilon^a \, k^b  ,
\end{equation}
where $C$ is a dimensionless constant. Through dimensional analysis it
is easy to see that one must have $a=2/3$ and $b= - 5/3$. We hence obtain
the famous $-5/3$ slope of the Kolmogorov energy power spectrum:
\begin{equation}
E(k)  = C \, \epsilon^{2/3} \, k^{-5/3}  .
\end{equation}
The constant $C$ is universal in Kolmogorov's theory, but cannot be
computed from first principles. Experiment and numerical simulations
give $C \simeq 1.5$ \citep{Pope2000}.

\begin{figure}
\sidecaption
\resizebox{7.3cm}{!}{\includegraphics{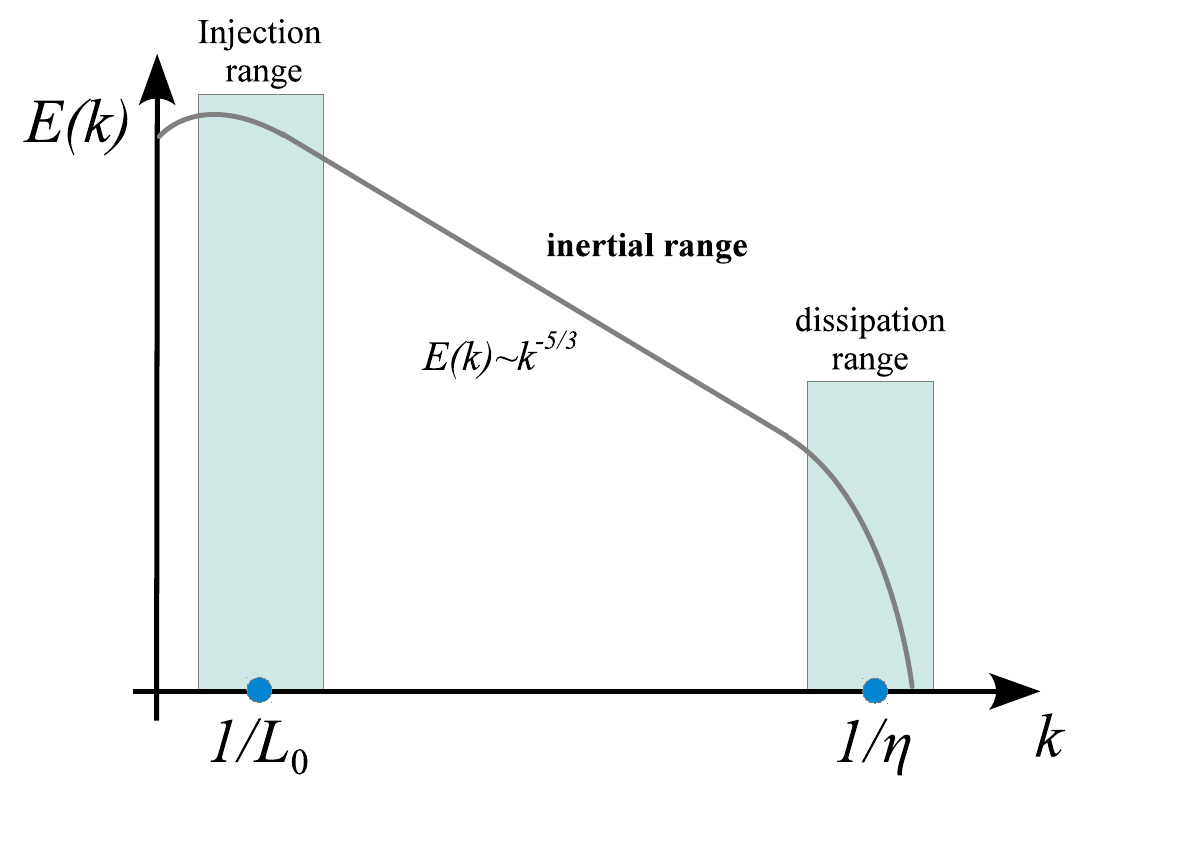}}
\caption{Schematic energy spectrum of Kolmogorov turbulence.}
\end{figure}

Actually, if we recall Kolmogorov's first similarity hypothesis, it
makes the stronger claim that the statistics for all small scale
motion is universal. This means that also the dissipation part of the
turbulence must have a universal form. To include this in the
description of the spectrum, we can for example write
\begin{equation}
E(k) =  C \, \epsilon^{2/3} \, k^{-5/3} f_\eta (k\eta),
\end{equation}
where $f_\eta (k\eta)$ us a universal function with $f_\eta(x) =1$ for
$x\ll 1$, and with $f_\eta(x) \to 0$ for $x\to \infty$. This function
has to be determined experimentally or numerically. A good fit to
different results is given by
\begin{equation}
f_\eta(x) = \exp \left( -\beta [ (x^4 + c^4)^{1/4} - c]\right),
\end{equation}
with $\beta_0 \sim 5.2$ and $c \sim 0.4$ \citep{Pope2000}.

\section{Eulerian hydrodynamics}

Many physical theories are expressed as partial differential equations
(PDEs), including some of the most fundamental laws of nature, such as
fluid dynamics (Euler and Navier Stokes equations), electromagnetism
(Maxwell's equations) or general relativity/gravity (Einstein's field
equations).  Broadly speaking, partial differential equations (PDE)
are equations describing relations between partial derivatives of a
dependent variable with respect to several independent variables.
Unlike for ordinary differential equations (ODEs), there is no simple
unified theory for PDEs. Rather, there are different types of PDEs
which exhibit special features \citep{Renardy2004}.

The Euler equations, which will be the focus of this section, are
so-called hyperbolic conservation laws. They are non-linear, because
they contain non-linear terms in the unknown functions and/or its
partial derivatives. We note that a full characterization of the
different types of PDEs goes beyond the scope of these lecture notes.

\subsection{Solution schemes for PDEs}

Unfortunately, for partial differential equations one cannot give a
general solution method that works equally well for all types of
problems. Rather, each type requires different approaches, and certain
PDEs encountered in practice may even be best addressed with special
custom techniques built by combining different elements from standard
techniques. Important classes of solution schemes include the following:

\begin{itemize}

\item {\bf Finite difference methods:} Here the differential operators
  are approximated through finite difference approximations, usually
  on a regular (cartesian) mesh, or some other kind of structured mesh
  (for example a polar grid). An example we already previously
  discussed is Poisson's equation treated with iterative (multigrid)
  methods.

\item {\bf Finite volume methods:} These may be seen as a subclass of
  finite difference methods. They are particularly useful for
  hyperbolic conservation laws. We shall discuss examples for this
  approach in applications to fluid dynamics later in this section.

\item {\bf Spectral methods:} Here the solution is represented by a
  linear combination of functions, allowing the PDE to be transformed
  to algebraic equations or ordinary differential equations. Often
  this is done by applying Fourier techniques. For example, solving
  the Poisson equation with FFTs, as we discussed earlier, is a
  spectral method.

\item {\bf Method of lines:} This is a semi-discrete approach where
  all derivatives except for one are approximated with finite
  differences. The remaining derivative is then the only one left, so
  that the remaining problem forms a set of ordinary differential
  equations (ODEs). Very often, this approach is used in
  time-dependent problems. One here discretizes space in terms of a
  set of $N$ points ${x_i}$, and for each of these points one obtains
  an ODE that describes the time evolution of the function at this
  point. The PDE is transformed in this way into a set of $N$ coupled
  ODEs. For example, consider the heat diffusion equation in one
  dimension,
\begin{equation}
\frac{\partial u}{\partial t} + \lambda \frac{\partial^2 u}{\partial x^2} = 0.
\end{equation}
If we discretize this into a set of points that are spaced $h$ apart,
we obtain $N$ equations
\begin{equation}
\frac{{\rm d} u_i}{{\rm d} t} + \lambda \frac{u_{i+1} + u_{i-1} - 2 u_i}{h^2} = 0.
\end{equation}
These differential equations can now be integrated in time as an ODE
system. Note however that this is not necessarily stable. Some
problems may require upwinding, i.e.~asymmetric forms for the finite
difference estimates to recover stability.

\item {\bf Finite element methods:} Here the domain is subdivided into
  ``cells'' (elements) of fairly arbitrary shape. The solution is then
  represented in terms of simple, usually polynomial functions on the
  element, and then the PDE is transformed to an algebraic problem for
  the coefficients in front of these simple functions. This is hence
  similar in spirit to spectral methods, except that the expansion is
  done in terms of highly localized functions on an element by element
  basis, and is truncated already at low order.

\end{itemize}

In practice, many different variants of these basic methods exist, and
sometimes also combinations of them are used.

\subsection{Simple advection}

First-order equations of hyperbolic type are particularly useful for
introducing the numerical difficulties that then also need to be
addressed for more complicated non-linear conservation laws
\citep[e.g.][]{Toro1997,LeVeque2002,Stone2008}. The simplest equation
of this type is the {\em advection equation} in one dimension.  This
is given by
\begin{equation}
\frac{\partial u}{\partial t} + v \cdot \frac{\partial u}{\partial x} = 0,
\end{equation}
where $u=u(x,t)$ is a function of $x$ and $t$, and $v$ is a constant
parameter. This equation is hyperbolic because the so-called
coefficient matrix\footnote{A linear system of first-order PDEs can be
written in the generic form 
\begin{equation}
\frac{\partial u_i}{\partial t} 
+ \sum_j A_{ij}\frac{\partial u_i}{\partial x_j} = 0,
\end{equation}
where $A_{ij}$ is the coefficient matrix.
} is real and trivially diagonalizable.

\begin{figure}
\sidecaption
\hfill\resizebox{7.4cm}{!}{\includegraphics{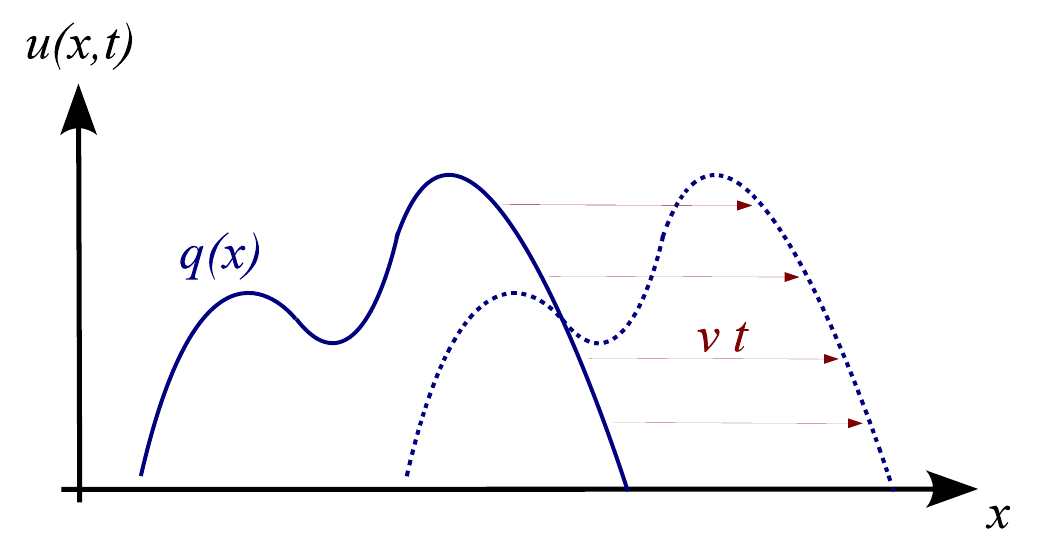}}
\caption{Simple advection with constant velocity to the right.}
\label{fig_advect}
\end{figure}

If we are given any function $q(x)$, then 
\begin{equation}
u(x,t) = q(x-vt)
\end{equation}
is a solution of the PDE, as one can easily check. We can interpret
$u(x,t=0) = q(x)$ as initial condition, and the solution at a later
time is then an exact copy of $q$, simply translated by $v\,
t$ along the $x$-direction, as shown in Fig.~\ref{fig_advect}.

\begin{figure}
\sidecaption
\resizebox{7.3cm}{!}{\includegraphics{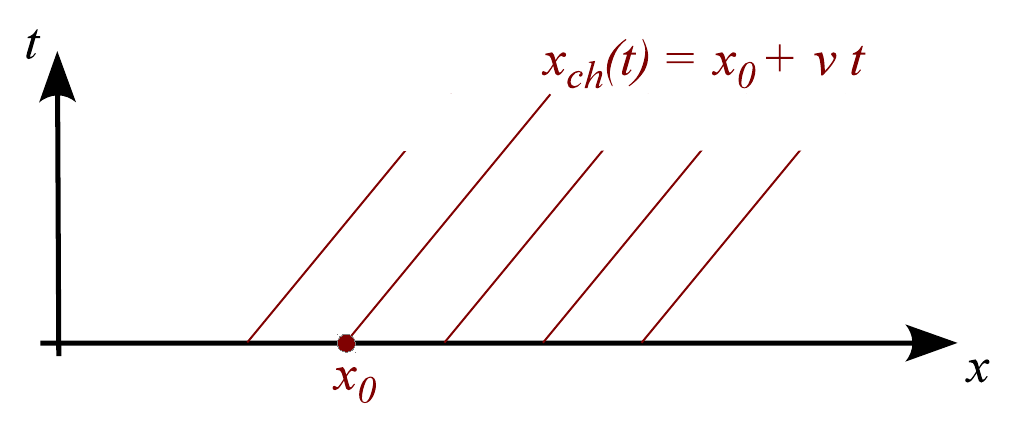}}
\caption{A set of flow characteristics for advection to the right with
constant velocity $v$.}
\label{fig_ch}
\end{figure}

Points that start at a certain coordinate $x_0$ are advected to a new
location $x_{\rm ch}(t) = v t + x_0$. These so-called {\em
  characteristics} (see Fig.~\ref{fig_ch}), which can be viewed as
mediating the propagation of information in the system, are straight
lines, all oriented in the downstream direction. Note that
``downstream'' refers to the direction in which the flow goes, whereas
``upstream'' is from where the flow comes.

Let's now assume we want to solve the advection problem
numerically. (Strictly speaking this is of course superfluous as we
have an analytic solution in this case, but we want to see how well a
numerical technique would perform here.) We can approach this with a
straightforward discretization of $u$ on a special mesh, using for
example the method of lines. This gives us:
\begin{equation}
\frac{{\rm d}u_i}{{\rm d} t} + v \frac{u_{i+1} - u_{i-1}}{2h} = 0.
\end{equation}
If we go one step further and also discretize the time derivative with a
simple Euler scheme, we get
\begin{equation}
u_i^{(n+1)} = u_i^{(n)} -  v \frac{u_{i+1}^{(n)} - u_{i-1}^{(n)}}{2h} \Delta t.
\end{equation}
This is a complete update formula which can be readily applied to a
given initial state on the grid. The big surprise is that this turns
out to be quite violently unstable! For example, if one applies this
to the advection of a step function, one invariably obtains strong
oscillatory errors in the downstream region of the step, quickly
rendering the numerical solution into complete garbage.  What is the
reason for this fundamental failure?

\begin{itemize}
\item First note that all characteristics (signals) propagate
  downstream in this problem, or in other words, information strictly
  travels in the flow direction in this problem.

\item But, the information to update $u_i$ is derived both from the
  upstream ($u_{i-1}$) and the downstream ($u_{i+1}$) side. 

\item According to how the information flows, $u_i$ should not really
  depend on the downstream side at all, which in some sense is
  causally disconnected. So let's try to get rid off this dependence
  by going to a one-sided approximation for the spatial derivative, of
  the form:
\begin{equation}
\frac{{\rm d}u_i}{{\rm d} t} + v \frac{u_{i} - u_{i-1}}{h} = 0.
\end{equation}
This is called {\em upwind differencing}. Interestingly, now the
stability problems are completely gone!

\item But there are still some caveats to observe: First of all, the
  discretization now depends on the sign of $v$. For negative $v$, one
  instead has to use
\begin{equation}
\frac{{\rm d}u_i}{{\rm d} t} + v \frac{u_{i+1} - u_{i}}{h} = 0.
\end{equation}
The other is that the solution is not advected in a perfectly faithful
way, instead it is quite significantly smoothed out, through a process
one calls {\em numerical diffusion}.

\end{itemize}

We can actually understand where this strong diffusion in the
1st-order upwind scheme comes from. To this end, let's rewrite the
upwind finite difference approximation of the spatial derivative as
\begin{equation}
  \frac{u_{i} - u_{i-1}}{h} =  \frac{u_{i+1} - u_{i-1}}{2h} - \frac{u_{i+1} - 2 u_{i} + u_{i-1}}{2h}.
\end{equation}
Hence our stable upwind scheme can also be written as
\begin{equation}
  \frac{{\rm d}u_i}{{\rm d} t} + v \frac{u_{i+1} - u_{i-1}}{2h} = \frac{vh}{2}\, \frac{u_{i+1} - 2 u_{i} + u_{i-1}}{h^2}.
\end{equation}
But recall from equation~(\ref{eqn2ndder}) that
\begin{equation}
  \left( \frac{\partial^2 u}{\partial x^2}\right)_i  \simeq  \frac{u_{i+1} - 2 u_{i} + u_{i-1}}{h^2},
\end{equation}
so if we define a diffusion constant $D = (vh)/2$, we are effectively
solving the following problem,
\begin{equation}
  \frac{\partial u}{\partial t} + v \cdot \frac{\partial u}{\partial x} = D   \frac{\partial^2 u}{\partial x^2},
\end{equation}
and not the original advection problem. The diffusion term on the
right hand side is here a byproduct of the numerical algorithm that we
have used. We needed to add this numerical diffusion in order to
obtain stability of the integration.

Note however that for better grid resolution, $h \to 0$, the diffusion
becomes smaller, so in this limit one obtains an ever better
solution. Also note that the diffusivity becomes larger for larger
velocity $v$, so the faster one needs to advect, the stronger the
numerical diffusion effects become.

Besides the upwinding requirement, integrating a hyperbolic
conservation law with an explicit method in time also requires the use
of a sufficiently small integration timestep, not only to get
sufficiently good accuracy, but also for reasons of {\em
  stability}. In essence, there is a maximum timestep that may be used
before the integration brakes down. How large can we make this
timestep? Again, we can think about this in terms of information
travel. If the timestep exceeds $\Delta t_{\rm max} = h/v$, then the
updating of $u_i$ would have to include information from $u_{i-2}$,
but if we don't do this, the updating will likely become unstable.

This leads to the so-called {\em Courant-Friedrichs-Levy} (CFL) timestep
condition \citep{Courant1928}, which for this problem takes the form
\begin{equation}
\Delta t \le \frac{h}{v}.
\end{equation}
This is a necessary but not sufficient condition for any explicit
finite different approach of the hyperbolic advection equation. For
other hyperbolic conservation laws, similar CFL-conditions apply.

\runinhead{Hyperbolic conservation laws}
We now consider a hyperbolic conservation law, such as the
continuity equation for the mass density of a fluid:
\begin{equation}
\frac{\partial \rho}{\partial t} + \nabla \cdot (\rho \vec{v}) = 0.
\end{equation}
We see that this is effectively the advection equation, but with a
spatially variable velocity $\vec{v} = \vec{v}(\vec{x})$. Here
$\vec{F} = \rho \vec{v}$ is the mass flux.

\begin{figure}
\sidecaption
\resizebox{6.8cm}{!}{\includegraphics{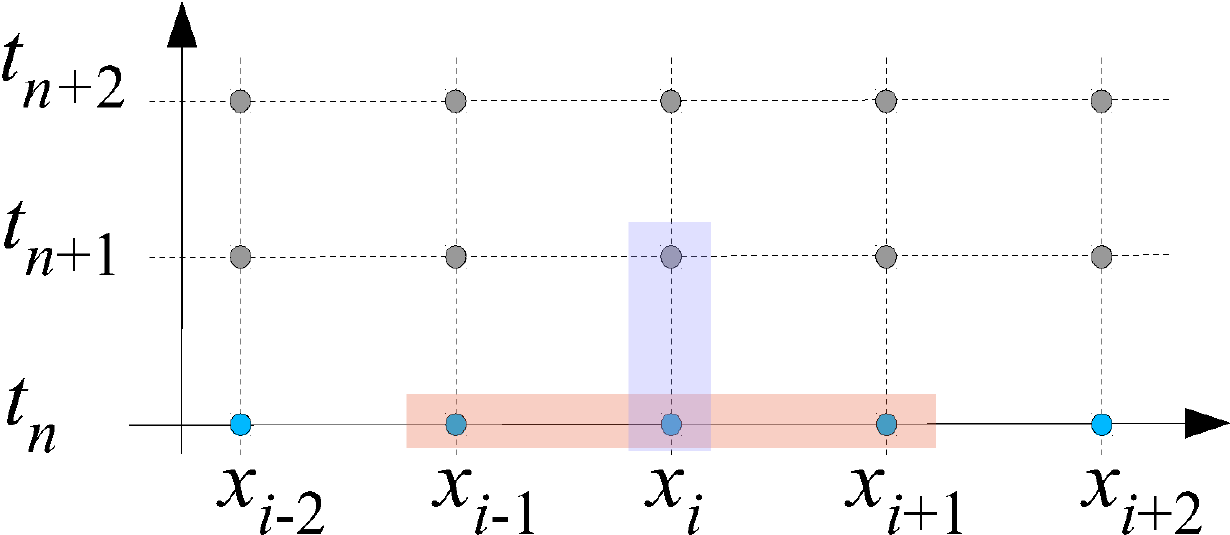}}
\caption{A discretization scheme for the continuity equation in one
  spatial dimension. The red and blue boxes mark the stencils that are
  applied for calculating the spatial and time derivatives. }
\end{figure}

Let's study the problem in one spatial dimension, and consider a
discretization both of the $x$- and $t$-axis.  This corresponds to
\begin{equation}
\frac{\rho_i^{(n+1)} - \rho_i^{(n)} }{\Delta t} + \frac{F_{i+1}^{(n)} - F_{i-1}^{(n)}}{2 \Delta x} = 0,
\end{equation}
leading to the update rule
\begin{equation}
\rho_i^{(n+1)} = \rho_i^{(n)} + \frac{\Delta t}{2 \Delta x} \left( F_{i-1}^{(n)} - F_{i+1}^{(n)}\right).
\end{equation}
This is again found to be highly unstable, for the same reasons as in
the plain advection problem: we have not observed in `which direction
the wind blows', or in other words, we have ignored in which direction
the local characteristics point. For example, if the mass flux is to
the right, we know that the characteristics point also to the
right. The upwind direction is therefore towards negative $x$, and by
using only this information in making our spatial derivative
one-sided, we should be able to resurrect stability.

Now, for the mass continuity equation identifying the local
characteristics is quite easy, and in fact, their direction can simply
be inferred from the sign of the mass flux. However, in more general
situations for systems of non-linear PDEs, this is far less
obvious. Here we need to use a so-called Riemann solvers to give us
information about the local solution and the local characteristics
\citep{Toro1997}. This then also implicitly identifies the proper
upwinding that is needed for stability.

\subsection{Riemann problem}

The Riemann problem is an initial value problem for a hyperbolic
system, consisting of two piece-wise constant states (two half-spaces)
that meet at a plane at $t=0$. The task is then to solve for the
subsequent evolution at $t>0$.

An important special case is the Riemann problem for the Euler
equations (i.e. for ideal gas dynamics). Here the left and right
states of the interface, can, for example, be uniquely specified by
giving the three ``primitive'' variables density, pressure and
velocity, viz.
\begin{equation}
U_L = \left(
\begin{array}{c}
\rho_L \\
P_L\\
\vec{v}_L
\end{array}
\right),
\;\;\;\;
U_R = \left(
\begin{array}{c}
\rho_R \\
P_R\\
\vec{v}_R
\end{array}
\right).
\end{equation}
Alternatively one can also specify density, momentum density, and
energy density.  For an ideal gas, this initial value problem can be
solved analytically \citep{Toro1997}, modulo an implicit equation
which requires numerical root-finding, i.e.~the solution cannot be
written down explicitly. The solution always contains characteristics
for three self-similar waves, as shown schematically in
Fig.~\ref{fig_riemann_waves}. Some notes on this:

\begin{figure}
\sidecaption
\resizebox{7.5cm}{!}{\includegraphics{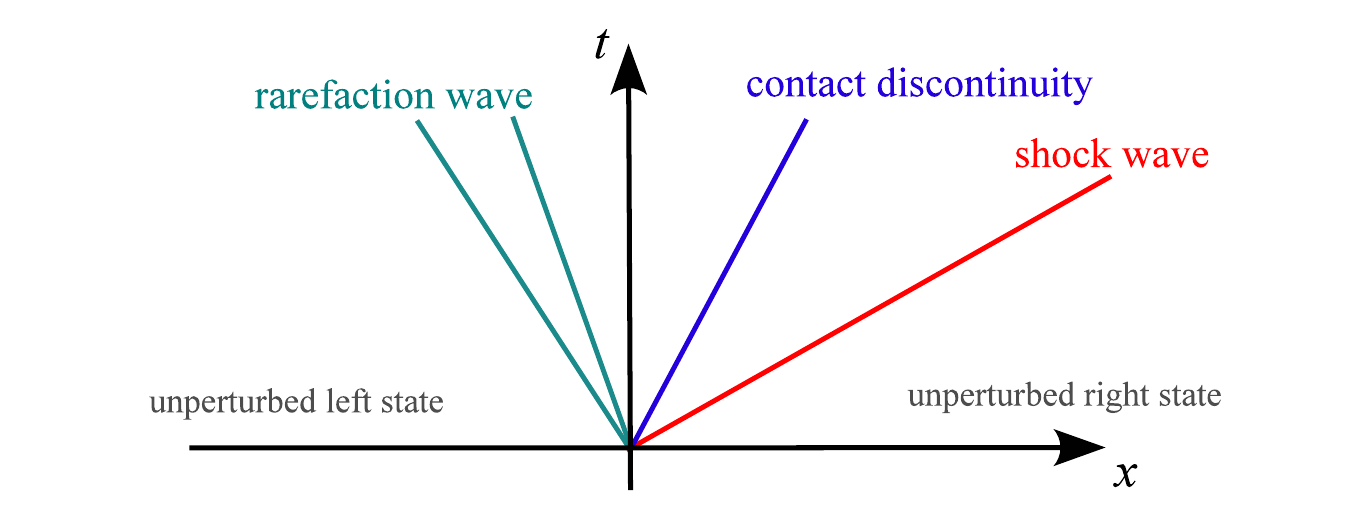}}
\caption{Wave structure of the solution of the Riemann problem. The
  central contact wave separates the original fluid
  phases. On the left and the right, there is either a shock or a
  rarefaction wave.} \label{fig_riemann_waves}
\end{figure}

\begin{itemize}
\item The middle wave is always present and is a contact wave that
  marks the boundary between the original fluid phases from the left
  and right sides.
\item The contact wave is sandwiched between a shock or a rarefaction
  wave on either side (it is possible to have shocks on both sides, or
  rarefactions on both sides, or one of each).  The rarefaction wave
  is not a single characteristic but rather a rarefaction fan with a
  beginning and an end.
\item These waves propagate with constant speed. If the solution is
  known at some time $t>0$, it can also be obtained at any other time
  through a suitable scaling transformation.  An important corollary
  is that at $x=0$, the fluid quantities $(\rho^\star, P^\star,
  \vec{v}^\star)$ are {\em constant in time} for $t > 0$.
\item For $\vec{v}_L = \vec{v}_R = 0$, the Riemann problem simplifies
  and becomes the `Sod shock tube' problem.

\end{itemize}

\begin{figure}
\hfill\resizebox{11cm}{!}{\includegraphics{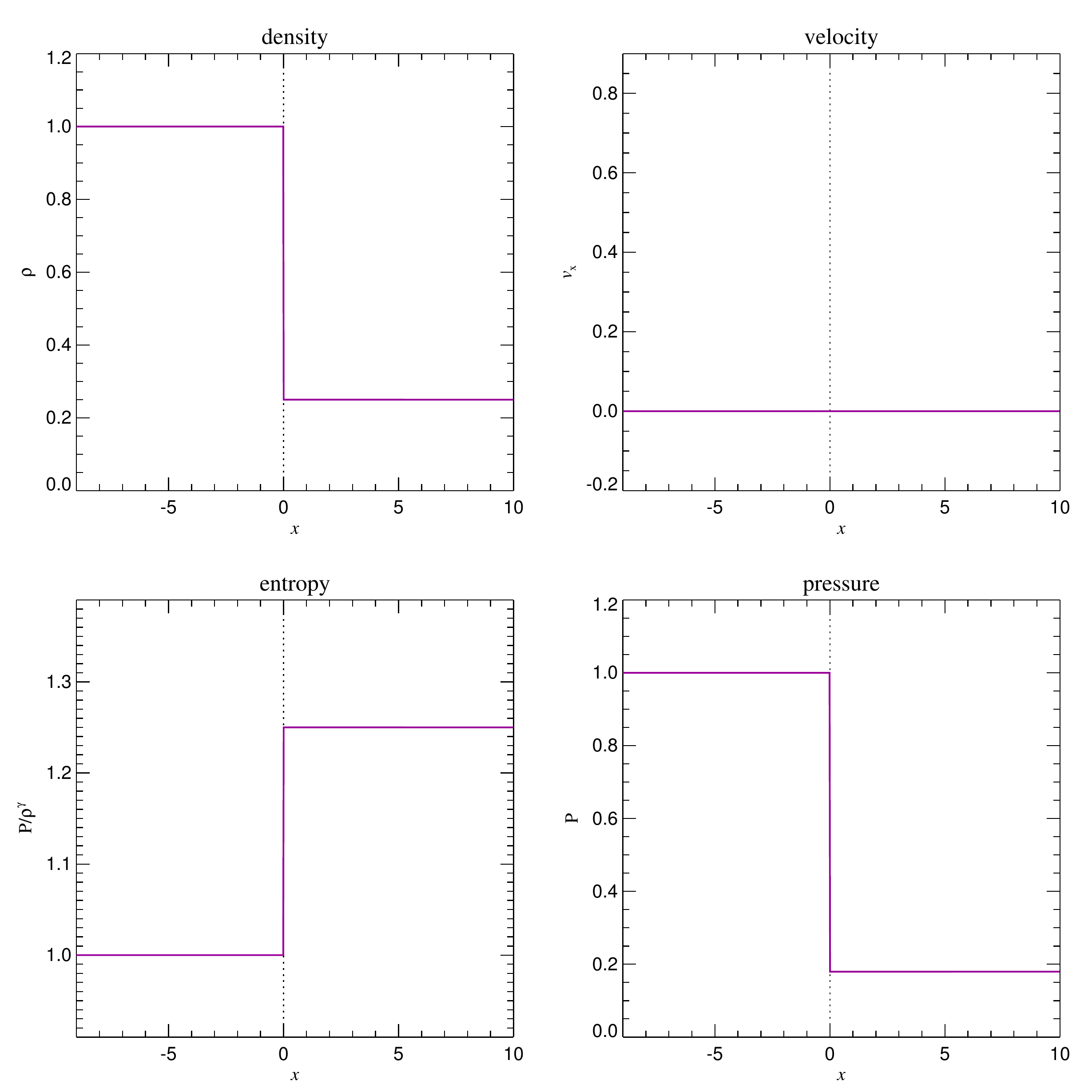}}
\caption{Initial state of an example Riemann problem, composed of two
  phases in different states that are brought into contact at $x=0$ at
  time $t=0$. (Since $v_x=0$, the initial conditions are actually an
  example of the special case of a Sod shock-tube problem.)}
\label{fig_initial_state}
\end{figure}

\begin{figure}
\hfill\resizebox{11cm}{!}{\includegraphics{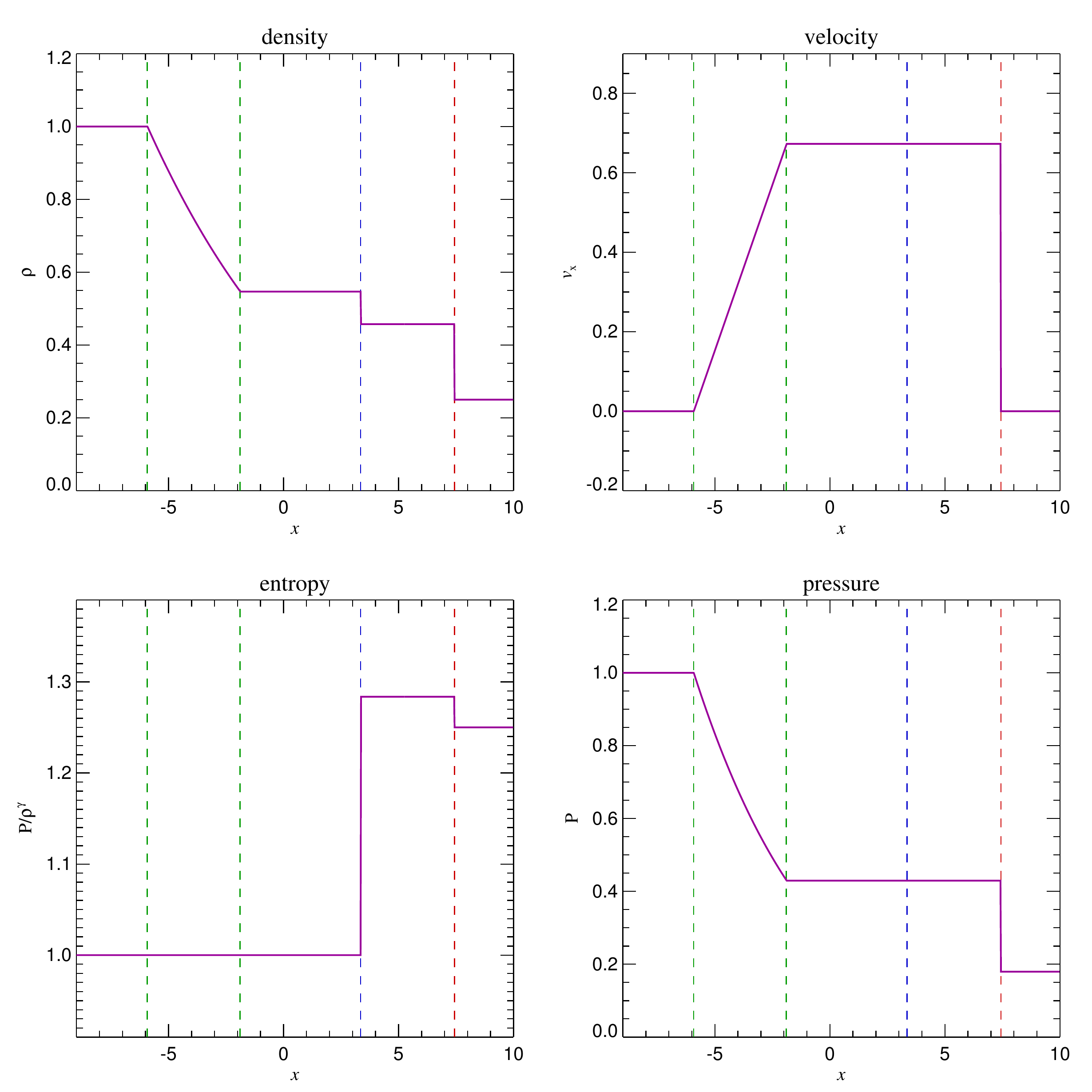}}
\caption{Evolved state at $t=5.0$ of the initial fluid state displayed
  in Fig.~\ref{fig_initial_state}. The blue dashed line marks the
  position of the contact wave, the green dashed lines give the
  location of the rarefaction fan, and the red dashed line marks the
  shock.} \label{fig_evolved_state}
\end{figure}

Let's consider an example how this wave structure looks in a real
Riemann problem.  We consider, for definiteness, a Riemann problem
with $\rho_L= 1.0$, $P_L =1.0$, $v_L = 0$, and $\rho_R= 0.25$, $P_R
=0.1795$, $v_R = 0$ (which is of Sod-shock type). The adiabatic
exponent is taken to be $\gamma = 1.4$. We hence deal at $t=0.0$ with
the initial state displayed in Figure~\ref{fig_initial_state}.  After
time $t=5.0$, the wave structure formed by a rarefaction to the left
(location marked in green), a contact in the middle (blue) and a shock
to the right (red) can be nicely seen in
Figure~\ref{fig_evolved_state}.

Some general properties of the waves appearing in the Riemann problem
can be summarized as follows:
\begin{itemize}
\item \emph{Shock:} This is a sudden compression of the fluid,
  associated with an irreversible conversion of kinetic energy to
  heat, i.e.~here entropy is produced. The density, normal velocity
  component, pressure, and entropy all change discontinuously at a
  shock.

\item \emph{Contact discontinuity:} This traces the original
  separating plane between the two fluid phases that have been brought
  into contact. Pressure as well as the normal velocity are constant
  across a contact, but density, entropy and temperature can jump.

\item \emph{Rarefaction wave:} This occurs when the gas (suddenly)
  expands. The rarefaction wave smoothly connects two states over a
  finite spatial region; there are no discontinuities in any of the
  fluid variables.

\end{itemize}

\subsection{Finite volume discretization} \label{SecEulerDisc}

Let's now take a look how Riemann solvers can be used in the finite
volume discretization approach to the PDEs of fluid dynamics. Recall
that we can write our hyperbolic conservation laws as
\begin{equation}
\frac{\partial \vec{U}}{\partial t} + \nabla\cdot \vec{F} = 0.
\end{equation}
Here $\vec{U}$ is a state vector and $\vec{F}$ is the flux vector. For
example, the Euler equations of section~\ref{SecEulerEqns} can be
written in the form
\begin{equation}
\vec{U} = \left(\begin{array}{c}
\rho\\
\rho\vec{v}\\
\rho e
\end{array}
\right),
\;\;\;\;\;\;\;
\vec{F} = \left(
\begin{array}{c}
\rho\vec{v}\\
\rho\vec{v}\vec{v}^{T} + P\\
(\rho e + P)\vec{v}
\end{array}
\right),
\end{equation}
with the specific energy $e = u + \vec{v}^2 / 2$ and $u$ being the
thermal energy per unit mass. The ideal gas equation gives the
pressure as $P = (\gamma -1 ) \rho u$ and provides a closure for the
system. 

In a finite volume scheme, we describe the system through the averaged
state over a set of finite cells. These cell averages are defined as
\begin{equation}
\vec{U}_i = \frac{1}{V_i} \int_{{\rm cell}\; i} \vec{U}(\vec{x})\, {\rm d} V.
\end{equation}
Let's now see how we could divise an update scheme for these
cell-averaged quantities.

\begin{enumerate}
\item We start by integrating the conservation law over a cell, and
  over a finite interval in time:
\begin{equation}
  \int_{x_{i-\frac{1}{2}}}^{x_{i+\frac{1}{2}}} {\rm d}x 
  \int_{t_n}^{t_{n+1}} {\rm d}t \left(\frac{\partial \vec{U}}{ \partial t} + \frac{\partial \vec{F}}{\partial x}\right) = 0. 
\end{equation}

\item This gives
\begin{equation}
\int_{x_{i-\frac{1}{2}}}^{x_{i+\frac{1}{2}}} {\rm d}x 
\left[ \vec{U}(x, t_{n+1}) - \vec{U}(x, t_{n}) \right]
+ 
\int_{t_n}^{t_{n+1}} {\rm d}t \left[
\vec{F}(x_{i+\frac{1}{2}}, t)
-\vec{F}(x_{i-\frac{1}{2}}, t)
 \right]  = 0.
\end{equation}
In the first term, we recognize the definition of the cell average:
\begin{equation}
\vec{U}_i^{(n)} \equiv \frac{1}{\Delta x} 
\int_{x_{i-\frac{1}{2}}}^{x_{i+\frac{1}{2}}}  \vec{U}(x, t_{n})
{\rm d}x .
\end{equation}
Hence we have
\begin{equation}
\Delta x  
\left[ \vec{U}_i^{(n+1)} - \vec{U}_i^{(n)} \right]
+ 
\int_{t_n}^{t_{n+1}} {\rm d}t \left[
\vec{F}(x_{i+\frac{1}{2}}, t)
-\vec{F}(x_{i-\frac{1}{2}}, t)
 \right]  = 0.
\end{equation}

\item Now, $\vec{F}(x_{i+\frac{1}{2}},t)$ for $t>t_n$ is given by the
  solution of the Riemann problem with left state $\vec{U}_i^{(n)}$
  and right state $\vec{U}_{i+1}^{(n)}$. At the interface, this solution
  is {\em independent} of time. We can hence write
\begin{equation}
\vec{F}(x_{i+\frac{1}{2}},t) = \vec{F}^\star_{i+\frac{1}{2}},
\end{equation}
where $\vec{F}^\star_{i+\frac{1}{2}} = \vec{F}_{\rm
  Riemann}(\vec{U}_i^{(n)}, \vec{U}_{i+1}^{(n)})$ is a short-hand
notation for the corresponding Riemann solution sampled at the
interface. Hence we now get
\begin{equation}
\Delta x  
\left[ \vec{U}_i^{(n+1)} - \vec{U}_i^{(n)} \right]
+ 
\Delta t \left[
\vec{F}^\star_{i+\frac{1}{2}}
-\vec{F}^\star_{i-\frac{1}{2}}
 \right]  = 0.
\end{equation}
Or alternative, as an explicit update formula:
\begin{equation}
\vec{U}_i^{(n+1)} = \vec{U}_i^{(n)}  + \frac{\Delta
  t}{\Delta x} \left[ \vec{F}^\star_{i-\frac{1}{2}}
  -\vec{F}^\star_{i+\frac{1}{2}} \right].
\end{equation}
The first term in the square bracket gives the flux that flows from
left into the cell, the second term is the flux out of the cell on its
right side. The idea to use the Riemann solution in the updating step
is due to Godunov, that's why such schemes are often called \emph{
  Godunov schemes}.

\end{enumerate}

It is worthwhile to note that we haven't really made any approximation
in the above (yet). In particular, if we calculate $\vec{F}_{\rm
  Riemann}$ analytically (and hence exactly), then the above seems to
account for the correct fluxes for arbitrarily long times. So does
this mean that we get a perfectly accurate result even for very large
timesteps? This certainly sounds too good to be true, so there must be
a catch somewhere.

Indeed, there is. First of all, we have assumed that the Riemann
problems are independent of each other and each describe infinite
half-spaces. This is not true once we consider finite volume cells,
but it is still ok for a while as long $t_{n+1}$ is close enough to
$t_n$ such that the waves emanating in one interface have not yet
arrived at the next interface left or right. This then leads to a
CFL-timestep criterion, were $\Delta t \le \Delta x / c_{\rm max}$ and
$c_{\rm max}$ is the maximum wavespeed.

Another point is more subtle and comes into play when we consider more
than one timestep. We assumed that the $\vec{U}_i^{(n)}$ describe
piece-wise constant states which can then be fed to the Riemann solver
to give us the flux. However, even when this is true initially, we
have just seen that after one timestep it will not be true anymore. By
ignoring this in the subsequent timestep (which is done by performing
an averaging step that washes out the cell substructure that developed
as part of the evolution during the previous timestep) we make some
error.

\subsection{Godunov's method and Riemann solvers}

It is useful to introduce another interpretation of common
finite-volume discretizations of fluid dynamics, so-called
Reconstruct-Evolve-Average (REA) schemes. We also use this here for a
short summary of Godunov's important method, and the way Riemann
solvers come into play in it.

\begin{figure}
  \sidecaption
  \resizebox{7.4cm}{!}{\includegraphics{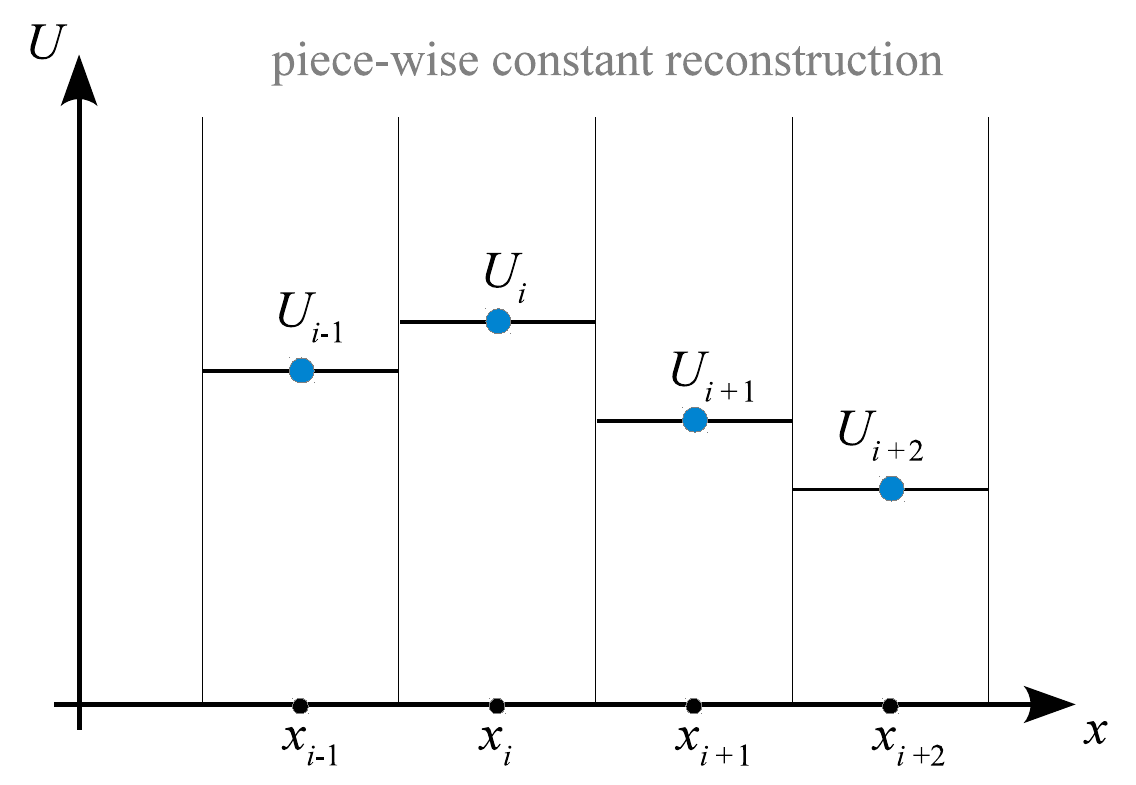}}
  \caption{Piece-wise constant states of a fluid forming the simplest
    possible reconstruction of its state based on a set of discrete
    values $U_i$ known at spatial positions $x_i$.}
\label{fig_piecewise}
\end{figure}

An REA update scheme of a hydrodynamical system discretized on a mesh
can be viewed as a sequence of three steps:
\begin{enumerate}
\item \emph{Reconstruct:} Using the cell-averaged quantities (as shown
  in Fig.~\ref{fig_piecewise}), this defines the run of these
  quantities everywhere in the cell. In the sketch, a piece-wise
  constant reconstruction is assumed, which is the simplest procedure
  one can use and leads to 1st order accuracy.
\item \emph{Evolve:} The reconstructed state is then evolved forward
  in time by $\Delta t$. In Godunov's approach, this is done by
  treating each cell interface as a piece-wise constant initial value
  problem which is solved with the Riemann solver exactly or
  approximately. This solution is formally valid as long as 
  the waves emanating from opposite sides of a cell do not yet
  start to interact. In practice, one therefore needs to limit the
  timestep $\Delta t$ such that this does not happen.
\item \emph{Average:} The wave structure resulting from the evolution
  over timestep $\Delta t$ is spatially averaged in a conservative
  fashion to compute new states $\vec{U}^{n+1}$ for each cell.
  Fortunately, the averaging step does not need to be done explicitly;
  instead it can simply be carried out by accounting for the fluxes
  that enter or leave the control volume of the cell.  Then the whole
  cycle repeats again.
\end{enumerate}

What is needed for the {\em evolve} step is a prescription to either
exactly or approximately solve the Riemann problem for a piece-wise
linear left and right state that are brought into contact at time
$t=t_n$. Formally, this can be written as
\begin{equation}
\vec{F}^{\star} =  \vec{F}_{\rm Riemann}(\vec{U}_L, \vec{U}_R).
\end{equation}
In practice, a variety of approximate Riemann solvers $\vec{F}_{\rm
  Riemann}$ are commonly used in the literature \citep{Rusanov1961,
  Harten1983, Toro1997, Miyoshi2005}. For the ideal gas and for
isothermal gas, it is also possible to solve the Riemann problem
exactly, but not in closed form (i.e.~the solution involves an
iterative root finding of a non-linear equation).

\ \\
\noindent There are now two main issues left:
\begin{itemize}
\item
How can this be extended to multiple spatial dimensions?
\item How can it be extended such that a higher order integration
  accuracy both in space and time is reached?
\end{itemize}
We'll discuss these issues next.

\subsection{Extensions to multiple dimensions}

So far, we have considered  {\em one-dimensional} hyperbolic
conservation laws of the form
\begin{equation}
\partial_t \vec{U} + \partial_x \vec{F(U)} =0 , \label{eqn1}
\end{equation}
where $\partial_t$ is a short-hand notation for $\partial_t =
\frac{\partial}{\partial t}$, and similarly $\partial_x =
\frac{\partial}{\partial x}$.
For example, for isothermal gas with soundspeed 
$c_s$, 
the state vector $\vec{U}$ and flux vector $\vec{F}(\vec{U})$
are given as
\begin{equation}
\vec{U} =\left( 
\begin{array}{c}
\rho \\
\rho u \\ 
\end{array}
\right),
\;\;\;
\vec{F} =\left( 
\begin{array}{c}
\rho u \\
\rho u^2 + \rho c_s^2 \\ 
\end{array}
\right) ,
\end{equation}
where $u$ is the velocity in the $x$-direction.

In three dimensions, the PDEs describing a fluid become considerably
more involved. For example, the Euler equations for an ideal gas are
given in explicit form as
\begin{equation}
\partial_t
\left(
\begin{array}{c}
\rho \\
\rho u \\ 
\rho v \\ 
\rho w \\
\rho e \\ 
\end{array}
\right)
+
\partial_x
\left(
\begin{array}{c}
\rho u \\
\rho u^2 + P  \\ 
\rho uv \\ 
\rho uw \\
\rho u(\rho e + P) \\ 
\end{array}
\right)
+
\partial_y
\left(
\begin{array}{c}
\rho v \\
\rho uv \\ 
\rho v^2 + P \\ 
\rho vw \\
\rho v(\rho e + P) \\ 
\end{array}
\right)
+
\partial_z
\left(
\begin{array}{c}
\rho w \\
\rho uw  \\ 
\rho vw \\ 
\rho w^2+P \\
\rho w(\rho e + P) \\ 
\end{array}
\right)
 =0 ,
\end{equation}
where $e = e_{\rm therm} + (u^2+v^2+w^2)/2$ is the total specific
energy per unit mass, $e_{\rm therm}$ is the thermal energy per unit
mass, and $P=(\gamma-1) \rho\, e_{\rm therm}$ is the pressure.  These
equations are often written in the following notation:
\begin{equation}
  \partial_t \vec{U} + \partial_x \vec{F}  + \partial_y \vec{G}  + \partial_z \vec{H} =0 . \label{eqn2}
\end{equation}
Here the functions $\vec{F(U)}$, $\vec{G(U)}$ and $\vec{H(U)}$ give
the flux vectors in the $x$-, $y$- and $z$-direction, respectively.

\subsubsection{Dimensional splitting}

Let us now consider the three dimensionally split problems derived
from equation (\ref{eqn2}):
\begin{equation}
\partial_t \vec{U} + \partial_x \vec{F} =0,
\label{eqnA}
\end{equation}
\begin{equation}
\partial_t \vec{U} + \partial_y \vec{G} =0,
\label{eqnB}
\end{equation}
\begin{equation}
\partial_t \vec{U} + \partial_z \vec{H} =0.
\label{eqnC}
\end{equation}
Note that the vectors appearing here have still the same
dimensionality as in the full equations. They are {\em augmented}
one-dimensional problems, i.e.~the transverse variables still appear
but spatial differentiation happens only in one direction. Because of
this, these additional transverse variables do not make the 1D problem
more difficult compared to the `pure' 1D problem considered earlier,
but the fluxes appearing in them still need to be included.

Now let us assume that he have a method to solve/advance each of these
one-dimensional problems. We can for example express this formally
through time-evolution operators ${\cal X}(\Delta t$), ${\cal
  Y}(\Delta t)$, and ${\cal Z}(\Delta t)$, which advance the solution
by a timestep $\Delta t$.  Then the full time advance of the system
can for example be approximated by
\begin{equation}
\vec{U}^{n+1} \simeq 
{\cal Z} ( \Delta t)
{\cal Y} ( \Delta t)
{\cal X} ( \Delta t)
\vec{U}^n  .
\end{equation}
This is one possible dimensionally split update scheme. In fact, this
is the exact solution if equations (\ref{eqnA})-(\ref{eqnB}) represent
the linear advection problem, but for more general non-linear
equations it only provides a first order approximation. However,
higher-order dimensionally split update schemes can also be easily
constructed. For example, in two-dimensions,
\begin{equation}
\vec{U}^{n+1} =
\frac{1}{2}
[ {\cal X} ( \Delta t)
{\cal Y} ( \Delta t) + 
{\cal Y} ( \Delta t)
{\cal X} ( \Delta t)]
\vec{U}^n
\end{equation}
and
\begin{equation}
\vec{U}^{n+1} =
{\cal X} ( \Delta t / 2)
{\cal Y} ( \Delta t) 
{\cal X} ( \Delta t / 2)
\vec{U}^n
\end{equation}
are second-order accurate. Similarly, for three dimensions the scheme
\begin{equation}
\vec{U}^{n+1} =
{\cal X} ( \Delta t/2)
{\cal Y} ( \Delta t/2) 
{\cal Z} ( \Delta t)
{\cal Y} ( \Delta t/2)
{\cal X} ( \Delta t/2)
\vec{U}^n
\end{equation}
is second-order accurate. As a general rule of thumb, the time
evolution operators have to be applied alternatingly in reverse order
to reach second-order accuracy. We see that the dimensionless
splitting reduces the problem effectively to a sequence of
one-dimensional solution operations which are applied to
multi-dimensional domains. Note that each one-dimensional operator
leads to an update of $\vec{U}$, and is a complete step for the
corresponding augmented one-dimensional problem. Gradients, etc., that
are needed for the next step then have to be recomputed before the
next time-evolution operator is applied. In practical applications of
mesh codes, these one-dimensional solves are often called {\em
  sweeps}.

\subsubsection{Unsplit schemes}

\begin{figure}
\hfill\begin{minipage}[c]{6cm}%
\resizebox{6.0cm}{!}{\includegraphics{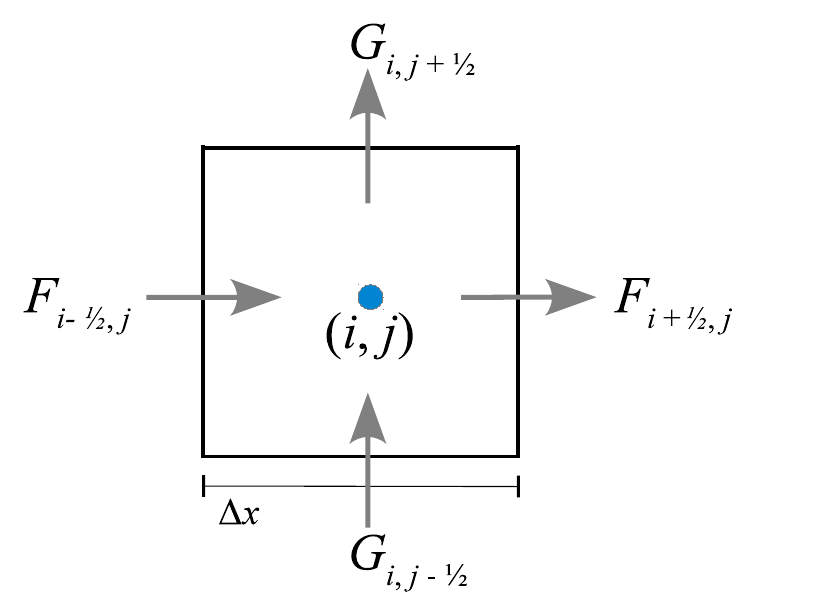}}%
\end{minipage}%
\begin{minipage}[c]{5cm}%
\resizebox{4.8cm}{!}{\includegraphics{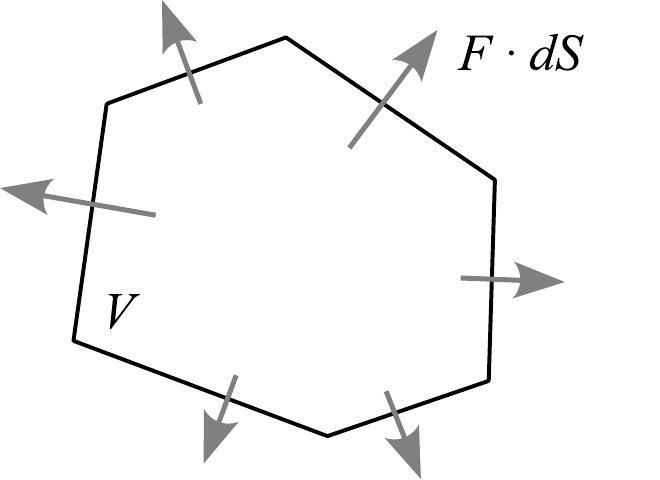}}
\end{minipage}
\caption{Sketch of unsplit finite-volume update schemes. On the left,
  the case of a structured Cartesian grid is shown, the case on the
  right is for an unstructured grid.}
\label{fig_unsplit}
\end{figure}

In an unsplit approach, all flux updates of a cell are applied
simultaneously to a cell, not sequentially. This is for example
illustrated in 2D in the situations depicted in
Figure~\ref{fig_unsplit}.  The unsplit update of cell $i,j$ in the
Cartesian case is then given by
\begin{equation}
U_{i,j}^{n+1} = U_{i,j}^n + 
\frac{\Delta t}{\Delta x}\left( \vec{F}_{i-\frac{1}{2}, j} -
  \vec{F}_{i+\frac{1}{2}, j}\right)
 + 
\frac{\Delta t}{\Delta y}\left( \vec{G}_{i, j-\frac{1}{2}} -
  \vec{G}_{i, j +\frac{1}{2}}\right).
\end{equation}

Unsplit approaches can also be used for irregular shaped cells like
those appearing in unstructured meshes (see
Fig.~\ref{fig_unsplit}). For example, integrating over a cell of
volume $V$ and denoting with $\vec{U}$ the cell average, we can write
the cell update with the divergence theorem as
\begin{equation}
\vec{U}^{n+1} = \vec{U}^n - \frac{\Delta t}{V}\int \vec{F}\, \cdot \vec{dS},
\end{equation}
where the integration is over the whole cell surface, with outwards
pointing face area vectors $\vec{dS}$.

\subsection{Extensions for high-order accuracy} \label{SecHighOrderExt}

We should first clarify what we mean with higher order
schemes. Loosely speaking, this refers to the convergence rate of a
scheme in smooth regions of a flow. For example, if we know the
analytic solution $\rho(x)$ for some problem, and then obtain a
numerical result $\rho_i$ at a set of $N$ points at locations $x_i$,
we can ask what the typical error of the solution is. One possibility
to quantify this would be through a L1 error norm, for example in the
form
\begin{equation}
L1 = \frac{1}{N} \sum_i | \rho_i  - \rho(x_i)|,
\end{equation}
which can be interpreted as the average error per cell. If we now
measure this error quantitatively for different resolutions of the
applied discretization, we would like to find that L1 decreases with
increasing $N$. In such a case our numerical scheme is converging, and
provided we use sufficient numerical resources, we have a chance to
get below any desired absolute error level. But the {\em rate of
  convergence} can be very different between different numerical
schemes when applied to the same problem. If a method shows a $L1
\propto N^{-1}$ scaling, it is said to be first-order accurate; a
doubling of the number of cells will then cut the error in half. A
second-order method has $L1 \propto N^{-2}$, meaning that a doubling
of the number of cells can actually reduce the error by a factor of
4. This much better convergence rate is of course highly desirable. It
is also possible to construct schemes with still higher convergence
rates, but they tend to quickly become much more complex and
computationally involved, so that one eventually reaches a point of
diminishing return, depending on the specific type of problem. But the
extra effort one needs to make to go from first to second-order is
often very small, sometimes trivially small, so that one basically
should always strive to try at least this.

\begin{figure}
\sidecaption
\resizebox{6.5cm}{!}{\includegraphics{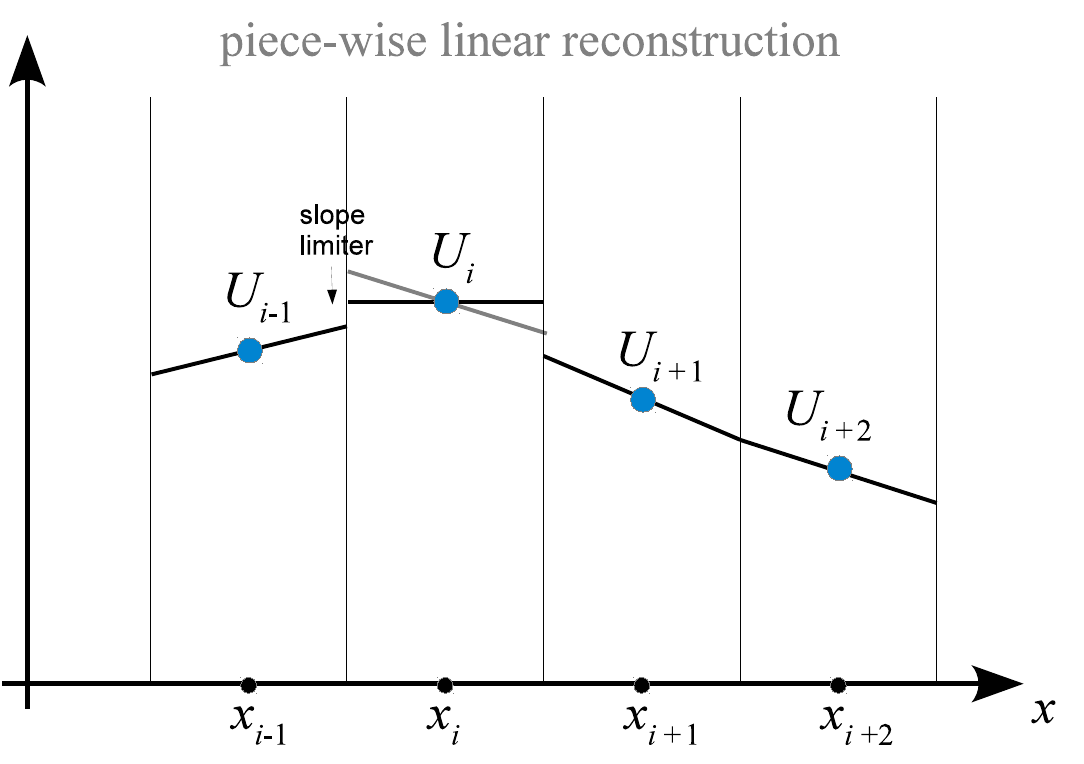}}
\caption{Piece-wise linear reconstruction scheme applied to a fluid
  state represented through a regular mesh.}
\end{figure}

A first step in constructing a 2nd order extension of Godunov's method
is to replace the piece-wise constant with a piece-wise linear
reconstruction. This requires that one first estimates gradients for
each cell (usually by a simple finite difference formula). These are
then slope-limited if needed such that the linear extrapolations of
the cell states to the cell interfaces do not introduce new
extrema. This slope-limiting procedure is quite important; it needs to
be done to avoid that real fluid discontinuities introduce large
spurious oscillations into the fluid.

Given slope limited gradients, for example $\nabla\rho$ for the
density, one can then estimate the left and right states adjacent to
an interface $x_{i+\frac{1}{2}}$ by spatial extrapolation from the
centers of the cells left and right from the interface:
\begin{eqnarray}
\rho_{i+\frac{1}{2}}^{L} & = & \rho_i + (\nabla \rho)_i \,
\frac{\Delta x}{2}  , \\
\rho_{i+\frac{1}{2}}^{R} & = & \rho_{i+1} - (\nabla \rho)_{i+1} \, \frac{\Delta x}{2}.
\end{eqnarray}
The next step would in principle be to use these states in the Riemann
solver. In doing this we will ignore the fact that our reconstruction
has now a gradient over the cell; instead we still pretend that the
fluid state can be taken as piece-wise constant left and right of the
interface as far as the Riemann solver is concerned. However, it turns
out that the spatial extrapolation needs to be augmented with a
temporal extrapolation one half timestep into the future, such that
the flux estimate is now effectively done in the middle of the
timestep. This is necessary both to reach second-order accuracy in
time and also for stability reasons.  Hence we really need to use
\begin{eqnarray}
\rho_{i+\frac{1}{2}}^{L} & = & \rho_i + (\nabla \rho)_i \,
\frac{\Delta x}{2} + \left(\frac{\partial \rho}{\partial t}\right)_i \,
\frac{\Delta t}{2}, \\
\rho_{i+\frac{1}{2}}^{R} & = & \rho_{i+1} - (\nabla \rho)_{i+1} \,
\frac{\Delta x}{2}
 + \left(\frac{\partial \rho}{\partial t}\right)_{i+1} \,
\frac{\Delta t}{2},
\end{eqnarray}
for extrapolating to the interfaces. More generally, this has to be
done for the whole state vector of the system, i.e.
\begin{eqnarray}
\vec{U}_{i+\frac{1}{2}}^{L} & = & \vec{U}_i + (\partial_x \vec{U})_i \,
\frac{\Delta x}{2} + (\partial_t \vec{U})_i \,
\frac{\Delta t}{2} ,\\
\vec{U}_{i+\frac{1}{2}}^{R} & = & \vec{U}_{i+1} - (\partial_x \vec{U})_{i+1} \,
\frac{\Delta x}{2} + (\partial_t \vec{U})_{i+1} \,
\frac{\Delta t}{2}  .
\end{eqnarray}
Note that here the quantity $(\partial_x \vec{U})_i$ is a
(slope-limited) {\em estimate} of the gradient in cell~$i$, based on
finite-differences plus a slope limiting procedure. Similarly, we
somehow need to estimate the time derivative encoded in $(\partial_x
\vec{U})_i$. How can this be done? One way to do this is to exploit
the Jacobian matrix of the Euler equations. We can write the Euler
equations as
\begin{equation}
\partial_t \vec{U} = -\partial_x \vec{F}(\vec{U}) =
- \frac{\partial\vec{F}}{\partial \vec{U}}\, \partial_x \vec{U} =
-\vec{A}(\vec{U})\, \partial_x\vec{U} ,
\end{equation}
where $\vec{A(U)}$ is the Jacobian matrix. Using this, we can simply
estimate the required time-derivative based on the spatial
derivatives:
\begin{equation}
(\partial_t \vec{U})_i = 
-\vec{A}(\vec{U}_i)\, (\partial_x\vec{U})_i .
\end{equation}
Hence the extrapolation can be done as
\begin{eqnarray}
\vec{U}_{i+\frac{1}{2}}^{L} & = & \vec{U}_i + 
\left[
\frac{\Delta x}{2} -
\frac{\Delta t}{2} \vec{A}(\vec{U}_i)
\right]   (\partial_x \vec{U})_i ,
\\
\vec{U}_{i+\frac{1}{2}}^{R} & = & \vec{U}_{i+1} + 
\left[-
\frac{\Delta x}{2} -
\frac{\Delta t}{2} \vec{A}(\vec{U}_{i+1})
\right]   (\partial_x \vec{U})_{i+1}  .
\end{eqnarray}
This procedure defines the so-called MUSCL-Hancock scheme
\citep{Leer1984,Toro1997,Leer2006}, which is a 2nd-order accurate
extension of Godunov's method.

Higher-order extensions such as the piece-wise parabolic method (PPM)
start out with a higher order polynomial reconstruction. In the case
of PPM, parabolic shapes are assumed in each cell instead of
piece-wise linear states. The reconstruction is still guaranteed to be
conservative, i.e.~the integral underneath the reconstruction recovers
the total values of the conserved variables individually in each
cell. So-called ENO and WENO schemes \citep[e.g.][]{Balsara2009} use
yet higher-order polynomials to reconstruct the state in a
conservative fashion. Here many more cells in the environment need to
be considered (i.e.~the so-called {\em stencil} of these methods is
much larger) to robustly determine the coefficients of the
reconstruction. This can for example involve a least-square fitting
procedure \citep{OllivierGooch1997}.

\section{Smoothed particle hydrodynamics}

Smoothed Particle Hydrodynamics (SPH) is a technique for approximating
the continuum dynamics of fluids through the use of particles, which
may also be viewed as interpolation points
\citep[SPH;][]{Lucy1977,Gingold1977,Monaghan1992,Springel2010b}.  The
principal idea of SPH is to treat hydrodynamics in a completely
mesh-free fashion, in terms of a set of sampling
particles. Hydrodynamical equations of motion are then derived for
these particles, yielding a quite simple and intuitive formulation of
gas dynamics. Moreover, it turns out that the particle representation
of SPH has excellent conservation properties.  Energy, linear
momentum, angular momentum, mass, and entropy (if no artificial
viscosity operates) are all simultaneously conserved. In addition,
there are no advection errors in SPH, and the scheme is fully Galilean
invariant, unlike alternative mesh-based Eulerian techniques.  Due to
its Lagrangian character, the local resolution of SPH follows the mass
flow automatically, a property that is convenient in representing the
large density contrasts often encountered in astrophysical problems.
 
\subsection{Kernel interpolation}

At the heart of smoothed particle hydrodynamics lie so-called kernel
interpolants. In particular, we use a kernel summation interpolant for
estimating the density, which then determines the rest of the basic
SPH equations through the variational formalism.

For any field $F(\vec{r})$, we may define a smoothed interpolated
version, $F_s(\vec{r})$, through a convolution with a kernel
$W(\vec{r},h)$:
\begin{equation}
F_s(\vec{r}) = \int F(\vec{r}')\, W(\vec{r} - \vec{r}',h)\,{\rm
d}\vec{r}'.
\label{eqk1}
\end{equation}
Here $h$ describes the characteristic width of the kernel, which is
normalized to unity and approximates a Dirac $\delta$-function in the
limit $h\to 0$. We further require that the kernel is symmetric and
sufficiently smooth to make it at least differentiable twice. One
possibility for $W$ is a Gaussian. However, most current SPH
implementations are based on kernels with a finite support. Usually a
cubic spline is adopted with $W(r, h)= w(\frac{r}{2h})$, and
\begin{equation}
w_{\rm 3D}(q) =\frac{8}{\pi} \left\{
\begin{array}{ll}
1-6 q^2 + 6 q^3, &
0\le  q \le\frac{1}{2} ,\\
2\left(1-q\right)^3, & \frac{1}{2}< q \le 1 ,\\
0 , & q>1 ,
\end{array}
\right.
\end{equation}
in three-dimensional normalization, but recent work also considered
various alternative kernels \citep[][]{Read2010, Dehnen2012}. Through
Taylor expansion, it is easy to see that the above kernel interpolant
is second-order accurate for regularly distributed points due to the
symmetry of the kernel.

\begin{figure}
\resizebox{5cm}{!}{\includegraphics{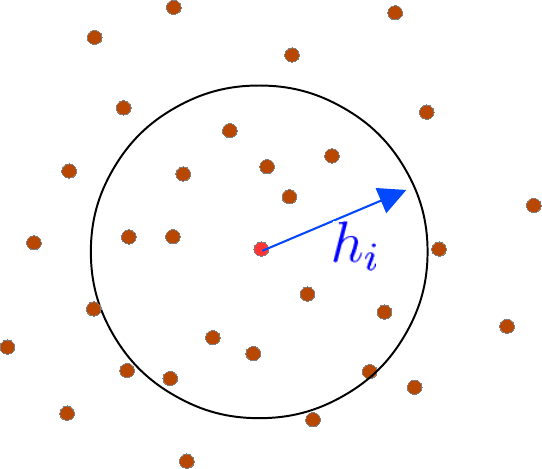}}
\resizebox{7cm}{!}{\includegraphics{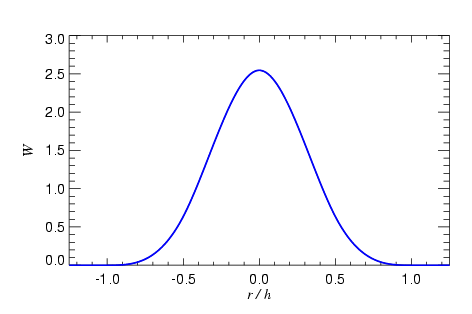}}
\caption{Kernel interpolation with a B-spline kernel.}
\end{figure}

Suppose now we know the field at a set of points $\vec{r}_i$, i.e.
$F_i=F(\vec{r}_i)$. The points have an associated mass $m_i$ and
density $\rho_i$, such that $V_i \sim m_i/\rho_i$ is
their associated finite volume element. Provided the points
sufficiently densely sample the kernel volume, we can approximate the
integral in Eqn.~(\ref{eqk1}) with the sum
\begin{equation}
F_s(\vec{r}) \simeq \sum_j \frac{m_j}{\rho_j}\,F_j\,
W(\vec{r} - \vec{r}_j,h).
\label{Eqkernelsum}
\end{equation}
This is effectively a Monte-Carlo integration, except that thanks to
the comparatively regular distribution of points encountered in
practice, the accuracy is better than for a random
distribution of the sampling points.  In particular, for points in one
dimension with equal spacing $d$, one can show that for $h=d$ the sum
of Eqn.~(\ref{Eqkernelsum}) provides a second order accurate
approximation to the real underlying function. Unfortunately, for the
irregular yet somewhat ordered particle configurations encountered in
real applications, a formal error analysis is not straightforward. It
is clear however, that at the very least one should have $h\ge d$,
which translates to a minimum of $\sim 33$ neighbors in 3D if a
Cartesian point distribution is assumed.

Importantly, we see that the estimate for $F_s(\vec{r})$ is defined
everywhere (not only at the underlying points), and is differentiable
thanks to the differentiability of the kernel, albeit with a
considerably higher interpolation error for the derivative.  Moreover,
if we set $F(\vec{r}) = \rho(\vec{r})$, we obtain
\begin{equation}
  \rho_s(\vec{r})  \simeq \sum_j  m_j  W (\vec{r}-\vec{r}_j, h),  \label{eqnDensEst}
\end{equation}
yielding a density estimate just based on the particle coordinates and
their masses.  In general, the smoothing length can be made variable
in space, $h=h(\vec{r},t)$, to account for variations in the sampling
density.  This adaptivity is one of the key advantages of SPH and is
essentially always used in practice. There are two options to
introduce the variability of $h$ into Eqn.~(\ref{eqnDensEst}). One is
by adopting $W(\vec{r}-\vec{r}_j, h(\vec{r}))$ as kernel, which
corresponds to the so-called `scatter' approach
\citep{Hernquist1989}. It has the advantage that the volume integral
of the smoothed field recovers the total mass, $\int
\rho_s(\vec{r})\,{\rm d}\vec{r} = \sum_i m_i$. On the other hand, the
so-called `gather' approach, where we use $W(\vec{r}-\vec{r}_j,
h(\vec{r}_i))$ as kernel in Eqn.~(\ref{eqnDensEst}), requires only
knowledge of the smoothing length $h_i = h(\vec{r}_i)$ for estimating
the density of particle $i$, which leads to computationally convenient
expressions when the variation of the smoothing length is consistently
included in the SPH equations of motion. Since the density is only
needed at the coordinates of the particles and the total mass is
conserved anyway (since it is tied to the particles), it is not
important that the volume integral of the gather form of
$\rho_s(\vec{r})$ exactly equals the total mass.

In the following we drop the subscript $s$ for indicating the smoothed
field, and adopt as SPH estimate of the density of particle $i$ the
expression
\begin{equation}
\rho_i  = \sum_{j=1}^N  m_j \, W (\vec{r}_i-\vec{r}_j, h_i).
\end{equation}
It is clear now why kernels with a finite support are preferred. They
allow the summation to be restricted to the $N_{\rm ngb}$ neighbors
that lie within the spherical region of radius $2h$ around the target
point $\vec{r}_i$, corresponding to a computational cost of order
${\cal O}(N_{\rm ngb}\,N)$ for the full density estimate.  Normally
this number $N_{\rm ngb}$ of neighbors within the support of the
kernel is approximately (or exactly) kept constant by choosing the
$h_i$ appropriately. $N_{\rm ngb}$ hence represents an important
parameter of the SPH method and needs to be made large enough to
provide sufficient sampling of the kernel volumes.  Kernels like the
Gaussian on the other hand would require a summation over all
particles $N$ for every target particle, resulting in a ${\cal
  O}(N^2)$ scaling of the computational cost.

If SPH was really a Monte-Carlo method, the accuracy expected from the
interpolation errors of the density estimate would be rather
problematic. But the errors are much smaller because the particles do
not sample the fluid in a Poissonian fashion. Instead, their distances
tend to equilibrate due to the pressure forces, which makes the
interpolation errors much smaller \citep{Price2012}. Yet, they remain
a significant source of error in SPH and are ultimately the primary
origin of the noise inherent in SPH results \citep{Bauer2012}.

Even though we have based most of the above discussion on the density,
the general kernel interpolation technique can also be applied to
other fields, and to the construction of differential operators. For
example, we may write down a smoothed velocity field and take its
derivative to estimate the local velocity divergence, yielding:
\begin{equation}
(\nabla\cdot\vec{v})_i = \sum_j \frac{m_j}{\rho_j} \vec{v}_j \cdot \nabla_i
  W(\vec{r}_i-\vec{r}_j, h) .
\end{equation}
However, an alternative estimate can be obtained by considering the
identity $\rho \nabla\cdot \vec{v} = \nabla(\rho\vec{v}) -
\vec{v}\cdot \nabla\rho$, and computing kernel estimates for the two
terms on the right hand side independently. Their difference then
yields
\begin{equation}
(\nabla\cdot\vec{v})_i = \frac{1}{\rho_i} 
\sum_j m_j (\vec{v}_j  - \vec{v}_i) \cdot \nabla_i W(\vec{r}_i-\vec{r}_j, h) .
\label{eqnVelDiv}
\end{equation}
This pair-wise formulation turns out to be more accurate in
practice. In particular, it has the advantage of always providing a
vanishing velocity divergence if all particle velocities are equal.

\subsection{SPH equations of motion}

The Euler equations for inviscid gas dynamics in Lagrangian 
form are given by
\begin{equation}
\frac{{\rm d}\rho}{{\rm d}t} + \rho \,\nabla\cdot \vec{v} = 0 ,
\end{equation}
\begin{equation}
\frac{{\rm d}\vec{v}}{{\rm d}t} + \frac{\nabla P}{\rho} = 0 ,
\end{equation}
\begin{equation}
\frac{{\rm d}u}{{\rm d}t} + \frac{P}{\rho} \nabla \cdot \vec{v} = 0 ,
\end{equation}
where ${\rm d}/{\rm d}t = \partial / \partial t + \vec{v}\cdot \nabla$
is the convective derivative. This system of partial differential
equations expresses conservation of mass, momentum and
energy. \citet{Eckart1960} has shown that the Euler equations for an
inviscid ideal gas follow from the Lagrangian
\begin{equation}
L = \int \rho\left(\frac{\vec{v}^2}{2} - u\right)  \,{\rm d}V .
\end{equation}
This opens up an interesting route for obtaining discretized equations
of motion for gas dynamics. Instead of working with the continuum
equations directly and trying to heuristically work out a set of
accurate difference formulas, one can discretize the Lagrangian and
then derive SPH equations of motion by applying the variational
principals of classical mechanics \citep{Springel2002}. Using a
Lagrangian also immediately guarantees certain conservation laws and
retains the geometric structure imposed by Hamiltonian dynamics on
phase space.
 
We start by discretizing the Lagrangian in terms of fluid particles of
mass $m_i$, yielding
\begin{equation}
L_{\rm SPH} = \sum_i \left(\frac{1}{2}m_i {\vec{v}_i^2} - m_i
u_i \right) , \label{eqndisctLgr}
\end{equation}
where it has been assumed that the thermal energy per unit mass of a
particle can be expressed through an entropic function $A_i$ of the
particle, which simply labels its specific thermodynamic entropy. The
pressure of the particles is
\begin{equation}
P_i = A_i \rho_i^\gamma  = (\gamma-1) \rho_i u_i,
\end{equation}
where $\gamma$ is the adiabatic index.  Note that for isentropic flow
(i.e.~in the absence of shocks, and without mixing or thermal
conduction) we expect the $A_i$ to be constant. We hence define $u_i$,
the thermal energy per unit mass, in terms of the density estimate as
\begin{equation}
u_i(\rho_i) =  A_i \frac{\rho_i^{\gamma-1}}{\gamma-1} . \label{eqU}
\end{equation}
This raises the question of how the smoothing lengths $h_i$ needed for
estimating $\rho_i$ should be determined.  As we discussed above, we
would like to ensure adaptive kernel sizes, meaning that the number of
points in the kernel should be approximately constant. In much of the
older SPH literature, the number of neighbors was allowed to vary
within some (small) range around a target number. Sometimes the
smoothing length itself was evolved with a differential equation in
time, exploiting the continuity relation and the expectation that
$\rho h^3$ should be approximately constant.  In case the number of
neighbors outside the kernel happened to fall outside the allowed
range, $h$ was suitably readjusted, at the price of some errors in
energy conservation.

A better method is to require that the mass in the kernel volume
should be constant, viz.
\begin{equation}
\rho_i  h_i^3 = const  \label{eqnconst}
\end{equation}
for three dimensions.  Since $\rho_i = \rho_i(\vec{r}_1, \vec{r}_2,
\ldots \vec{r}_N, h_i)$ is only a function of the particle coordinates
and of $h_i$, this equation implicitly defines the function $h_i =
h_i(\vec{r}_1, \vec{r}_2, \ldots \vec{r}_N)$ in terms of the particle
coordinates.

We can then proceed to derive the equations of motion from
\begin{equation}
\frac{{\rm d}}{{\rm d} t}
\frac{\partial L}{\partial \dot {\vec{r}}_i} -
\frac{\partial L}{\partial \vec{r}_i} = 0.
\end{equation} 
This  first gives
\begin{equation}
 m_i \frac{\dd \vec{v}_i}{\dd t} 
= - 
\sum_{j=1}^{N} m_j \frac{P_j}{\rho_j^2} \,\frac{\partial \rho_j}{\partial
\vec{r}_i},
\end{equation}
where the derivative $\partial \rho_j/\partial \vec{r}_i$ stands for
the total variation of the density with respect to the coordinate
$\vec{r}_i$, including any variation of $h_j$ this may entail. We can
hence write
\begin{equation}
\frac{\partial \rho_j}{\partial \vec{r}_i}  =
\vec{\nabla}_i \rho_j +  
\frac{\partial \rho_j}{\partial h_j}
\frac{\partial h_j}{\partial \vec{r}_i} ,
\label{eqA1}
\end{equation}
where the smoothing length is kept constant in the first derivative on
the right hand side (in our notation, the Nabla operator $\nabla_i
= \partial / \partial \vec{r}_i$ means differentiation with respect to
$\vec{r}_i$ holding the smoothing lengths constant).  On the other
hand, differentiation of $\rho_j h_j^3 = const$ with respect to
$\vec{r}_i$ yields
\begin{equation}
 \frac{\partial \rho_j}{\partial h_j}
\frac{\partial h_j}{\partial \vec{r}_i} 
\left[ 1+ \frac{3\,\rho_j}{h_j} \left(\frac{\partial \rho_j}{\partial
h_j}\right)^{-1} \right]  = - \vec{\nabla}_i\rho_j .
\label{eqA2}
\end{equation}
Combining equations (\ref{eqA1}) and (\ref{eqA2}) we then find
\begin{equation}
 \frac{\partial \rho_j}{\partial \vec{r}_i}  =
\left( 1+\frac{h_j}{3\rho_j} \frac{\partial \rho_j}{\partial h_j}    
\right)^{-1} \vec{\nabla}_i \rho_j \label{eqA3} .
\end{equation}
 Using 
\begin{equation} 
\nabla_i{ \rho_j} = m_i \nabla_i W_{ij}(h_j)
+\delta_{ij}\sum_{k=1}^N m_k \nabla_i W_{ki}(h_i) \, ,  \label{eqA4}
\end{equation}
 we finally
obtain the equations of motion 
\begin{equation}
\frac{\dd \vec{v}_i}{\dd t} = -
\sum_{j=1}^N m_j \left[ f_i \frac{P_i}{\rho_i^2} \nabla_i W_{ij}(h_i)
+ f_j \frac{P_j}{\rho_j^2} \nabla_i W_{ij}(h_j) \right], 
\label{eqnmot} 
\end{equation}
 where the $f_i$ are defined by 
\begin{equation}
 f_i = \left[ 1 +
\frac{h_i}{3\rho_i}\frac{\partial \rho_i}{\partial h_i} \right]^{-1} ,
\label{eqA5}
\end{equation} 
and the abbreviation $W_{ij}(h)= W(|\vec{r}_{i}-\vec{r}_{j}|, h)$ has
been used. Note that the correction factors $f_i$ can be easily
calculated alongside the density estimate, all that is required is an
additional summation to get $\partial \rho_i/ \partial \vec{r}_i$ for
each particle. This quantity is in fact also useful to get the correct
smoothing radii by iteratively solving $\rho_i h_i^3=const$ with a
Newton-Raphson iteration \citep{Springel2002}.

The equations of motion~(\ref{eqnmot}) for inviscid hydrodynamics are
remarkably simple. In essence, we have transformed a complicated
system of partial differential equations into a much simpler set of
ordinary differential equations.  Furthermore, we only have to solve
the momentum equation explicitly.  The mass conservation equation as
well as the total energy equation (and hence the thermal energy
equation) are already taken care of, because the particle masses and
their specific entropies stay constant for reversible gas
dynamics. However, later we will introduce an artificial viscosity
that is needed to allow a treatment of shocks. This will introduce
additional terms in the equation of motion and requires the time
integration of one thermodynamic quantity per particle, which can
either be chosen as entropy or thermal energy.  Indeed, the above
formulation can also be equivalently expressed in terms of thermal
energy instead of entropy. This follows by taking the time derivative
of Eqn.~(\ref{eqU}), which first yields
\begin{equation}
\frac{{\rm d}u_i}{{\rm d} t}
= \frac{P_i}{\rho_i^2} \sum_j \vec{v}_j \cdot \frac{\partial
\rho_i}{\partial \vec{r}_j} .
\end{equation}
Using equations (\ref{eqA3}) and (\ref{eqA4}) then gives the evolution
of the thermal energy as
\begin{equation}
\frac{{\rm d}u_i}{{\rm d} t}
=  f_i \frac{P_i}{\rho_i^2} \sum_j m_j (\vec{v}_i - \vec{v}_j) \cdot
\vec{\nabla}W_{ij}(h_i), \label{eqDuDt}
\end{equation}
which needs to be integrated along the equation of motion if one wants
to use the thermal energy as independent thermodynamic variable. There
is no difference however to using the entropy; the two are completely
equivalent in the variational formulation.

Note that the above formulation readily fulfills the conservation laws
of energy, momentum and angular momentum. This can be shown based on
the discretized form of the equations, but it is also manifest due to
the symmetries of the Lagrangian that was used as a starting point.
The absence of an explicit time dependence gives the energy
conservation, the translational invariance implies momentum
conservation, and the rotational invariance gives angular momentum
conservation.

\subsection{Artificial Viscosity} \label {SecVisco}

Even when starting from perfectly smooth initial conditions, the gas
dynamics described by the Euler equations may readily produce true
discontinuities in the form of shock waves and contact
discontinuities. At such fronts the differential form of the Euler
equations breaks down, and their integral form (equivalent to the
conservation laws) needs to be used.  At a shock front, this yields
the Rankine-Hugoniot jump conditions that relate the upstream and
downstream states of the fluid.  These relations show that the
specific entropy of the gas always increases at a shock front,
implying that in the shock layer itself the gas dynamics can no longer
be described as inviscid. In turn, this also implies that the
discretized SPH equations we derived above can not correctly describe
a shock, simply because they keep the entropy strictly constant.

One thus must allow for a modification of the dynamics at shocks and
somehow introduce the necessary dissipation. This is usually
accomplished in SPH by an artificial viscosity. Its purpose is to
dissipate kinetic energy into heat and to produce entropy in the
process. The usual approach is to parameterize the artificial
viscosity in terms of a friction force that damps the relative motion
of particles.  Through the viscosity, the shock is broadened into a
resolvable layer, something that makes a description of the dynamics
everywhere in terms of the differential form possible.  It may seem a
daunting task though to somehow tune the strength of the artificial
viscosity such that just the right amount of entropy is generated in a
shock. Fortunately, this is however relatively unproblematic. Provided
the viscosity is introduced into the dynamics in a conservative
fashion, the conservation laws themselves ensure that the right amount
of dissipation occurs at a shock front.

What is more problematic is to devise the viscosity such that it is
only active when there is really a shock present. If it also operates
outside of shocks, even if only at a weak level, the dynamics may
begin to deviate from that of an ideal gas.

The viscous force is most often added to the equation of motion as 
\begin{equation}
 \left. \frac{\dd \vec{v}_i}{\dd t}\right|_{\rm visc} 
=
-\sum_{j=1}^N m_j \Pi_{ij} \nabla_i\overline{W}_{ij} \, ,
\label{eqnvisc}
\end{equation}
where
\begin{equation}
 \overline{W}_{ij}= \frac{1}{2}\left[ W_{ij}(h_i) +
W_{ij} (h_j)\right] 
\end{equation} denotes
a symmetrized kernel, which some researchers prefer to define as
$\overline{W}_{ij}= W_{ij}( [h_i+h_j]/2)$.  Provided the viscosity
factor $\Pi_{ij}$ is symmetric in $i$ and $j$, the viscous force between
any pair of interacting particles will be antisymmetric and along the
line joining the particles. Hence linear momentum and angular momentum
are still preserved. In order to conserve total energy, we need to
compensate the work done against the viscous force in the thermal
reservoir, described either in terms of entropy,
\begin{equation}
\left. \frac{\dd A_i}{\dd t}\right|_{\rm visc} =
\frac{1}{2}\frac{\gamma-1}{\rho_i^{\gamma-1}}\sum_{j=1}^N m_j \Pi_{ij}
\vec{v}_{ij}\cdot\nabla_i \overline{W}_{ij} \,,
\label{eqnentropy}
\end{equation}
or in terms of thermal energy per unit mass,
\begin{equation}
\left. \frac{\dd u_i}{\dd t}\right|_{\rm visc} =  \frac{1}{2}
\sum_{j=1}^N m_j
 \Pi_{ij} \vec{v}_{ij}
\cdot \nabla_i \overline{W}_{ij}\, ,
\label{eqnu-as}
\end{equation}
where $\vec{v}_{ij}= \vec{v}_i - \vec{v}_j$. There is substantial
freedom in the detailed parametrization of the viscosity $\Pi_{ij}$.
The most commonly used formulation of
the viscosity is
\begin{equation}
\label{eqvisc}
\Pi_{ij}=\left\{
\begin{tabular}{cl}
${\left[-\alpha c_{ij} \mu_{ij} +\beta \mu_{ij}^2\right]}/{\rho_{ij}}$ &
\mbox{if
$\vec{v}_{ij}\cdot\vec{r}_{ij}<0$} \\
0 & \mbox{otherwise}, \\
\end{tabular}
\right. 
 \end{equation}
 with 
\begin{equation}
\mu_{ij}=\frac{h_{ij}\,\vec{v}_{ij}\cdot\vec{r}_{ij} }
{\left|\vec{r}_{ij}\right|^2 + \epsilon h_{ij}^2}.\label{egnMu} 
\end{equation}
Here $h_{ij}$ and $\rho_{ij}$ denote arithmetic means of the
corresponding quantities for the two particles $i$ and $j$, with
$c_{ij}$ giving the mean sound speed, and $\vec{r}_{ij}\equiv
\vec{r}_i - \vec{r}_j$.  The strength of the viscosity is regulated by
the parameters $\alpha$ and $\beta$, with typical values in the range
$\alpha\simeq 0.5-1.0$ and the frequent choice of $\beta=2\,\alpha$.
The parameter $\epsilon\simeq 0.01$ is introduced to protect against
singularities if two particles happen to get very close.

In this form, the artificial viscosity is basically a combination of a
bulk and a von Neumann-Richtmyer viscosity.  Historically, the
quadratic term in $\mu_{ij}$ has been added to the original
Monaghan-Gingold form to prevent particle penetration in high Mach
number shocks.  Note that the viscosity only acts for particles that
rapidly approach each other, hence the entropy production is always
positive definite.

\subsection{New trends in SPH}

Smoothed particle hydrodynamics is a remarkably versatile and simple
approach for numerical fluid dynamics. The ease with which it can
provide a large dynamic range in spatial resolution and density, as
well as an automatically adaptive resolution, are unmatched in
Eulerian methods.  At the same time, SPH has excellent conservation
properties, not only for energy and linear momentum, but also for
angular momentum. The latter is not automatically guaranteed in
Eulerian codes, even though it is usually fulfilled at an acceptable
level for well-resolved flows. When coupled to self-gravity, SPH
conserves the total energy exactly, which is again not manifestly true
in most mesh-based approaches to hydrodynamics.  Finally, SPH is
Galilean-invariant and free of any errors from advection alone, which
is another advantage compared to Eulerian mesh-based approaches.

Thanks to its completely mesh-free nature, SPH can easily deal with
complicated geometric settings and large regions of space that are
completely devoid of particles. Implementations of SPH in a numerical
code tend to be comparatively simple and transparent. At the same
time, the scheme is characterized by remarkable robustness. For
example, negative densities or negative temperatures, sometimes a
problem in mesh-based codes, can not occur in SPH by
construction. Although shock waves are broadened in SPH, the
properties of the post-shock flow are correct.

The main disadvantage of SPH is its limited accuracy in
multi-dimensional flows \citep[e.g.][]{Agertz2007, Bauer2012}. One
source of noise originates in the approximation of local kernel
interpolants through discrete sums over a small set of nearest
neighbors. While in 1D the consequences of this noise tend to be
reasonably benign, particle motion in multiple dimensions has a much
higher degree of freedom. Here the mutually repulsive forces of
pressurized neighboring particle pairs do not easily cancel in all
dimensions simultaneously, especially not given the errors of the
discretized kernel interpolants. As a result, some `jitter' in the
particle motions readily develops, giving rise to velocity noise up to
a few percent of the local sound speed. This noise seriously messes up
the accuracy that can be reached with the technique, especially for
subsonic flow, and also leads to a slow convergence rate.

Another problem is the relatively high numerical viscosity of SPH. To
reduce the numerical viscosity of SPH in regions away from shocks,
several studies have recently advanced the idea of keeping the
functional form of the artificial viscosity, but making the viscosity
strength parameter $\alpha$ variable in time \citep{Morris1997,
  Dolag2005_turbulence, Rosswog2005}.  Adopting $\beta = 2\alpha$, one
may for example evolve the parameter $\alpha$ individually for each
particle with an equation such as
\begin{equation}
\frac{{\rm d}\alpha_i}{{\rm d}t} = - \frac{\alpha_i - \alpha_{\rm
    max}}{\tau_i} + S_i ,
\end{equation}
where $S_i$ is some source function meant to ramp up the viscosity
rapidly if a shock is detected, while the first term lets the
viscosity exponentially decay again to a prescribed minimum value
$\alpha_{\rm min}$ on a timescale $\tau_i$. So far, mostly simple
source functions like $S_i = \max[-(\nabla\cdot\vec{v})_i, 0]$ and
timescales $\tau_i \simeq h_i/c_i$ have been explored and the
viscosity $\alpha_i$ has often also been prevented from becoming
higher than some prescribed maximum value $\alpha_{\rm max}$. It is
clear that the success of such a variable $\alpha$ scheme depends
critically on an appropriate source function. The form above can still
not distinguish purely adiabatic compression from that in a shock, so
is not completely free of creating unwanted viscosity. More advanced
parameterizations that try to address this problem have therefore also
been developed \citep{Cullen2010}.

Particularly problematic in SPH are also fluid instabilities across
contact discontinuities, such as Kelvin-Helmholtz instabilities. These
are usually found to be suppressed in their growth. Recent new
formulations of SPH try to improve on this either through different
kernel shapes combined with a much larger number of smoothing
neighbors \citep[e.g.][]{Read2012}, through artificial thermal
conduction at contact discontinuities to reduce pressure force errors
and spurious surface tension there \citep[e.g.][]{Price2008}, or by
alluding to a pressure-based formulation where the density is
estimated only indirectly from a kernel-interpolated pressure field
\citep{Hopkins2013}. These developments appear certainly promising.
At the moment many new ideas for improved SPH formulations are still
advanced, and new implementations are published regularly. While many
problems of SPH have been addresses by these new schemes, so far they
have not yet been able to cure the relatively large gradient errors in
SPH, suggesting that the convergence rate of them is still lower than
that of comparable mesh-based approaches \citep[e.g.][]{Hu2014}.

\section{Moving-mesh techniques}

\subsection{Differences between Eulerian and Lagrangian techniques}

It has become clear over recent years that both Lagrangian SPH and
Eulerian AMR techniques suffer from fundamental limitations that make
them inaccurate in certain regimes.  Indeed, these methods sometimes
yield conflicting results even for basic calculations that only
consider non-radiative hydrodynamics \citep[e.g.][]{Frenk1999,
  Agertz2007, Tasker2008, Mitchell2008}.  SPH codes have comparatively
poor shock resolution, offer only low-order accuracy for the treatment
of contact discontinuities, and suffer from subsonic velocity
noise. Worse, they appear to suppress fluid instabilities under
certain conditions, as a result of a spurious surface tension and
inaccurate gradient estimates across density jumps. On the other hand,
Eulerian codes are not free of fundamental problems either.  They do
not produce Galilean-invariant results, which can make their accuracy
sensitive to the presence of bulk velocities
\citep[e.g.][]{Wadsley2008,Tasker2008}.  Another concern lies in the
preference of certain spatial directions in Eulerian hydrodynamics,
which can make poorly resolved disk galaxies align with the coordinate
planes \citep{Dubois2014}.

There is hence substantial motivation to search for new hydrodynamical
methods that improve on these weaknesses of the SPH and AMR
techniques. In particular, we would like to retain the accuracy of
mesh-based hydrodynamical methods (for which decades of experience
have been accumulated in computational fluid dynamics), while at the
same time we would like to outfit them with the Galilean-invariance
and geometric flexibility that is characteristic of SPH. The principal
idea for achieving such a synthesis is to allow the mesh to move with
the flow itself. This is in principle an obvious and old idea
\citep{Braun1995, Gnedin1995, Whitehurst1995, Mavripilis1997, Xu1997,
  Hassan1998, Pen1998, Trac2004}, but one fraught with many practical
difficulties.  In particular, mesh tangling (manifested in `bow-tie'
cells and hourglass like mesh motions) is the traditional problem of
such attempts to simulate multi-dimensional hydrodynamics in a
Lagrangian fashion.

\subsection{Voronoi tessellations}

We here briefly describe a new formulation of continuum hydrodynamics
based on an unstructured mesh that overcomes many of these problems
\citep{Springel2010}. The mesh is defined as the Voronoi tessellation
of a set of discrete mesh-generating points, which are in principle
allowed to move freely.  For the given set of points, the Voronoi
tessellation of space consists of non-overlapping cells around each of
the sites such that each cell contains the region of space closer to
it than to any of the other sites.  Closely related to the Voronoi
tessellation is the Delaunay tessellation, the topological dual of the
Voronoi diagram. Both constructions can also be used for natural
neighbor interpolation and geometric analysis of cosmic structures
\citep[e.g.][]{Weygaert1994, Pelupessy2003, vandeWeygaert2009}. In 2D,
the Delaunay tessellation for a given set of points is a triangulation
of the plane, where the points serve as vertices of the triangles.
The defining property of the Delaunay triangulation is that each
circumcircle around one of the triangles of the tessellation is not
allowed to contain any of the other mesh-generating points in its
interior. This empty circumcircle property distinguishes the Delaunay
triangulation from the many other triangulations of the plane that are
possible for the point set, and in fact uniquely determines the
triangulation for points in general position.  Similarly, in three
dimensions, the Delaunay tessellation is formed by tetrahedra that are
not allowed to contain any of the points inside their circumspheres.

\begin{figure}
\sidecaption
\resizebox{7.5cm}{!}{\includegraphics{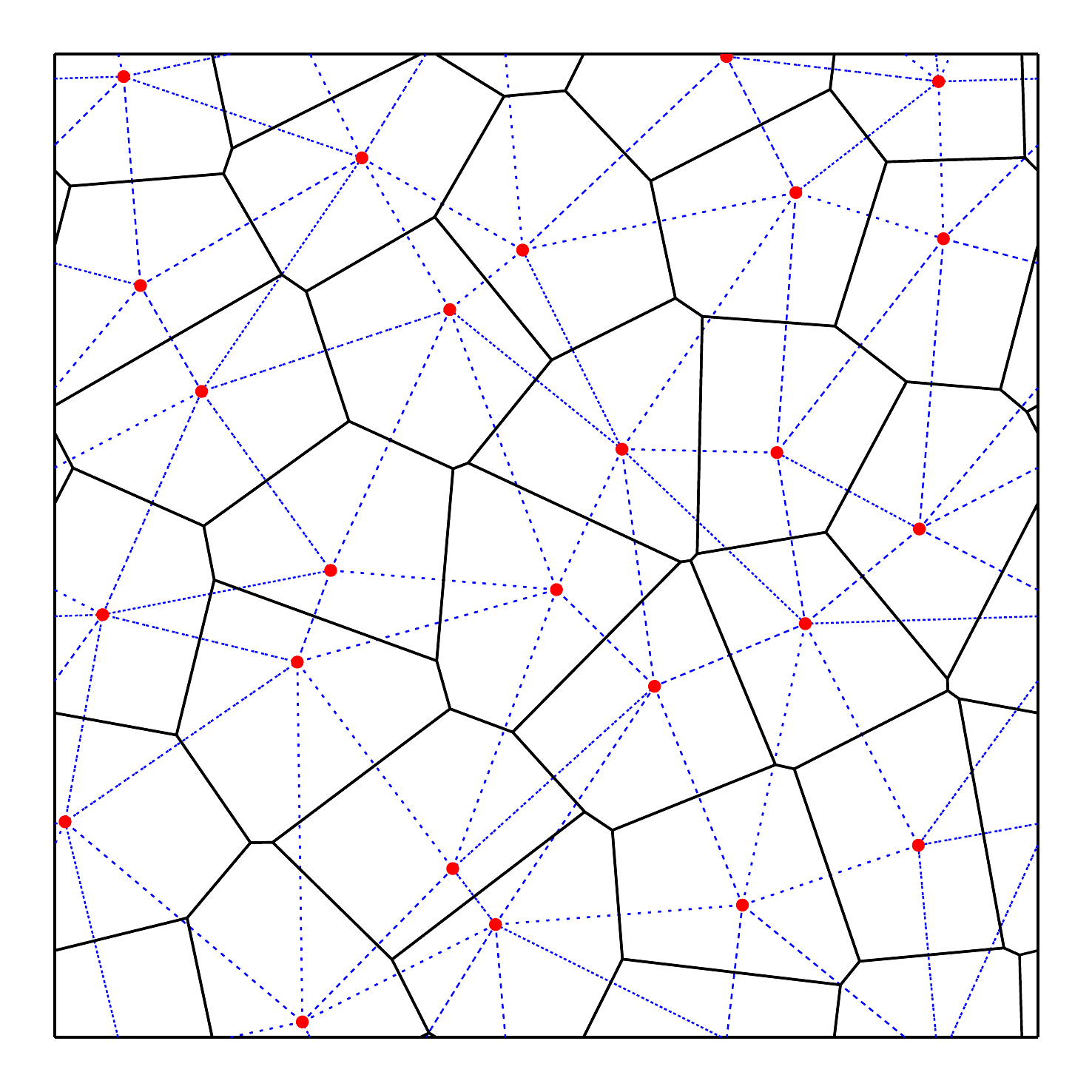}}
\caption{Example of a Voronoi and Delaunay tessellation in 2D, with periodic
  boundary conditions. The red circles show the generating points of the
  Voronoi tessellation, which is drawn with solid lines. Its topological dual,
  the Delaunay triangulation, is overlaid with thin dashed lines.
  \label{FigVoronoiExample}}
\end{figure}

As an example, Figure~\ref{FigVoronoiExample} shows the Delaunay and
Voronoi tessellations for a small set of points in 2D, enclosed in a
box with imposed periodic boundary conditions. The midpoints of the
circumcircles around each Delaunay triangle form the vertices of the
Voronoi cells, and for each line in the Delaunay diagram, there is an
orthogonal face in the Voronoi tessellation.

The Voronoi cells can be used as control volumes for a finite-volume
formulation of hydrodynamics, using the same principal ideas for
reconstruction, evolution and averaging (REA) steps that we have
discussed earlier in the context of Eulerian techniques. However, as
we will discuss below it is possible to consistently include the mesh
motion in the formulation of the numerical steps, allowing the
REA-scheme to become Galilean-invariant.  Even more importantly, due
to the mathematical properties of the Voronoi tessellation, the mesh
continuously deforms and changes its topology as a result of the point
motion, without ever leading to the dreaded mesh-tangling effects that
are the curse of traditional moving mesh methods.

\subsection{Finite volume hydrodynamics on a moving mesh}
\label{SecHydro}

As already discussed earlier in section~\ref{SecEulerDisc}, the Euler
equations are conservation laws for mass, momentum and energy that
take the form of a system of hyperbolic partial differential
equation. They can be written in the compact form
\begin{equation}
\frac{\partial\vec{U}}{\partial t} +
\vec{\nabla}\cdot \vec{F} = 0,  \label{EqnEuler}
\end{equation}  
which emphasizes their character as conservation laws.

Following the \emph{finite-volume} strategy, we describe the fluid's
state by the cell-averages of the conserved quantities for these
cells. In particular, integrating the fluid over the volume $V_i$ of
cell $i$, we can define the total mass $m_i$, momentum $p_i$ and
energy $E_i$ contained in the cell as follows,
\begin{equation}
\vec{Q}_i = \left(
\begin{array}{c}
m_i\\
\vec{p}_i\\
E_i 
\end{array}
\right)
= \int_{V_i} \vec{U}\,{\rm d} V .
\end{equation} 
With the help of the Euler equations, we can calculate the rate of
change of $\vec{Q}_i$ in time. Converting the volume integral over the
flux divergence into a surface integral over the cell results in
\begin{equation}
\frac{{\rm d}\vec{Q}_i}{{\rm d}t}
= -\int_{\partial V_i} \left[ \vec{F}(\vec{U}) - \vec{U}
\vec{w}^T\right] {\rm d}\vec{n} .
\label{EqQevol}
\end{equation} 
Here $\vec{n}$ is an outward normal vector of the cell surface, and
$\vec{w}$ is the velocity with which each point of the boundary of the
cell moves. In Eulerian codes, the mesh is taken to be static, so that
$\vec{w}=0$, while in a fully Lagrangian approach, the surface would
move at every point with the local flow velocity,
i.e.~$\vec{w}=\vec{v}$. In this case, the right hand side of
equation~(\ref{EqQevol}) formally simplifies, because then the first
component of $\vec{Q}_i$, the mass, stays fixed for each
cell. Unfortunately, it is normally not possible to follow the
distortions of the shapes of fluid volumes exactly in
multi-dimensional flows for a reasonably long time, or in other words,
one cannot guarantee the condition $\vec{w}=\vec{v}$ over the entire
surface. In this case, one needs to use the general formula of
equation~(\ref{EqQevol}). As an aside, we note that one conceptual
problem of SPH is that these surface fluxes due to the $\vec{w}$-term
are always ignored.

The cells of our finite volume discretization are polyhedra with flat
polygonal faces (or lines in 2D). Let $\vec{A}_{ij}$ describe the
oriented area of the face between cells $i$ and $j$ (pointing from $i$
to $j$).  Then we can define the averaged flux across the face $i$-$j$
as
\begin{equation}
\vec{{F}}_{ij} = \frac{1}{A_{ij}} \int_{A_{ij}} \left[ \vec{F}(\vec{U})
- \vec{U} \vec{w}^T\right] {\rm d}\vec{A}_{ij},
\end{equation}  
and the Euler equations in finite-volume form become
\begin{equation} 
\frac{{\rm d}\vec{Q}_i}{{\rm d}t} = - \sum_j A_{ij} \vec{F}_{ij}.
\end{equation}  
We obtain a manifestly conservative time discretization of this equation
by writing it as
\begin{equation} 
\vec{Q}_i^{(n+1)} = \vec{Q}_i^{(n)} - \Delta t \sum_j A_{ij}
\vec{{\hat F}}_{ij}^{(n+1/2)}, 
\label{eqnupdate}
\end{equation} 
where the $\vec{{\hat F}}_{ij}$ are now an appropriately time-averaged
approximation to the true flux $\vec{F}_{ij}$ across the cell face.
The notation $\vec{Q}_i^{(n)}$ is meant to describe the state of the
system at step $n$.  Note that $\vec{{\hat F}}_{ij}$ = $-\vec{{\hat
    F}}_{ji}$, i.e.~the discretization is manifestly conservative.

\begin{figure}[t]
\centering
\resizebox{11cm}{!}{\includegraphics{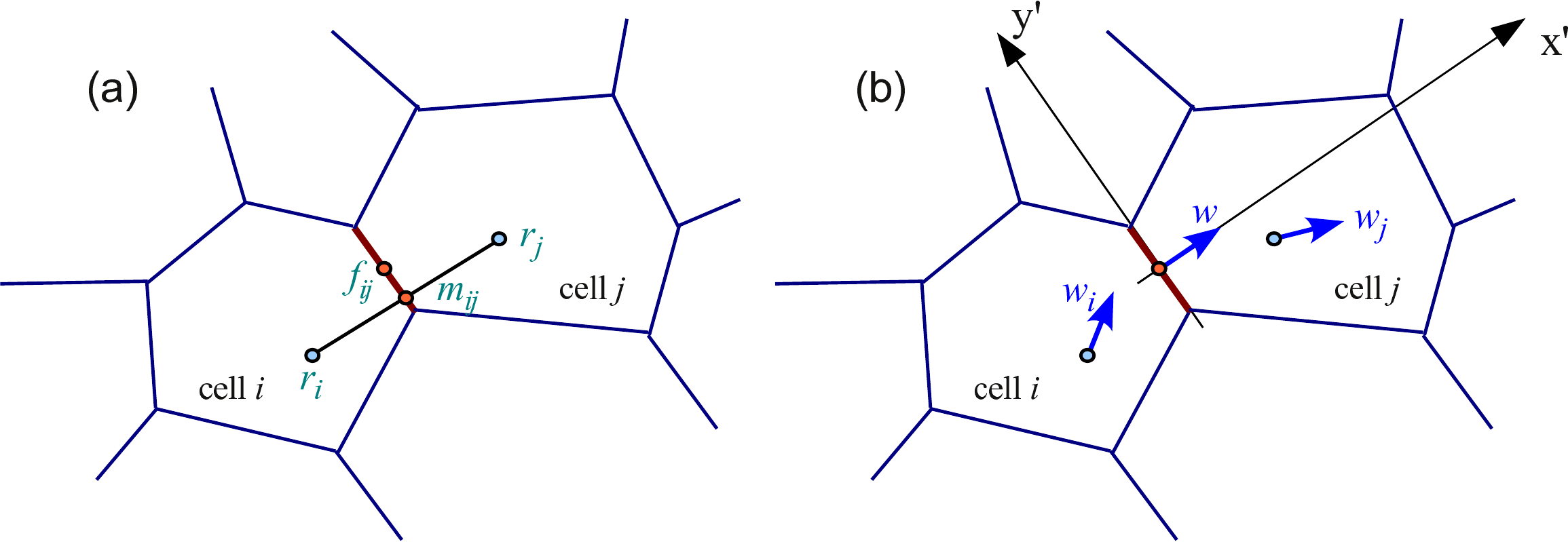}}
\caption{Sketch of a Voronoi mesh and the relevant geometric quantities that
  enter the flux calculation across a face. In (a), we show the the
  mesh-generating points $\vec{r}_{i}$ and $\vec{r}_{j}$ of two cells $i$ and
  $j$. The face between these two cells has a center-of-mass vector
  $\vec{f}_{ij}$, which in general will be offset from the mid-point $m_{ij}$
  of the two points. In (b), we illustrate the two velocity vectors $\vec{w}_{i}$
  and $\vec{w}_{j}$ associated with the mesh-generating points. These are normally
  chosen equal to the gas velocity in the cells, but other choices are allowed
  too. The motion of the mesh-generating points uniquely determines the motion
  of the face between the cells. Only the normal velocity $\vec{w}$ is however
  needed for the flux computation in the rotated frame $x'$, $y'$.
  \label{FigFluxSketch}}
\end{figure}

Evidently, a crucial step lies in obtaining a numerical estimate of
the fluxes $\vec{{\hat F}}_{ij}$.  We employ the MUSCL-Hancock scheme
\citep{Leer1984,Toro1997,Leer2006} already discussed in
section~\ref{SecHighOrderExt}, which is a well-known and relatively
simple approach for obtaining second-order accuracy in space and
time. This scheme is used in several state-of-the art Eulerian codes
\citep[e.g.][]{Fromang2006,Mignone2007,Cunningham2007}. In its basic
form, the MUSCL-Hancock scheme involves a slope-limited piece-wise
linear reconstruction step within each cell, a first order prediction
step for the evolution over half a timestep, and finally a Riemann
solver to estimate the time-averaged inter-cell fluxes for the
timestep. After the fluxes have been applied to each cell, a new
averaged state of the cells is constructed.  This sequence of steps in
a timestep hence follows the general REA approach.

Figure~\ref{FigFluxSketch} gives a sketch of the geometry involved in
estimating the flux across the face between two Voronoi cells. Truly
multidimensional Riemann solvers have been developed recently
\citep{Wendroff1999,Brio2001,Balsara2010}, but it is unclear whether
they can be readily adapted to our complicated face geometry. We
therefore follow the common approach and calculate the flux for each
face separately, treating it as an effectively one-dimensional
problem. Since we do not work with Cartesian meshes, we cannot use
operating splitting to deal with the individual spatial dimensions,
hence we apply an {\em unsplit} approach.  For defining the Riemann
problem normal to a cell face, we rotate the fluid state into a
suitable coordinate system with the $x'$-axis normal to the cell face
(see sketch).  This defines the left and right states across the face,
which we pass to an (exact) Riemann solver, following
\citet{Toro1997}.  Once the flux has been calculated with the Riemann
solver, we transform it back to the lab frame.  In order to obtain
Galilean-invariance and stable upwind behavior, the Riemann problem
needs to be solved {\em in the frame of the moving face}.

In the moving-mesh hydrodynamical scheme implemented in the {\small
  AREPO}
\footnote{Named after the enigmatic word {\small AREPO}
in the Latin palindromic sentence \emph{sator
arepo tenet opera rotas}, the `Sator Square'.} 
code \citep{Springel2010}, each timestep hence involves the following basic steps:
\begin{enumerate}
\item Calculate a new Voronoi tessellation based on the current
  coordinates $\vec{r}_i$ of the mesh generating points. This also
  gives the centers-of-mass $\vec{s}_i$ of each cell, their volumes
  $V_i$, as well as the areas $A_{ij}$ and centers $\vec{f}_{ij}$ of
  all faces between cells.
\item Based on the vector of conserved fluid variables $\vec{Q}_i$
  associated with each cell, calculate the `primitive' fluid variables
  $\vec{W}_i=(\rho_i, \vec{v}_i, P_i)$ for each cell.
\item Estimate the gradients of the density, of each of the velocity
  components, and of the pressure in each cell, and apply a
  slope-limiting procedure to avoid overshoots and the introduction of
  new extrema.
\item Assign velocities $\vec{w}_i$ to the mesh generating points.
\item Evaluate the Courant criterion and determine a suitable timestep size
  $\Delta t$.
\item For each Voronoi face, compute the flux $\vec{{\hat F}}_{ij}$ across it
  by first determining the left and right states at the midpoint of the face
  by linear extrapolation from the cell midpoints, and by predicting these
  states forward in time by half a timestep. Solve the Riemann problem in a
  rotated frame that is moving with the speed of the face, and transform the
  result back into the lab-frame.
\item For each cell, update its conserved quantities with the total flux over
  its surface multiplied by the timestep, using equation
  (\ref{eqnupdate}). This yields the new state vectors $\vec{Q}_i^{(n+1)}$ of
  the conserved variables at the end of the timestep.
\item Move the mesh-generating points with their assigned velocities
  for this timestep.
\end{enumerate}

Full details for each of these different steps as well as test
problems can be found in \citet{Springel2010}. Recently, a number of
science applications involving fairly large calculations with {\small
  AREPO} have been carried out that demonstrate the practical
advantages of this technique for applications in galaxy formation and
evolution
\citep[e.g.][]{Greif2011a,Greif2011b,Vogelsberger2012,
Vogelsberger2013, Vogelsberger2014, Marinacci2014, Pakmor2014}.

\section{Parallelization techniques and current computing trends}

Modern computer architectures offer ever more computational power that
we ideally would like to use to their full extent for scientific
purposes, in particular for simulations in astrophysics. However,
unlike in the past, the speed of individual compute cores, which may
be viewed as serial computers, has recently hardly grown any more (in
stark contrast to the evolution a few years back). Instead, the number
of cores on large supercomputers has started to increase
exponentially. Even on laptops and cell-phones, multi-core computers
have become the norm rather than the exception.

However, most algorithms and computer languages are constructed around
the concepts of a serial computer, in which a stream of operations is
executed sequentially. This is also how we typically think when we
write computer code. In order to exploit the power available in modern
computers, one needs to change this approach and adopt parallel
computing techniques. Due to the large variety of computer hardware,
and the many different technical concepts for devising parallel
programs, we can only scratch the surface here and make a few basic
remarks about parallelization, and some basic techniques that are
currently in wide use for it. The interested student is encouraged to
read more about this in textbooks and/or in online resources.

\subsection{Hardware overview}

Let's start first with a schematic overview over some of the main
characteristics and types of current computer architectures.

\subsubsection{Serial computer}
\begin{figure}
\sidecaption
\resizebox{4cm}{!}{\includegraphics{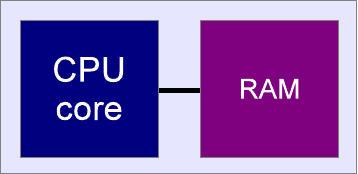}}
\caption{Simple schematic sketch of a serial computer -- most 
traditional computer languages are formulated for this type of
machine.} 
\label{fig_serial}
\end{figure}

The traditional model of a computer consists of a central processing
unit (CPU), capable of executing a sequential stream of load, store,
and compute operations, attached to a random access memory (RAM) used
for data storage, as sketched in Fig.~\ref{fig_serial}. Branches and
jumps in this stream are possible too, but at any given time, only one
operation is carried out. The operating system may still provide the
illusion that several programs are executed concurrently, but in this
case this is reached by time slicing the single compute resource.

Most computer languages are built around this model; they can be
viewed as a means to create the stream of serial operations in a
convenient way. One can in principle also by-pass the computer
language and write down the machine instructions directly (assembler
programming), but fortunately, modern compilers make this unnecessary
in scientific applications (except perhaps in very special
circumstances where extreme performance tuning is desired).

\subsubsection{Multi-core nodes}

\begin{figure}
\sidecaption
\resizebox{4.2cm}{!}{\includegraphics{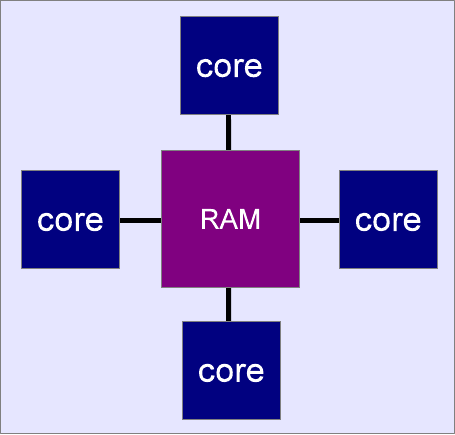}}
\caption{Multi-core computer with shared memory.}
\label{fig_multicore}
\end{figure}

It is possible to attach multiple CPUs to the same RAM (see
Fig.~\ref{fig_multicore}), and, especially in recent times, computer
vendors have started to add multiple cores to individual CPUs. On each
CPU and each core of a CPU, different programs can be executed
concurrently, allowing real parallel computations.

In machines with uniform memory access, the individual cores can
access the memory with the same speed, at least in principle. In this
case the distinction between a CPU and a core can become confusing
(and is in fact superfluous at some level), because it is ambiguous
whether ``CPU'' refers to a single core, or all the cores on the same
die of silicon (it's hence best to simple speak about cores to avoid
any confusion).

\subsubsection{Multi-socket compute nodes}

\begin{figure}
\hfill\resizebox{9cm}{!}{\includegraphics{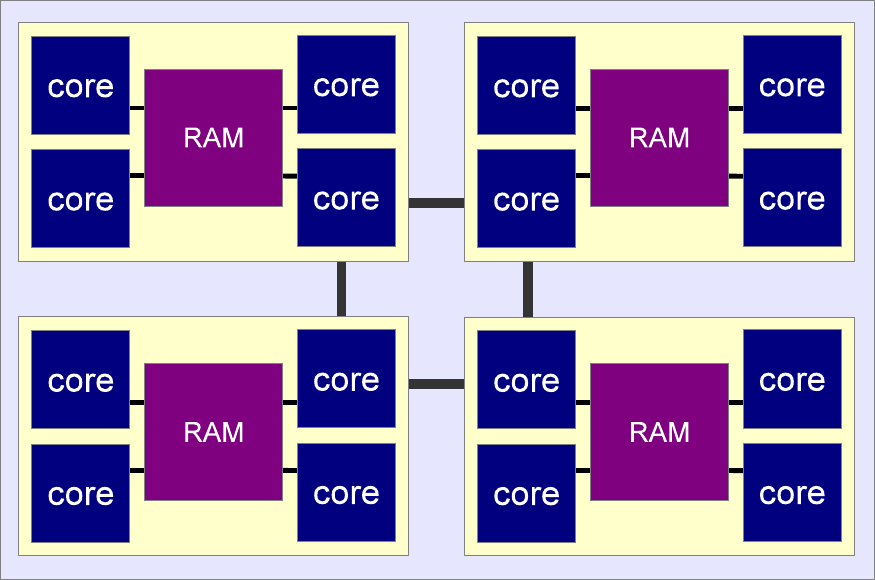}}
\caption{Non-uniform memory access computer. Here multiple sockets
  contain several processor dies, each with multiple cores. The total
  memory is split up into banks, which can be accessed with maximum speed
  by the processor associated with it, and with a reduced speed from
  different processors.}
\label{fig_numa}
\end{figure}

Most powerful compute servers feature these days a so-called NUMA
(non-uniform memory access) architecture. Here the full main memory is
accessible by all cores, but not all parts of it with the same
speed. The compute nodes usually feature individual multi-core CPUs,
each with a dedicated memory bank attached (see
Fig.~\ref{fig_numa}). Read and write operations to this part of
physical memory are fastest, while accessing the other memory banks is
typically noticeably slower and often involves going through special,
high-bandwidth interprocessor bus systems.

In such machines, maximum compute performance is only reached when the
data that is worked on by a core resides on the ``right'' memory
bank. Fortunately, the operating system will normally try to help with
this by satisfying memory requests out of the closest part of physical
memory, if possible. It is then also advantageous to tell the
operating system to ``pin'' execution of a process or thread to a
certain physical core, so that it is not rescheduled to run on another
core from which the already allocated data may be accessible only with
slower speed.

\subsubsection{Compute clusters}

\begin{figure}
\hfill\resizebox{10cm}{!}{\includegraphics{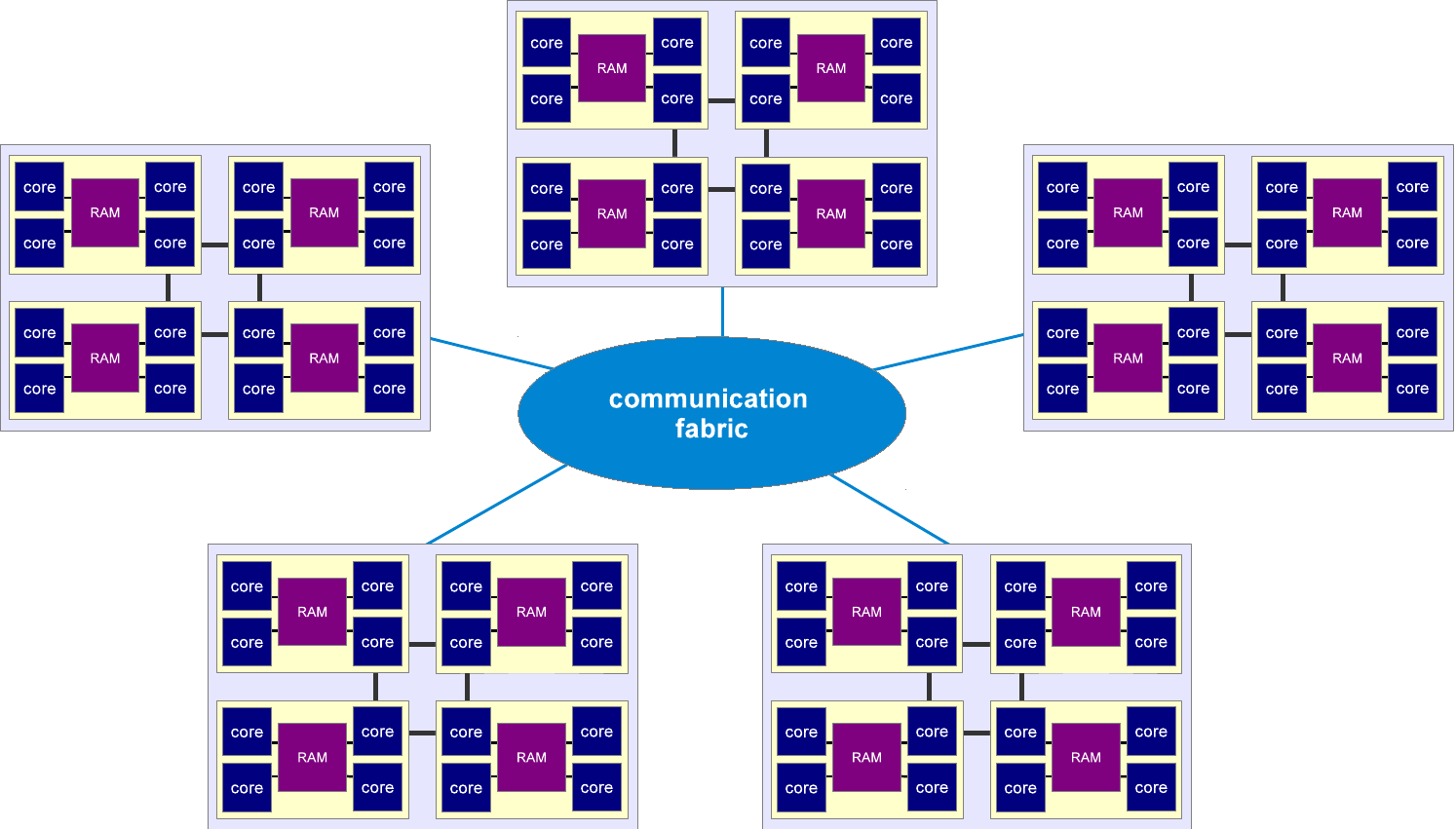}}
\caption{In large high-performance supercomputers, one typically
  connects a large number of powerful compute nodes (often of NUMA
  type) through a very fast dedicated communication network.}
\label{fig_cluster}
\end{figure}

Very powerful supercomputers used in the field of high-performance
computing (HPC) can be formed by connecting many compute nodes through
a fast communication network, as sketched in
Figure~\ref{fig_cluster}. This can be standard gigabit ethernet in
some cases, but usually much faster (and more expensive) communication
networks such as infiniband are employed. The leading supercomputers
in the world are of this type, and currently typically reach several
$10^5$ cores in total, with the first machines surpassing even $10^6$
cores. Towards the end of the decade, when exaflop machines (capable
of carrying out $10^{18}$ floating point operations per second) are
expected, this may even grow to $10^8$ or beyond. How to use these
machines efficiently for \emph{interesting} science problems, which
tend to be tightly coupled and not amenable to unlimited levels of
computational concurrency, is however still an unsolved problem.

\subsubsection{Device computing}

A comparatively new trend in scientific computing is to augment
classical compute nodes with special accelerator cards that are
particularly tuned to floating point calculations. These cards have
much simpler, less flexible compute cores, but the transistors saved
on implementing chip complexity can be spent on building more powerful
compute engines that can execute many floating point operations in
parallel. Graphics processing units (GPUs) have been originally
developed with such a design just for the vector operations necessary
to render graphics, but now their streaming processors can also be
used for general purpose calculations. For certain applications, GPUs
can be much faster than ordinary CPUs, but programming them is harder.

\begin{figure}
\sidecaption
\resizebox{7.0cm}{!}{\includegraphics{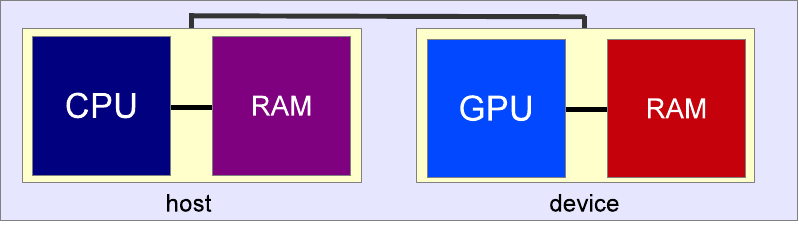}}
\caption{Hybrid compute node. An accelerator device (a GPU, or an
  Intel Xeon Phi card) is connected to an ordinary compute node through a
  fast bus system. Usually, host CPU and computing device each have
  their own RAM. }
\label{fig_gpu}
\end{figure}

In so-called hybrid compute nodes (Fig.~\ref{fig_gpu}), one has one or
several ordinary CPUs coupled to one or several GPUs, or other
accelerator cards such as the new Intel Xeon Phi. Of course, these
hybrid nodes can be clustered again with a fabric to form powerful
supercomputers. In fact, the fastest machines in the world are
presently of this type.

\subsubsection{Vector cores}

Another hardware aspect that should not be overlooked is that single
compute cores are actually increasingly capable to carry out so-called
vector instructions. Here a single instruction (such as addition,
multiplication, etc.) is applied to multiple data elements (a
vector). This is also a form of parallelization, allowing the
calculation throughput to be raised significantly. Below is an example
that calculates $x=a \cdot b$ element by element for 4-vectors $a$ and
$b$ using Intel's Advanced Vector Instructions (AVX). These can be
programmed explicitly through {\em intrinsics} in C, which are
basically individual machine instructions hidden as macros or function
calls within C. (Usually, one does not do this manually though, but
rather hopes the compiler emits such instructions somehow
automatically.)
\\

\begin{minipage}{11.0cm}
\begin{lstlisting}[frame=trBL]
  #include <xmmintrin.h>

  void do_stuff(void)
  {
    double a[4], b[4], res[4];

    __m256d x = _mm256_load_pd(a);
    __m256d y = _mm256_load_pd(b);

    x = _mm256_mul_pd(x, y);

    _mm256_store_pd(res, x);
  }
\end{lstlisting}
\end{minipage}

The current generation of the x86 processors from Intel/AMD features
SSE/AVX instructions that operate on vectors of up to 256 bits. This
means that 4 double-precision or 8 single precision operations can be
executed with such an instruction, roughly in the same speed that an
ordinary double or single-precision operation takes. So if these
instructions can be used in an optimum way, one achieves a speed-up by
a factor of up to 4 or 8, respectively. On the Intel Xeon Phi chips,
the vector length has already doubled again and is now 512 bits, hence
allowing another factor of 2 in the performance. Likely, we will see
even larger vector lengths in the near future.

\subsubsection{Hyperthreading}

A general problem in exploiting modern computer hardare to its full
capacity is that accessing main memory is very much slower than doing
a single floating point operation in a compute core (note that moving
data also costs more energy than doing a floating point calculation,
which is becoming an important consideration too). As a result, a
compute core typically spends a large fraction of its cycles waiting
for data to arrive from memory.

The idea of hyperthreading as implemented in CPUs by Intel and in
IBM's Power architecture is to use this wait time by letting the core
do some useful work in the meantime. This is achieved by
``overloading'' the compute core with several execution streams. But
instead of letting the operating system toggle between their
execution, the hardware itself can switch very rapidly between these
different ``hyperthreads''. Even though there is still some overhead
in changing the execution context from one thread to another, this
strategy can still lead to a substantial net increase in the
calculational throughput on the core. Effectively, to the operating
system and user it appears as if there are more cores (so called
virtual cores) than there are real physical cores. For example, the
IBM CPU on the Bluegene/Q machine has 16 physical cores with 4-fold
hyperthreading, yielding 64 virtual cores. One may then start 64
threads in the user application. Compared with just starting 16
threads, one will then not get four times the performance, but still
perhaps 2.1 times the performance or so, depending on the particular
application.

\subsection{Amdahl's law}

Before we discuss some elementary parallelization techniques, it is
worthwhile to point out a fundamental limit to the parallel speed-up
that may be reached for a given program. We define the speed up here
as the ratio of the total execution time without parallelization
(i.e.~when the calculation is done in serial) to the total execution
time obtained when the parallelization is enabled.
 
Suppose we have a program that we have successfully parallelized. In
practice, this parallelization is never going to be fully
perfect. Normally there are parts of the calculation that remain
serial, either for algorithmic reasons, due to technical limitations,
or we considered them unimportant enough that we have not bothered to
parallelize those too. Let us call the residual serial fraction $f_s$,
i.e.~this is the fraction of the execution time spent in the
corresponding code parts when the program is executed in ordinary
serial mode.

Then Amdahl's law \citep{Amdahl1967} gives the maximum parallel speed
up as
\begin{equation}
\mbox{maximum parallel speed up} = \frac{1}{f_s}.
\end{equation}
This is simply because in the most optimistic case we can at most
assume that our parallelization effort has been perfect, so that the
time spent in the parallel parts approaches zero for a sufficiently
large number of cores. The serial time remains unaffected by this,
however, and does not shrink at all. The lesson is a bit sobering:
Achieving large parallel speed-ups, say beyond a factor of 100 or so,
also requires extremely tiny residual serial fractions. This is
sometimes very hard to reach in practice, and for certain problems, it
may even be impossible.

\subsection{Shared memory parallelization}

Shared memory parallelization can be used to distribute a
computational work-load on the multiple available compute cores that
have access to the same memory, which is where the data resides.  UNIX
processes are {\em isolated} from each other -- they usually have
their own protected memory, preventing simple joint work on the same
memory space (data exchange requires the use of files, sockets, or
special devices such as /dev/shm). But, a process may be split up into
multiple execution paths, called {\em threads}. Such threads share all
the resources of the parent process (memory, files, etc.), and they
are the ideal vehicle for efficient shared memory parallelization.

Threads can be created and destroyed manually, e.g.~with the
pthreads-libary of the POSIX standard. Here is a simple example how
this could be achieved:
\\

\begin{minipage}{11.0cm}
\begin{lstlisting}[frame=trBL]
  #include <pthread.h>

  void do_stuff(void)
  {
    int i, threadid = 1;
    pthread_attr_r attr;
    pthread_t mythread;
    pthread_attr_init(&attr);
    pthread_create(mythread, &attr, evaluate, &threadid);

    for(i = 0; i < 100; i++)
      some_expensive_calculation(i);
  }

  void *evaluate(void *p)
  { 
    int i;

    for(i = 100; i < 200; i++)
      some_expensive_calculation(i);
  }
\end{lstlisting}
\end{minipage}

Here the two loops in lines 11/12 and 19/20 will be carried out
simultaneously (i.e.~in parallel) by two different threads of the same
parent process.  While certainly doable in principle this style of
parallel programming is a bit cumbersome if one has to do it regularly
-- fortunately, there is a convenient alternative (see below) where
much of the thread logistics is carried out by the compiler.

\subsubsection{OpenMP and its fork-join model }

A simpler approach for shared memory programming is to use the OpenMP
standard, which is a language/compiler extension for C/C++ and
Fortran. It allows the programmer to give simple hints to the
compiler, instructing it which parts can be executed in parallel
sections. OpenMP then automatically deals with the thread creation and
destruction, and completely hides this nuisance from the
programmer. As a result, it becomes possible to parallelize a code
with minimal modifications, and the modified program can still be
compiled and executed without OpenMP as a serial code. Here is how the
example from above would like like in OpenMP:
\\

\begin{minipage}{11.0cm}
\begin{lstlisting}[frame=trBL]
  #include <omp.h>

  void do_stuff(void)
  {
    int i;

  #pragma omp parallel for
    for(i = 0; i < 200; i++)
      some_expensive_calculation(i);
  }
\end{lstlisting}
\end{minipage}

This is obviously a lot simpler. We see here an example of so-called
loop-level parallelism with OpenMP.  In practice, one simply puts a
special directive for the compiler in front of the loop.  That's
basically all. The OpenMP compiler will then automatically wake up all
available threads at the beginning of the loop (the ``fork''), it will
then distribute the loop iterations evenly onto the different threads,
and they are then executed concurrently. Finally, once all loop
iterations have completed, the threads are put to sleep again, and
only the master thread continues in serial fashion. Note that this
will only work correctly if there are {\em no dependencies} between
the different loop iterations, or in other words, the order in which
they are carried out needs to be unimportant. If everything goes well,
the loop is then executed faster by a factor close to the number of
threads.

This illustrates the central idea of OpenMP, which is to let the
programmer identify sections in a code that can be executed in
parallel and annotate these to the compiler. Whenever such a section
is encountered, the program execution is split into a number of threads
that work in a team in parallel on the work of the section. Often,
this work is a simple loop whose iterations are distributed evenly on
the team, but also more general parallel sections are possible. At the
end of the parallel section, the threads join again onto the master
thread, the team is dissolved, and serial execution resumes until the
next parallel section starts.  Normally, the number of threads used in
each parallel section is constant, but this can also be changed
through calls of OpenMP runtime library functions. In order for this
to work in practice, one has to do a few additional things:
\begin{itemize}
\item The code has to be compiled with an OpenMP capable
  compiler. This feature often needs to be enabled with a special
  switch, e.g.~with gcc, 
{\small
\begin{verbatim}
    gcc -fopenmp ...
\end{verbatim}
}
needs to be used.
\item 
For some more advanced OpenMP features accessible through calls of
OpenMP-library functions, one should include the OpenMP header file
{\small
\begin{verbatim}
    #include <omp.h>
\end{verbatim}}

\item In order to set the number of threads that are used, one should
  set the {\tt OMP \_NUM\_THREADS} environment variable before the
  program is started. Depending on the shell that is used (here bash
  is assumed), this can be done for example through
{\small
\begin{verbatim}
    export OMP_NUM_THREADS=8
\end{verbatim}
}
  in which case 8 threads would be allocated by OpenMP. Normally one
  should then also have at least eight (virtual) cores available.  The
  {\tt omp\_get\_num\_} \\ {\tt threads()} function call can be used inside a
  program to check how many threads are available.

\end{itemize}

\subsubsection{Race conditions}

When OpenMP is used, one can easily create hideous bugs if different
threads are allowed to modify the same variable at the same
time. Which thread wins the ``race'' and gets to modify a variable
first is essentially undetermined in OpenMP (note that the exact
timings on a compute core can vary stochastically due to ``timing
noise'' originating in interruptions from the operating system), so
that subsequent executions may each time yield a different result and
seemingly produce non-deterministic behavior.  A simple example for
incorrect code with this problem is the following double-loop:
\\

\begin{minipage}{11.0cm}
\begin{lstlisting}[frame=trBL]
    int i, j;
  #pragma omp parallel for 
    for(i = 0; i < N; i++)
      {
        for(j = 0; j < N; j++)
          {
            do_stuff(i, j);
          }
      }
\end{lstlisting}
\end{minipage}

Here the simple OpenMP directive in the outer loop will instruct the
{\tt i}-loop to be split up between different threads. However, there
is only one variable for {\tt j}, {\em shared} by all the
threads. They are hence not able to carry out the inner loop
independent from each other! What is needed here is that each thread
gets its own copy of {\tt j}, so that the inner loop can be executed
independently. This can be achieved by either adding a {\tt private(j)}
clause to the OpenMP directive of the outer loop, like this:
\\

\begin{minipage}{11.0cm}
\begin{lstlisting}[frame=trBL]
    int i;
  #pragma omp parallel for private(j)
    for(i = 0; i < N; i++)
      {
        for(j = 0; j < N; j++)
          {
            do_stuff(i, j);
          }
      }
\end{lstlisting}
\end{minipage}

\noindent or by exploiting the scoping rules of C for the variable {\tt j},
declaring it in the loop body:
\\

\begin{minipage}{11.0cm}
\begin{lstlisting}[frame=trBL]
    int i;
  #pragma omp parallel for
    for(i = 0; i < N; i++)
      {
        int j;
        for(j = 0; j < N; j++)
          {
            do_stuff(i, j);
          }
      }
\end{lstlisting}
\end{minipage}

\subsubsection{Reductions}

Another common construct in code are reductions that build, e.g.,
large sums or products, such as attempted incorrectly in this example:
\\

\begin{minipage}{11.0cm}
\begin{lstlisting}[frame=trBL]
     int count = 0;
  #pragma omp parallel for 
     for(i = 0; i < N; i++)
       {
         if(complicated_calculation(i) > 0)
            count++;
       }
\end{lstlisting}
\end{minipage}

Again, even though here the loop is nicely parallelized by OpenMP, we
may nevertheless get an incorrect result for {\tt count}. This is
because the increment of this variable is not necessarily carried out
as a single instruction. It basically involves a read from memory, an
addition of 1, and a write back. If now two threads happen to arrive
at this statement at essentially the same time, they will both read
{\tt count}, increment it, and then write it back. But in this case
the variable will end up being incremented only by one unit and not by
two, because one of the threads is ignorant of the change of {\tt
  count} by the other and overwrites it. We have here another example
for a race conditions.

There are different solutions to this problem. One is to serialize the
increment of {\tt count} by putting a so-called lock around it. Since
we here have a simple increment of a variable, a particularly fast look -- a
so-called atomic instruction -- is possible. This can be done
through:
\\

\begin{minipage}{11.0cm}
\begin{lstlisting}[frame=trBL]
     int count = 0;
  #pragma omp parallel for 
     for(i = 0; i < N; i++)
       {
         if(complicated_calculation(i) > 0)
           {
  #pragma omp atomic
             count++;
           }
       }
\end{lstlisting}
\end{minipage}

But this can still cost substantial performance: If one or several
threads arrive at the statement protected by the atomic lock at the
same time, they have to wait and do the operation one after the other.

A better solution would be to have private variables for {\tt count}
for each thread, and only at the end of the parallel section 
add up the different copies to get the global sum. OpenMP can generate
the required code automatically, all that is needed is to add the
clause {\tt reduction(+:count)} to the directive for parallelizing the
loop:
\\

\begin{minipage}{11.0cm}
\begin{lstlisting}[frame=trBL]
     int count = 0;
  #pragma omp parallel for reduction(+:count)
     for(i = 0; i < N; i++)
       {
         if(complicated_calculation(i) > 0)
            count++;
       }
\end{lstlisting}
\end{minipage}

This shall suffice for giving a flavor of the style and the concepts of
OpenMP. A more detailed description of the OpenMP standard can for
example be found in various textbooks, and good online
tutorials\footnote{{\tt https://computing.llnl.gov/tutorials/openMP}}.

\subsection{Distributed memory parallelization with MPI}

To use multiple nodes in compute clusters, OpenMP is not
sufficient. Here one either has to use special programming languages
that directly support distributed memory models (for example UPC,
Co-Array Fortran, or Chapel), or one turns to the ``Message Passing
Interface'' (MPI). MPI has become the de-facto standard for
programming large-scale simulation code.

MPI offers library functions for exchanging messages between different
processes running on the same or different compute nodes. The compute
nodes do not necessarily have to be physically close, in principle
they can also be loosely connected over the internet (although for
tightly coupled problems the message latency makes this unattractive).
Most of the time, the same program is executed on all compute cores
(SPMD, ``single program multiple data''), but they operate on
different data such that the computational task is put onto many
shoulders and a parallel speed up is achieved.  Since the MPI
processes are isolated from each other, all data exchanges necessary
for the computations have to be explicitly programmed -- this makes
this approach much harder than, e.g., OpenMP. Often substantial
program modifications and algorithmic changes are needed for MPI.

Once a program has been parallelized with MPI, it may also be
augmented with OpenMP. Such hybrid parallel code may then be executed
in different ways on a cluster. For example, if one has two compute
nodes with 8 cores each, one could run the program with 16 MPI tasks,
or with 2 MPI tasks that each using 8 OpenMP threads, or with 4 MPI
tasks and 4 OpenMP threads each. It would not make sense to use 1 MPI
task and 16 OpenMP threads, however -- then only one of the two
compute nodes could be used.

\subsubsection{General structure of an MPI program}

A basic template of a simple MPI program in C looks as follows:
\\

\begin{minipage}{11.0cm}
\begin{lstlisting}[frame=trBL]
  #include <mpi.h>

  int main(int argc, char **argv)
  {
     MPI_Init(&argc, &argv);
      . 
      .
     /* now we can send/receive message to other MPI ranks */
      .
      .
     MPI_Finalize();
  }
\end{lstlisting}
\end{minipage}

\begin{itemize}
\item To compile this program, one would normally use a compiler
  wrapper, for example {\tt mpicc} instead of {\tt cc}, which 
  sets pathnames correctly such that the MPI header files and MPI
  library files are found by the compiler.
\item For executing the MPI program, one will normally use a start-up
  program such as {\tt mpirun} or {\tt mpiexec}. For example, the
  command
\begin{verbatim}
  mpirun -np 8 ./mycalc
\end{verbatim}
could be used to launch 8 instances of the program {\tt mycalc}.
\end{itemize}

If a normal serial program is augmented by {\tt MPI\_Init} in the
above fashion, and if it is started multiple times with {\tt mpirun
  -np X}, it will simply do multiple times exactly the same thing as
the corresponding serial program (unless they somehow synchronize
their work through modifying common files). To change this behavior
and achieve non-trivial parallelism, the execution paths taken in each
copy of the program need to become different. This is normally
achieved by making it explicitly depend on the {\em rank} of the MPI
task. All the $N$ processes of an MPI program form a so-called
communicator, and they are labelled with a unique rank-id, with the
values $0$, $1$, $2$, $\ldots$, $N-1$. MPI processes can then send and
receive message from different ranks using these IDs to identify each
other.

The first thing an MPI program normally does is therefore to find out
how many MPI processes there are in the ``world'', and what the rank
of the current instance of the program is. This is done with the
function calls
\\

\begin{minipage}{11.0cm}
\begin{lstlisting}[frame=trBL]
    int NTask, ThisTask;
 
    MPI_Comm_size(MPI_COMM_WORLD, &NTask); 
    MPI_Comm_rank(MPI_COMM_WORLD, &ThisTask);   
\end{lstlisting}
\end{minipage}

The returned integers {\tt NTask} and {\tt ThisTask} then contain the
number of MPI tasks and the rank of the current one, respectively.

\subsubsection{A simple point to point message}
With this information in hand, we can then exchange simple messages
between two different MPI ranks. For example, a send of a message
consisting of 50 integers from rank 5 to rank 7 could be programmed
like this\footnote{We note that normally one would of course not
  hard-code the rank numbers like this, but rather design the
  communication such that the program can run with different numbers
  of MPI tasks.}:
\\

\begin{minipage}{11.0cm}
\begin{lstlisting}[frame=trBL]
  int data[50], result[50]

  if(ThisTask == 5)
    MPI_Send(data, 50, MPI_INT,   // buffer, size, type
             7, 12345,            // destination, message tag
             MPI_COMM_WORLD);     // communicator id

  if(ThisTask == 7)
    MPI_Recv(result, 50, MPI_INT, // buffer, size, type
             5, 12345,            // destination, message tag
             MPI_COMM_WORLD, MPI_STATUS_IGNORE); //id, status
\end{lstlisting}
\end{minipage}

Here one sees the general structure of most send/recv calls, which
always decompose a message into an ``envelope'' and the ``data''. The
envelope describes the rank-id of sender/receiver, the size and type
of the message, and a user-specified tag (this is the `12345' here),
which can be used to distinguish messages of the same length.

Through the if-statements that depend on the local MPI rank, different
execution paths for sender and receiver are achieved in this
example. Note that if something goes wrong here, for example an MPI
rank posts a receive but the matching send does not occur, the program
will deadlock, where one or several of the MPI tasks gets stuck in
waiting in vain for messages that are not sent. This is one of the
many new types of bugs one has to cope with in distributed parallel
programs.

It is also possible to make MPI communications non-blocking and
achieve asynchronous communication in this way. The MPI-2 standard
even contains some calls for one-sided communication operations that
do not always require direct involvement of both the sending and
receiving sides.

\subsubsection{Collective communications}

The MPI standard knowns a large number of functions that can be used
to conveniently make use of commonly encountered communication
patterns. For example, there are calls for {\em broadcasts} which send
the same data to all other MPI tasks in the same communicator. There
are also {\em gather} and {\em scatter} operations that collect data
elements from all tasks, or distribute them as disjoint sets to the
other tasks. Finally, there are {\em reduction} function that allow
one to conveniently calculate sums, minima, maxima, etc., over
variables held by all MPI tasks in a communicator.

A detailed description of all these possibilities is way passed the
scope of these brief lecture notes. Please check out a textbook
\citep[e.g.][]{Pacheco1997} or some of the online
resources\footnote{For example: {\tt
    https://computing.llnl.gov/tutorials/mpi}} on MPI if you want to
get detailed information about MPI and start to program in it.

\bibliographystyle{apj}

\bibliography{notes_springel.bib}

\end{document}